\documentclass[review,12pt]{elsarticle}
\usepackage{graphicx}
\usepackage{amssymb}
\graphicspath{{figures/}}
\journal{Microelectronic-Engineering}
\begin{document}
\begin{frontmatter}
%
% title
%
\title{Focused electron beam induced deposition meets materials science}
%
% authors
%
\author{M.\ Huth\corref{cor1}}
\ead{michael.huth@physik.uni-frankfurt.de}
\author{F.\ Porrati}
\author{O.\, V.\ Dobrovolskiy}
\address{Institute of Physics, Goethe University, Max-von-Laue-Str.\ 1, 60438 Frankfurt am Main, Germany}
%
% abstract
%
\begin{abstract}
Focused electron beam induced deposition (FEBID) is a direct-write method for the fabrication of nanostructures whose lateral resolution rivals that of advanced electron lithography but is in addition capable of creating complex three-dimensional nano-architectures. Over the last decade several new developments in FEBID and focused electron beam induced processing (FEBIP) have led to a growing number of scientific contributions in solid state physics and materials science based on FEBID-specific materials and particular shapes and arrangements of the employed nanostructures. In this review an attempt is made to give a broad overview of these developments and the resulting contributions in various research fields encompassing mesoscopic physics with nanostructured metals at low temperatures, direct-write of superconductors and nano-granular alloys or intermetallic compounds and their applications, the contributions of FEBID to the field of metamaterials, and the application of FEBID structures for sensing of force or strain, dielectric changes or magnetic stray fields. The very recent development of FEBID towards simulation-assisted growth of complex three-dimensional nano-architectures is also covered. In the review particular emphasis is laid on conceptual clarity in the description of the different developments, which is reflected in the mostly schematic nature of the presented figures, as well as in the recurring final sub-sections for each of the main topics discussing the respective ''challenges and perspectives''.
\end{abstract}
%
% keywords
%
\end{frontmatter}
%
%\linenumbers
%
% main text
%
\tableofcontents
%
% Introduction
%
\section{Introduction}
Focused electron beam induced deposition (FEBID) is a direct-write approach for the fabrication of 2D- and 3D-nanostructures \cite{Randolph2006_febid_review, Utke2008_febid_review, Huth2012_febid_review}. Over the last decade FEBID or, more generally, focused electron beam induced processing (FEBIP) has developed from a rather exotic technique employed by a small number of specialist groups for a rather limited but important selection of applications, such as mask repair \cite{Edinger2014_mask_repair}, into a highly versatile technology for various materials research areas. These comprise amorphous and polycrystalline superconductors \cite{Makise2014_ebid_superconductivity, Sengupta2015_W_ebid_superconductivity, Winhold2014_Pb_superconductor}, magnetic materials \cite{DeTeresa2016_ebid_magnetic_review}, alloys and intermetallic compounds \cite{Che2005_FEBID_FePt_holography, Winhold2011_PtSi_alloy, Porrati2012_CoPt_alloy, Porrati2013_CoSi_alloy, Shawrav2014_AuFe_alloy, Porrati2015_CoFe_precursor, Porrati2016_FeSi_alloy}, multilayer structures \cite{Porrati2016_FeSi_alloy, Porrati2017_FeCoSi_multilayer} and metamaterials in which suitable materials combinations result in a desired functionality \cite{Dobrovolskiy2015_CoPt_treatment_H2_O2, DeTeresa2016_ebid_magnetic_review}. The latter approach, in particular, has opened a new pathway to the realization of different sensor applications \cite{Schwalb2010_strain_sensing, Huth2014_diel_sensing_theory, Huth2014_sensor_ttfca, Dukic2016_afm_sensor, Moczala2017_ebid_sensor}.

In this review an attempt is made to give an overview of important milestones along this development and, in parallel, keep a critical eye on the future perspective of this field. Important issues to be addressed will be material purity and reproducibility in the targeted material properties. Following this goal setting, this review will give a brief introduction into the basics of FEBID (see \cite{Utke2008_febid_review, Utke2012_book} for a comprehensive overview) and will then, in particular, address recent developments in the fabrication of all-metal FEBID structures, discuss FEBID approaches to direct-write superconductors and provide several examples of how alloys and intermetallic compounds can be obtained by FEBIP. In two additional sections, the potential of FEBID materials for different sensor application areas will be addressed and examples will be given for the use of FEBID materials towards the realization of metamaterials. A very recent development in FEBID is complex, high-resolution 3D-structure fabrication \cite{Fowlkes2016_febid_3D_simulation}, e.\,g.\ for plasmonics \cite{Winkler2017_3D_plasmonic}. This development holds great potential which is why it will also be covered in this review. Many aspects relating to the fabrication and characterization of magnetic FEBID materials have been discussed in depth in a very recent, excellent review by De Teresa and collaborators \cite{DeTeresa2016_ebid_magnetic_review}, therefore here magnetic FEBID structures are only covered with regard to 3D variants, their combination with superconductors and their application as sensor elements.
%
% Basics of focused electron beam induced deposition
%
\section{Basics of focused electron beam induced deposition}
\begin{figure}
\centering
\includegraphics[width=0.7\textwidth]{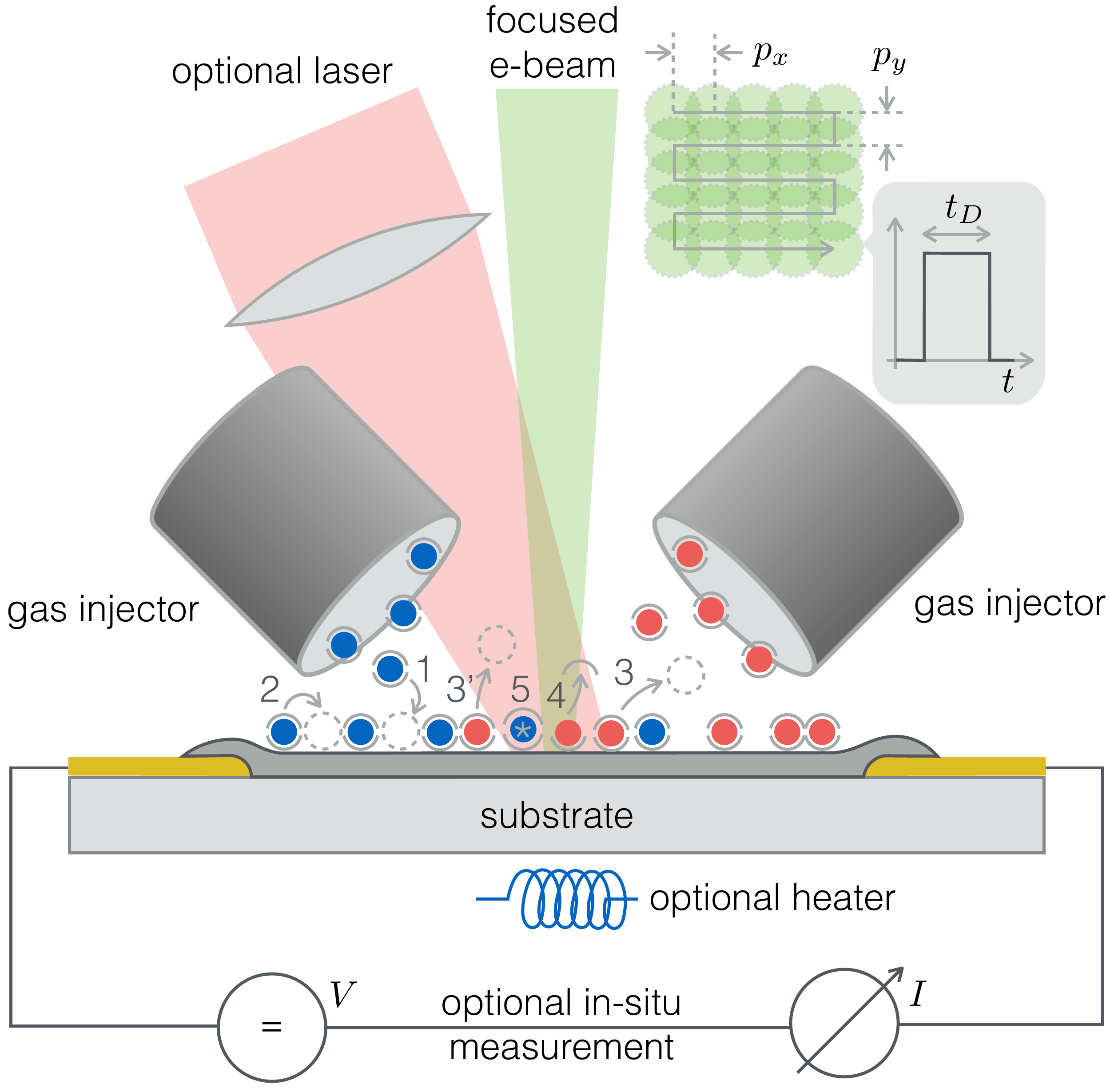}
\caption{Schematic of a SEM adapted for FEBIP with additional process equipment: (A) Steady-state or pulsed IR-laser heat supply for enhanced desorption of undesired organic dissociation products. (B) In-situ electrical characterization of the deposits by using substrate materials with pre-patterned electrodes. (C) Multi-component deposits by employing several precursor gas injection channels. Illustration of the FEBID process with two different precursor species. Precursor molecules (here: organometallic complex; blue or red: metal, gray: organic ligands) are supplied by the gas-injection system and physisorb (1) on the surface. Surface diffusion (2), thermally induced desorption (3') and electron-stimulated desorption (3) take place. Within the focus of the electron beam, adsorbed precursor molecules are dissociated followed by desorption of volatile organic ligands (4). Upper right: For pattern definition the electron beam is moved in a raster fashion (here: serpentine) over the surface and settles on each dwell point for a specified dwell time $t_D$. After one raster sequence is completed the process is repeated until a predefined number of repeated loops is reached. Neighboring dwell points have distances of $p_x$ and $p_y$ in $x$- and $y$-direction, respectively.}
\label{fig_intro_febip_scheme}
\end{figure}

The basic principle of focused electron beam induced deposition is simple. Provided by a gas injection system or an environmental chamber inside of an electron microscope, a precursor gas adsorbed on a surface is dissociated in the focus of an electron beam (see Fig.\,\ref{fig_intro_febip_scheme} for illustration). This brief description shows an apparent conceptual similarity to 3D printing, in particular if one considers the 3D writing capabilities of FEBID. Even the use of ''liquid ink'' precursors has been pioneered by the Hastings group and holds great potential for metallic nanostructure fabrication with very decent writing speeds \cite{Donev2009_Pt_liquid_precursor, Bresin2013_bimetallic_liquid_precursor}. A closer look, however, reveals the intrinsic complexity of the FEBID process. The electron-induced dissociation process is mostly triggered by low-energy electrons, i.\,e.\ the secondary electrons generated by the primary electrons (SE\,I) and also by the backscattered electrons (SE\,II). For the dissociation process several channels are available with strongly energy-dependent and precursor-specific cross-sections \cite{Thorman2015_dissociation_case_studies}. Which precursor to choose for a given application has to be carefully considered, as the growth process depends on several precursor-specific aspects, such as its vapor pressure at around room temperature, the adsorption characteristics of the precursor molecules and their stability under adsorption. By now several examples are documented of the intrinsic instability of several precursors under adsorption in dependence of the surface state of the underlying substrate \cite{Muthukumar2011_wco6_adsorption_dft, Muthukumar2012_dissociation_co2co8}. This is not to say that this instability cannot be turned into a profit. By proper pre-conditioning with the electron beam, a site-selective auto-dissociation of some precursor species, such as Fe(CO)$_5$ and Co$_2$(CO)$_8$ can be induced resulting in metallic magnetic nanostructures. Pioneered by the Marbach group this process has been termed EBISA (electron beam induced surface activation) \cite{Walz2010_ebisa}. Additional contributions to the growth characteristics relate to the surface mobility of the precursor molecules and their average residence time. As the last important criterion in precursor selection we mention the final composition of the deposits. The composition may depend strongly on the writing parameters (beam energy and current, precursor flux, writing strategy), as has often been observed for metal-carbonyl precursors, or else may be quite insensitive to variations of the writing process, such as in metal-organic complexes with cyclopentadienyl ligands. Most often the aim is to obtain deposits with high metal content which will be covered in more detail in the next section.

The practical adaption of the typically employed scanning electron microscopes (SEM) for more advanced electron beam induced processing has significantly progressed in the last couple of years, as also indicated schematically in Fig.\,\ref{fig_intro_febip_scheme}. The optional parallel use of more than one precursor gas injection channel has become routine in several laboratories. This will be discussed in more detail in a separate section on alloys and intermetallic compounds. Also, steady-state heating of the substrate \cite{Botman2009_purification_first_review} or, even more effective, pulse-heating of the substrate surface with an IR-laser synchronized to the electron beam dwell events \cite{Roberts2012_purification_pulsed_laser} have become valuable additions to the process arsenal in FEBIP with a goal to improve the metal content. This will be taken up again in the section on all-metal FEBID structures. Recently, the group of Fedorov at Georgia Tech has demonstrated that the use of a supersonic precursor gas jet can yield very high metal content deposits \cite{Henry2016_gasjet_pure_tungsten}. With a view to metallic deposits, the option for in-situ electrical characterization during or after growth has been shown to be an important added value. This is especially true with regard to semi-automatic deposition optimization approaches pioneered by us. They use the time-dependent conductance increase during growth as a parameter for feedback control of the writing process, viz.\ by suitably changing the dwell time and pitches employing a genetic algorithm \cite{Weirich2013_ga_first_pub, Winhold2014_ga_modeling}.
%
% Metallic FEBID structures
%
\section{Metallic FEBID structures}
For a wide range of application fields, such as nano-contact fabrication or mesoscopic structure preparation, the ability to obtain fully metallic deposits is in fact a condition to which FEBIP has to live up to. This is not an easy task to fulfill. In the vast majority of cases, precursor materials used in FEBID are adopted from chemical vapor deposition (CVD) and are not specifically designed to be efficiently and completely dissociated under electron irradiation. As a consequence, the elemental composition of FEBID materials very often shows substantial contributions from carbon, oxygen and other undesired dissociation products that interfere with good metallic conductivity; at least at low temperatures. Several developments over the last decade or so have brought significant progress into advancing FEBID to a state that it is now possible to get fully metallic structures of the elements Au, Pt, Fe, Co, Pb, and Co$_3$Fe alloy. These developments are (1) the optimization of the deposition processes for carbonyl precursors of Fe, Co and a hetero-nuclear precursor of Fe and Co, (2) the development of several post-growth purification protocols for Au, Pt and Co, (3) the combination of FEBID with selective-area atomic layer deposition of Pt and (4) the establishment of dual gas channel deposition techniques in which the precursor is combined with a reactive gas to directly yield metallic Au and Pt. These developments are reviewed in compact form in the following sub-sections. 
%
% As-grown metal structures
%
\subsection{As-grown metal structures}
Without doubt, the most convenient FEBID process for metal structure definition would be a single-step process. In addition, such a process should be quite forgiving of slight variations in the process parameters, such as, e.\,g.\ beam current and energy or precursor gas flow. This is, however, not the case and it is also not to be expected. The dissociation process is precursor-specific and strongly dependent on the beam energy. The residual gas contribution of the process chamber and the associated partial pressures have a strong influence on the final deposit composition. The ratio of electron and precursor flux, in conjunction with precursor residence time and diffusion coefficient determine the precursor density on the growing surface, which, in turn, determines the growth regime. Depending on the growth regime and the character of the dissociation products, the latter may be able to desorb before they are incorporated in the growing deposit or not. Because of this complex interplay of factors that influence the final composition of the deposit, it is mandatory to monitor and control as many parameters as possible in a FEBID process and to pay special attention to the reported process conditions, if an attempt is made to reproduce experiments reported in the literature.

In Tab.\,\ref{tab_precursors_metal_contents} an overview is given of the small subset of precursors so far used in FEBID for which metallic deposits can be obtained in a single-step process. By metallic we mean materials whose room temperature resistivity is clearly below the critical value set by the Mott-Ioffe-Regel criterion \cite{Hussey2004_mott_ioffe_regel}. In this case one can expect good conductivity also at low temperatures with, ideally, increasing values as the temperature is lowered. This latter point is particularly important with a view to the use of FEBID structure in mesoscopic physics for which sufficiently long dephasing times are mandatory.

\begin{table}[htb]
\small
\begin{tabular}{|l|l|l|l|l|}\hline\hline
Precursor & Max.\ metal content & Impurities & Microstructure & Ref \\ \hline \hline
Me$_3$CpMePt(IV) + O$_2$ & Pt, $\approx 100\,$at$\%$ & C & polycrystalline & \cite{Villamor2015_Pt_purification_direct_O2} \\
Me$_2$(tfac)Au(III) + H$_2$O & Au, $>90\,$at$\%$ & C, O & polycrystalline & \cite{Shawrav2016_Au_with_H2O} \\
W(CO)$_6$, gas jet & W, $95\,$at$\%$ & C, O & unknown & \cite{Henry2016_gasjet_pure_tungsten} \\
W(CO)$_6$ & W, $50\,$at$\%$ & C, O & amorphous & \cite{Sengupta2015_W_ebid_superconductivity} \\
Et$_4$Pb & Pb, $46\,$at$\%$ & C, O & granular & \cite{Winhold2014_Pb_superconductor} \\
Fe(CO)$_5$ & Fe, $76\,$at$\%$ & C, O & amorphous & \cite{Lavrijsen2011_Fe_febid_pure} \\
Fe$_2$(CO)$_9$ & Fe, $75\dots 80\,$at$\%$ & C, O & amorphous & \cite{Cordoba2016_Fe_nonacarbonyl} \\
Co$_2$(CO)$_8$ & Co, $93\,$at$\%$ & C, O & polycrystalline & \cite{Ramon2011_Co_febid_clean_highres} \\
HCo$_3$Fe(CO)$_{12}$ & Co$_3$Fe, $84\,$at$\%$ & C, O & granular & \cite{Porrati2015_CoFe_precursor} \\
AgO$_2$Me$_2$Bu & Ag, $75\,$at$\%$ & C, O & polycrystalline & \cite{Hoeflich2017_Ag_metal} \\
\hline\hline
\end{tabular}
\caption{Precursor materials for FEBID that result in metallic deposits, if process conditions for maximum metal content are used. See listed references for details.}
\label{tab_precursors_metal_contents}
\end{table}
The single-step deposition of metallic Pt by FEBID using Me$_3$CpMePt(IV) was pioneered by the gas chemistry group around Hans Mulders and Piet Trompenaars of FEI company \cite{Villamor2015_Pt_purification_direct_O2}, which was also the pacemaker for the post-growth purification activities of Pt and Au \cite{Mulders2014_review_purification, Mehendale2015_Au_purification_O2}, as is briefly reviewed in the next sub-section. By the parallel injection of molecular O$_2$ during deposition with the precursor Me$_3$CpMePt(IV), Villamor and collaborators could obtain void-free Pt deposits with resistivity values only about a factor of six larger than the bulk value of Pt \cite{Villamor2015_Pt_purification_direct_O2}. It was found that the O$_2$-to-precursor flux ratio had to be as large as $3.5\times 10^4$ to obtain the highest Pt metal content. This is in accordance with the results of other work which found that below a threshold value of the flux ratio pure Pt deposition does not occur \cite{Mehendale2015_Au_purification_O2}. These extreme flux ratios can only be realized by reducing the precursor flux to very small values while in parallel increasing the O$_2$ flow. As a result, the deposition yield is reduced to about $1.2\times 10^{-5}\,\mu$m$^3/$nC which renders this process most suitable for small deposit volumes. As is generally the case when O$_2$ is used together with Pt-deposits, the necessary dissociative chemisorption of O$_2$ and the delayed desorption of CO formed by the catalytically supported oxidation of the carbon has to be finely balanced to get the most efficient removal of unwanted carbon in the deposit (see also next sub-section). Inspired by this single-step approach, the group of Heinz Wanzenb\"ock developed a similar process for Au by employing the precursor Me$_2$(tfac)Au(III) together with water as oxidative enhancer \cite{Shawrav2016_Au_with_H2O}. In this case the water-to-precursor flux ratio was estimated to be $10$. Interestingly, the authors reported resistivity values as low as $8.8\,\mu\Omega$cm, i.\,e.\ only a factor of $4$ larger than the bulk value of Au.

A different approach was followed by Andrey Fedorov's group. By employing an inert carrier gas jet together with the precursor W(CO)$_6$ Henry and collaborators obtained in a single-step process FEBID structures with up to $95\,$at$\%$ of W \cite{Henry2016_gasjet_pure_tungsten}. So far resistivity values for these deposits have not been reported, but they are expected to be clearly in the metallic regime. Interestingly, Sengupta and co-workers found that an optimized standard FEBID process using W(CO)$_6$ can yield deposits with resistivity values just on the metallic side of the Mott-Ioffe-Regel criterion \cite{Sengupta2015_W_ebid_superconductivity}. These deposits become superconducting below $2\,$K (see section on superconductivity). In this case the metallic behavior is percolative in nature. Similar results were obtained by us using the precursor Et$_4$Pb in a single-step standard FEBID process. At metal contents up to $46\,$at$\%$ we found percolating metallic behavior with room temperature resistivity values of about $160\,\mu\Omega$cm that dropped by more than an order of magnitude under cooling, followed by a superconducting transition between $5$ and $7\,$K \cite{Winhold2014_Pb_superconductor}.

Another effective method was developed by the group of Philip Rack \cite{Roberts2012_purification_pulsed_laser}. Synchronizing laser pulses with the FEBID dwell events and optimizing the pulse duration and laser power, Roberts and collaborators obtained deposits from the precursor Me$_3$CpMePt(IV) with strongly reduced resistivity values of about $1000\,\mu\Omega$cm. The same method was also applied with the precursor W(CO)$_6$ \cite{Roberts2013_laebid_tungsten}. In both cases the improved metal content in the deposits was attributed to the enhanced by-product desorption caused by the short-time heating of the substrate surface induced by the laser pulses.

A particularly promising precursor group for metallic deposits in a single-step FEBID process are the carbonyls of Fe and Co. Apparently, an intrinsic instability of these precursor materials after loss of one or several of the carbonyl groups under electron irradiation provides favorable conditions for a rather complete dissociation and desorption of the carbonyl ligands \cite{Thorman2015_dissociation_case_studies}. In particular, the group of Jos\'e Mar\'ia De Teresa at Zaragoza developed optimized deposition protocols for the precursors Co$_2$(CO)$_8$, Fe(CO)$_5$ and Fe$_2$(CO)$_9$ resulting in metal contents well above $90\,$at$\%$ \cite{Ramon2011_Co_febid_clean_highres, Lavrijsen2011_Fe_febid_pure, Cordoba2016_Fe_nonacarbonyl}. Importantly, these protocols are suitable for high-resolution work and established a standard for FEBID nanostructure fabrication of Fe and Co for micromagnetic studies. In our work on the hetero-nuclear precursor HCo$_3$Fe(CO)$_{12}$ we found similar results, as metal contents above $80\,$at$\%$ can be obtained under beam conditions suitable for high-resolution work with resistivity values as low as $43\,\mu\Omega$cm \cite{Porrati2015_CoFe_precursor}.

As a last, very recent example with significant potential for plasmonics, we refer to work by Katja H\"oflich and collaborators \cite{Hoeflich2017_Ag_metal}. By using the precursor AgO$_2$Me$_2$Bu, which must be heated to $150^\circ$C and for which the substrate has to be kept at elevated temperature ($120^\circ$C) during the FEBID process, Ag metal contents as high as $75\,$at$\%$ could be obtained under optimized beam conditions.
%
% Post-growth purification and surface-activated growth
%
\subsection{Post-growth purification and surface-activated growth}
\label{sec_post_growth_purification}
\begin{figure}
\centering
\includegraphics[width=0.7\textwidth]{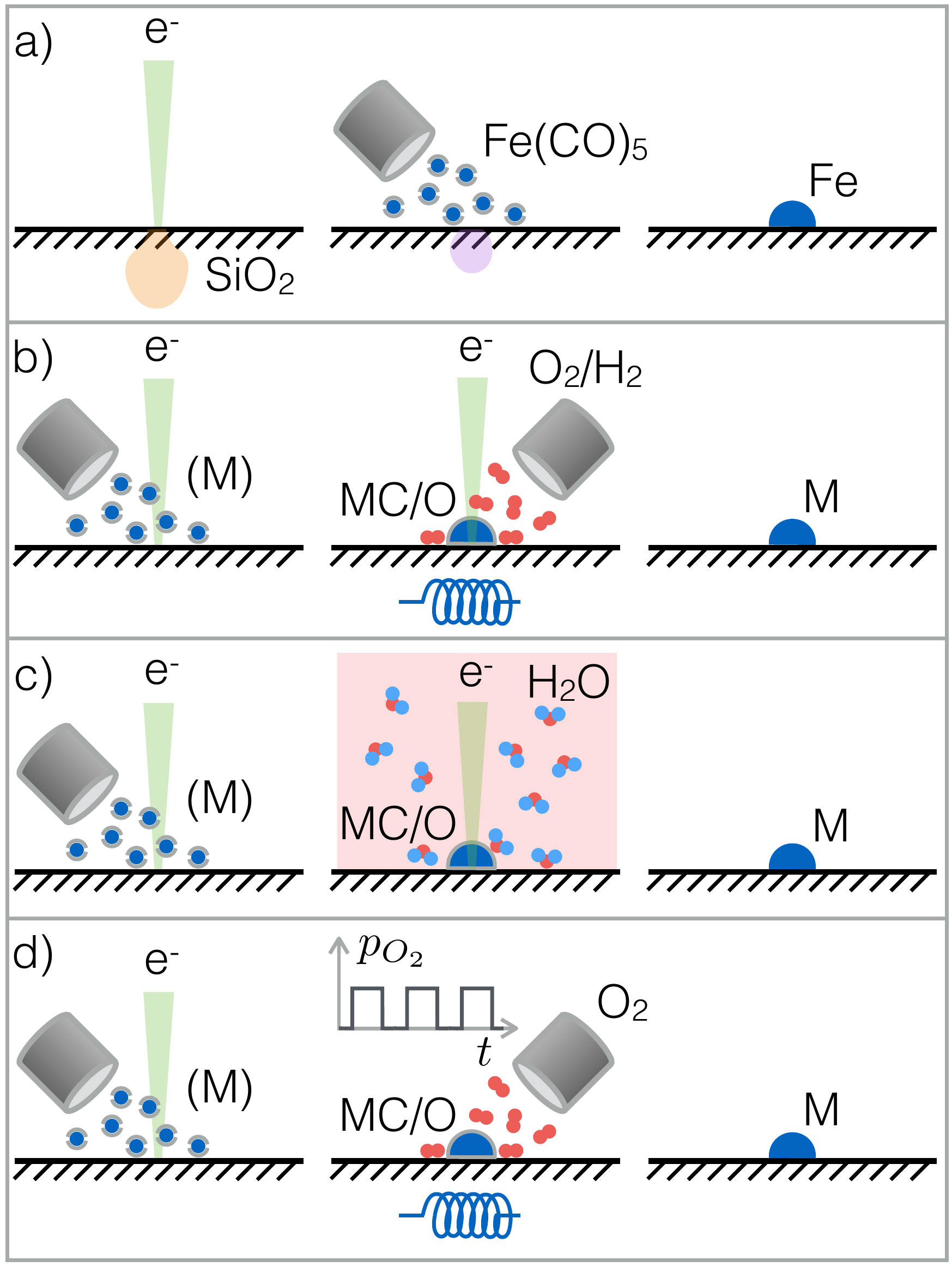}
\caption{Schematic of different post-growth purification approaches. (a) SiO$_2$ (native) is electron irradiated resulting in SiO$_2$-activation. Fe(CO)$_5$ dissociates spontaneously on top of the activated region, followed by auto-catalytic Fe growth. (b) Standard FEBID is used to obtain M = Au, Pt or Co nanostructures within a carbonaceous matrix. Electron irradiation in the presence of O$_2$ (M = Au or Pt) or H$_2$ (M = Co) leads to removal of matrix material. For M = Co in the presence of H$_2$ heating to about $250$--$300^\circ$C is necessary. (c) After a standard FEBID process to obtain M = Au or Pt, electron irradiation in an H$_2$O atmosphere inside an environmental SEM is used to purify the deposit. (d) Following standard FEBID of Pt within a carbonaceous matrix, pulsed oxygen exposure at a substrate temperature of about $150^\circ$C is used to purify the deposit without the need of electron irradiation.}
\label{fig_pgp_approaches}
\end{figure}

Deposits with improved metal content up to the level of complete removal of undesired dissociation products can also be obtained by post-growth purification. With a focus on Pt and Au, but also Fe and Co, different approaches have been developed which are schematically indicated in Fig.\,\ref{fig_pgp_approaches}. The figure also depicts an electron-induced surface activation (EBISA) approach that has been developed by Hubertus Marbach and collaborators \cite{Walz2010_ebisa, Porrati2011_Fe_nanowires_ebisa, Vollnhals2013_ebisa_Fe} and will be discussed first.
EBISA is a two-step process to grow clean metallic nanostructures, in particular Fe when using the precursor Fe(CO)$_5$  \cite{Walz2010_ebisa}. In the first step, the surface of a Si\,$(001)$ substrate with $300\,$nm thermally grown SiO$_x$ is locally activated by a highly focused electron beam whose raster pattern defines a latent image for the second step. According to Walz \emph{et al.} the surface site activation is a consequence of the formation of oxygen vacancies caused by electron-induced desorption of oxygen by a Knotek-Feibelman mechanism \cite{Walz2010_ebisa}. Following the activation, the precursor Fe(CO)$_5$ is introduced close to the substrate surface where it spontaneously decomposes at the activated sites forming Fe nano-clusters on which the growth  proceeds further by autocatalytic processes. An important prerequisite for EBISA to work is a very low background pressure in the SEM chamber requiring an ultra-high vacuum setup. If this condition is met, EBISA was shown to be also applicable for the growth of clean Fe nanostructures on TiO$_2$ (rutile phase) and porphyrin-layers on Ag\,$(111)$ \cite{Vollnhals2013_ebisa_Fe}.

We now turn to a brief description of different post-growth purification approaches for Pt-, Au- and Co-containing FEBID structures which have been developed over the last few years. As briefly reviewed by Hans Mulders in \cite{Mulders2014_review_purification}, first electron-beam stimulated purification experiments on nano-granular Pt(C) deposits from the precursor Me$_3$CpMePt(IV) have been inspired by two observations. First, it was shown by Botman and collaborators that annealing of Pt(C) deposits at temperatures above about $200^\circ$C is effective in removing the carbon matrix and resulting in structures with metal contents of up to $70\,$at$\%$ \cite{Botman2009_purification_first_review}. Second, in later work it was demonstrated that electron irradiation of Pt(C) is very effective in increasing the conductivity by up to four orders of magnitude as compared to as-grown deposits \cite{Porrati2011_PtC_irradiation, Plank2011_PtC_irradiation, Sachser2011_PtC_universal_conductance}. This was attributed to modest grain-size increases and the partial removal of carbon leading to strongly enhanced tunnel couplings between the Pt nano-crystallites. By using O$_2$, as reactive gas species, in conjunction with electron irradiation of as-grown Pt(C) FEBID structures, a complete purification to metallic Pt was demonstrated by several groups \cite{Mehendale2013_PtC_purification_O2, Plank2014_PtC_purification_O2, Lewis2015_purification_pt_oxygen_modeling}, as schematically depicted in Fig.\,\ref{fig_pgp_approaches}(b). The same oxygen-based approach was shown to be effective for Au-containing FEBID deposits \cite{Mehendale2015_Au_purification_O2}. Geier, Winkler and collaborators could show that using H$_2$O as reactive gas in an environmental SEM is very efficient in purifying Pt(C) and Au(C) FEBID structures \cite{Geier2014_Pt_purification_H2O, Winkler2017_3D_plasmonic} (see Fig.\,\ref{fig_pgp_approaches}(c)). A particular advantage of this approach is the shape-fidelity that can be reached by careful selection of suitable process conditions leading to pore-free and compact Pt and Au nanostructures. By employing H$_2$ as reactive gas we could show that oxidation-sensitive deposits with carbon and oxygen impurities, such as Co-C-O, can also be purified under post-growth irradiation \cite{Begun2015_Co_purification_H2}. For this process to work the substrate temperature has to be elevated to about $250^\circ$C. We could also demonstrate that the post-growth purification of Pt and Co can be combined to fabricate magnetic heterostructures with controlled magnetic anisotropy \cite{Dobrovolskiy2015_CoPt_treatment_H2_O2}. If a substrate heater is an available option for a SEM stage, Pt(C) purification can also be accomplished without electron irradiation. Roland Sachser and colleagues could show that the catalytic efficacy of Pt is sufficient to remove the carbon matrix in a pulsed-oxygen process starting to operate sufficiently fast at about $150^\circ$C \cite{Sachser2014_PtC_purification_pulsed_O2}. Dissociative chemisorption of O$_2$ on the Pt nano-grains facilitates the oxidation of carbon to CO. If the O$_2$ supply is periodically stopped, the formed CO can desorb from the surface. After $8$ to 12 pulse cycles with a duration of several minutes each, a complete removal of the carbon matrix is accomplished. As the pulsed-oxygen approach does not need electron-assisted activation, it can work over larger substrate surface areas. Tab.\,\ref{tab_postgrowth_purification} gives an overview of post-growth purification results obtained so far for a range of selected precursors.
\begin{table}
\small
\begin{tabular}{|l|l|l|l|l|}\hline\hline
Precursor & [Metal] (at$\%$) & Purification method & $\rho(300\,\mathrm{K})$ ($\mu\Omega$cm) & Ref \\ \hline \hline
Me$_3$CpMePt(IV) & $96$, --, -- & e-irradiation in O$_2$ & $70\pm 8$, $<350$, -- & \cite{Mehendale2013_PtC_purification_O2, Plank2014_PtC_purification_O2, Lewis2015_purification_pt_oxygen_modeling} \\
Me$_3$CpMePt(IV) & -- & e-irradiation in H$_2$O & -- & \cite{Geier2014_Pt_purification_H2O} \\
Me$_3$CpMePt(IV) & -- & pulsed O$_2$ at $150^\circ$C & $79.5$ & \cite{Sachser2014_PtC_purification_pulsed_O2} \\
Me$_2$(tfac)Au(III) & -- & e-irradiation in O$_2$ & $17\pm 2$ & \cite{Mehendale2015_Au_purification_O2} \\
Me$_2$(acac)Au(III) & $> 99$ & e-irradiation in H$_2$O & -- & \cite{Winkler2017_3D_plasmonic} \\
Co$_2$(CO)$_8$ & $85$, $92$ & e-irradiation in H$_2$ & $22.4$, -- & \cite{Begun2015_Co_purification_H2, Dobrovolskiy2015_CoPt_treatment_H2_O2} \\
\hline\hline
\end{tabular}
\caption{Selection of metallic FEBID materials obtained after applying different post-growth purification protocols. [Metal] denotes the metal content in the deposit after purification. $\rho(300\,\mathrm{K})$ refers to the room-temperature resistivity. A dash indicates that the respective values were not stated in the respective references.}
\label{tab_postgrowth_purification}
\end{table}
%
% Area-selective atomic layer deposition
%
\subsection{Area-selective atomic layer deposition}
\begin{figure}
\centering
\includegraphics[width=0.7\textwidth]{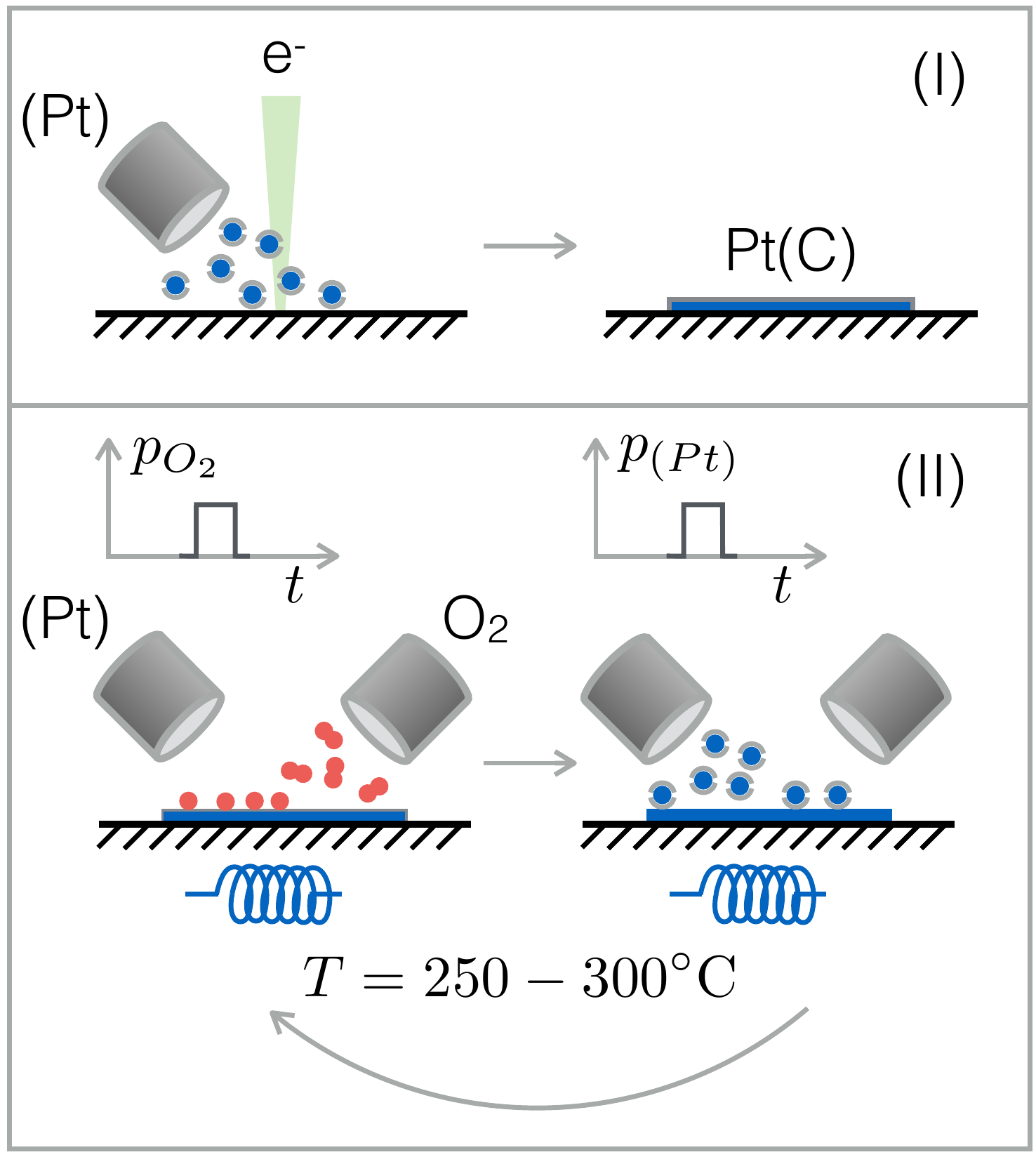}
\caption{Schematic of area-selective atomic layer deposition of Pt. By FEBID a Pt-containing seed layer is deposited (top), thus defining the lateral shape of the metallic Pt layer to be grown with ALD (bottom). (Pt) denotes the precursor Me$_3$CpMePt(IV).}
\label{fig_as_ald_principle}
\end{figure}

In 2010 Mackus and collaborators introduced a highly versatile new methodology for pure Pt thin film growth on the lateral nanoscale \cite{Mackus2010_Pt_asald_first, Mackus2012_Pt_asad_resolution, Mackus2013_Pt_asald_highres}. In area-selective atomic layer deposition (AS-ALD) the preparation of a Pt-containing seed layer with FEBID, that defines the desired lateral Pt thin film shape, is followed by the well established Pt ALD process that uses Me$_3$CpMePt(IV) as precursor and O$_2$ as reactive gas. The principle of AS-ALD is depicted in Fig.\,\ref{fig_as_ald_principle}.
In contradistinction to most other AS-ALD methods, the combinatorial FEBID/ALD approach relies on locally stimulating catalytic Pt ALD growth instead of using masking techniques for specific area deactivation. In addition, the FEBID seed layers is effective in overcoming the commonly observed nucleation problem of ALD metal layer growth. Thus, the combination of FEBID and AS-ALD merges the patterning capability of FEBID with the ability of ALD to deposit high purity and low resistive materials with good thickness control \cite{Altonen2003_Pt_ald} and room temperature resistivity values as low as $11\,\mu\Omega$cm \cite{Mackus2012_Pt_asad_resolution}. The pulsed-oxygen Pt(C) purification process described in the last sub-section is in fact closely related to the microscopic working principle of AS-ALD.

\begin{figure}
\centering
\includegraphics[width=0.7\textwidth]{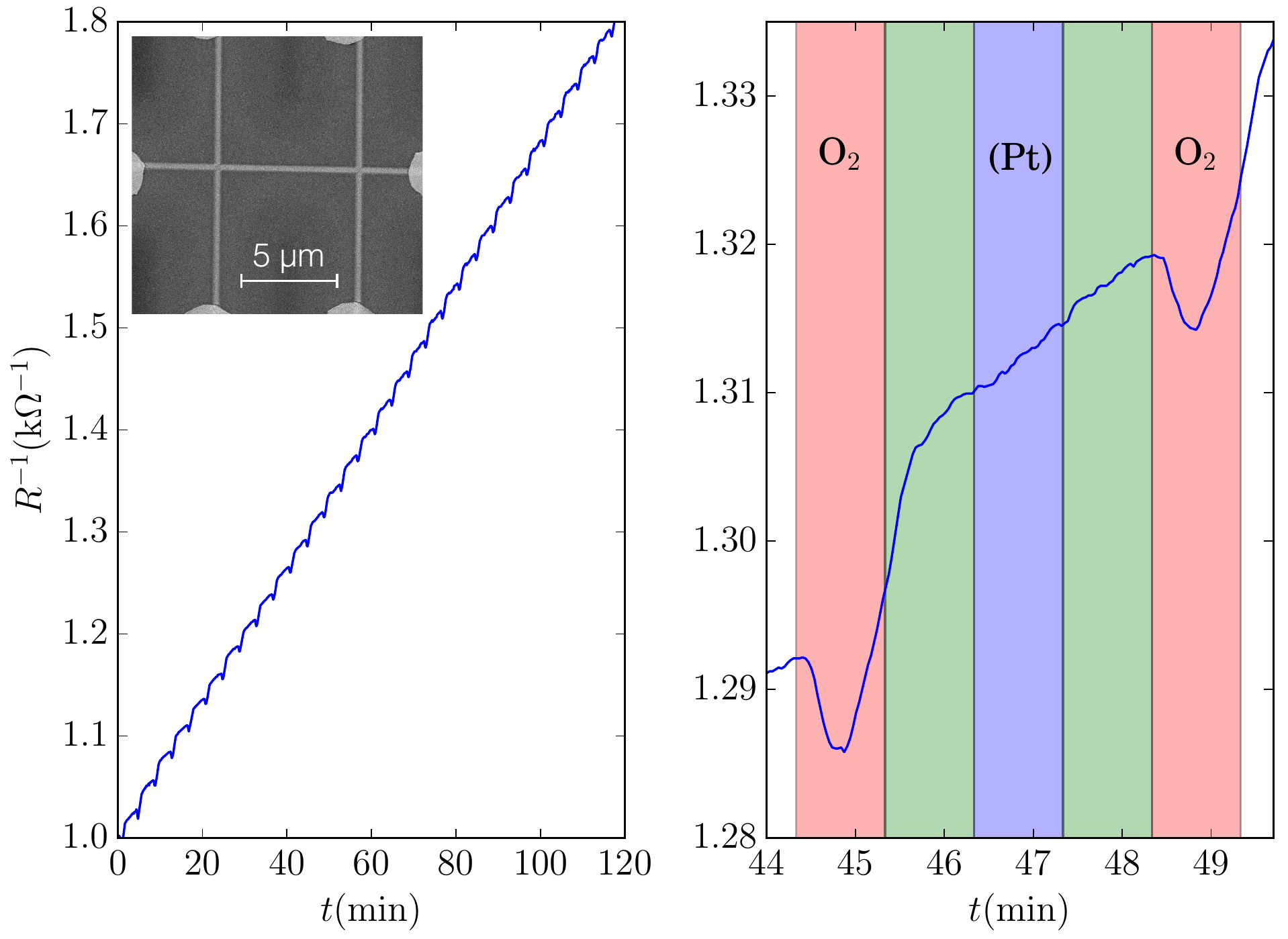}
\caption{Conductance of Pt layer as monitored during several cycles of ALD-growth of Pt (left). The ALD-growth proceeds on top of a Pt-containing FEBID seed layer in six-probe geometry. The inset shows an SEM image of the final Pt structure after about $100$ ALD cycles. A detailed view of the conductance progression during the different cycle phases reveals a mostly linear growth over time during the Pt-precursor feed and subsequent pump period (green), whereas the conductance first drops and then increases again during oxygen flow. See \cite{Diprima2017_asald_monitoring} for details.}
\label{fig_as_ald_conductance_monitor}
\end{figure}

\emph{In-situ} monitoring is a highly useful technique to follow the evolution of the electrical conductance of FEBID layers during or after growth, as well as during post-growth electron irradiation \cite{Porrati2011_PtC_irradiation}. In addition, conductance monitoring during growth has been shown to allow for semi-automatic growth optimization by using the conductance monitor signal as input to a genetic algorithm (GA) that adapts a set of deposition parameters, such as dwell time and pitches, until the rate of conductance increase per writing loop is maximized \cite{Weirich2013_ga_first_pub, Winhold2014_ga_modeling}. The same approach is also applicable to AS-ALD and the conductance monitor signal can be used to optimize the cycling times for the oxygen and precursor flow, as well as the duration of the pumping times between the gas-flow periods \cite{Diprima2017_asald_monitoring}. In Fig.\,\ref{fig_as_ald_conductance_monitor} we show an example for the conductance monitor signal during several ALD cycles showing a step-like behavior from which, e.\,g., the height increase per cycle can be deduced.
For FEBID-induced AS-ALD growth monitoring, so far established monitoring approaches, such as ellipsometry, X-ray photoelectron or UV spectroscopy and X-ray reflectivity, are not suitable, as they rely on the availability of sufficiently large lateral growth areas. Conductance monitoring is therefore particularly suited for AS-ALD on the lateral nanoscale. Considering the sub-$10\,$nm resolution structure obtained by Mackus \cite{Mackus2013_Pt_asald_highres}, the fabrication of point-contact structures on the nm-scale with controlled coupling strength appears feasible, if conductance monitoring during growth is employed.
%
% Challenges and perspectives
%
\subsection{Challenges and perspectives}
A metal content of, ideally, $100\,$at$\%$ is certainly the most important factor that has guided the development of the different FEBID-based approaches briefly reviewed in this section. However, additional factors have to be taken into account depending on the particular application which is addressed. These factors are, among others, the void-free nature of the deposits, the electronic properties of possible residual contaminants (e.\,g.\ magnetic), the compatibility of the approach with high-resolution writing conditions in FEBID, and the degree of co-deposit formation, as well as the electronic nature of the co-deposit, either directly after growth or after the respective processing steps, such as purification. In particular with a view to the future use of (3D) metallic FEBID structures in plasmonics, further efforts are necessary in developing processes that result in functional, metallic nanostructures with good shape-fidelity \cite{Winkler2017_3D_plasmonic}. Nevertheless, large progress has been made and this is expected to continue in the future, as more and more (pure) metals become accessible by the FEBID approach.
%
% Superconducting FEBID materials
%
\section{Superconducting FEBID materials}
By application of a magnetic field $H$ or by increasing the temperature $T$ the superconducting state of a material is suppressed. For bulk samples, it is the material itself that determines the superconductor's phase boundary, as represented in the $H-T$ phase diagram. Because of the very small surface-to-volume ratio in bulk materials, the sample topology can be mostly neglected. For mesoscopic samples, on the other hand, this ratio is large and the spatial development of the superconducting state depends strongly on the sample shape which imposes specific boundary conditions. Moreover, the Cooper pair state can be substantially modified by the properties of materials in close contact with the mesoscopic superconductor. Broadly speaking, on the mesoscopic scale one has the freedom to design confinement patterns for the Cooper pair condensate and its magnetic excitations (vortices) by using their quantum nature. Sophisticated nanostructuring is needed to fabricate such mesoscopic patterns and it is quite obvious that the development of a FEBID-based direct write approach for superconducting nanostructures would be of high interest.
%
% As-grown FEBID superconductors
%
\subsection{As-grown FEBID superconductors}
First research on the direct-write fabrication of superconducting nanostructures was done employing a focused ion beam (FIBID -- focused ion beam induced deposition) and not with FEBID. In 2004 Sadki and collaborators showed that type-II superconducting nanostructures of the approximate composition W$_{0.4}$Ga$_{0.2}$C$_{0.4}$ can be directly written using Ga-FIBID and W(CO)$_6$ as precursor \cite{Sadki2004_SC_WGa_FIB_first}. With a critical temperature $T_c$ of about $5.2\,$K, the exact value depending on the preparation conditions \cite{Li2008_SC_WGa_tuning}, this superconducting material turned out to be quite attractive for several follow up works. Its weak-coupling BCS-like behavior was confirmed in tunneling spectroscopy \cite{Guillamon2008_tunneling_WGa} and it was used in studies relating to vortex matter \cite{Guillamon2014_vortex_matter_WGa}, in spin-polarized Andreev reflection experiments with ferromagnetic counter electrodes prepared by FEBID \cite{Sangiao2011_SC_WGa_andreev_reflection} and in studies on induced odd-frequency triplet pairing in superconductor-ferromagnet proximity junctions \cite{Wang2010_odd_freq_sc_proximity_nanowire_first, Kompaniiets2014_proximity_triplet_sc1}. By using different precursors other superconducting materials prepared by Ga-FIBID were discovered, such as Mo-Ga-C-O with the precursor Mo(CO)$_6$ \cite{Weirich2014_SC_MoGaCO}, showing a maximum $T_c$ of $3.8\,$K for Mo$_{0.41}$Ga$_{0.26}$C$_{0.26}$O$_{0.07}$, and in Ga-C-O employing the purely organic aromatic precursor phenanthrene (C$_{14}$H$_{10}$) with a maximum $T_c$ of $7\,$K for Ga$_{0.27}$C$_{0.35}$O$_{0.38}$ \cite{Dhakal2010_SC_CGaO}.

Evidently, the W-based FIBID superconductor, which closely follows BCS theory and exhibits homogeneous superconducting properties under optimized growth conditions, is an attractive candidate for various areas of research in which the direct-write approach provides a unique advantage. However, there are limitations which have to be considered. First, the high degree of disorder in this system puts this material close to a disorder-induced metal-insulator transition, thus limiting its usability, as statistical fluctuations of the superconducting order parameter lead to spatial variations of the critical temperature \cite{Sadovskii1997_sc_in_disordered_systems}. Second, the application of a Ga focused ion beam for the deposition is always accompanied by ion beam induced etching. Simultaneous etching does not necessarily have to be regarded as a drawback for the preparation of FIBID superconductors, but it will limit the application of this technique with regard to the fabrication of, e.\,g., superconducting multilayer structures. This is why a direct-write approach of superconducting structures with a focused electron beam is even more attractive. It is also more challenging, since the integral metal content in Ga-FIBID structures tends to be higher than in FEBID structures employing the same precursor.

Nevertheless, within less than one year three independents works were published with clear evidence for superconductivity in FEBID structures employing the precursors Mo(CO)$_6$ in conjunction with H$_2$O leading to an amorphous Mo-C-O phase that showed a broad onset to superconductivity at about $10\,$K in resistance measurements \cite{Makise2014_ebid_superconductivity}, W(CO)$_6$ resulting in superconducting nanowires with an onset of superconductivity at $2\,$K under carefully optimized deposition conditions \cite{Sengupta2015_W_ebid_superconductivity} and tetraethyllead which resulted in metallic and superconducting Pb-C-O deposits with a maximum $T_c$ of $7.3\,$K for the composition Pb$_{0.44}$C$_{0.31}$O$_{0.25}$, a value which corresponds to the critical temperature of bulk Pb \cite{Winhold2014_Pb_superconductor}. So far no follow up works using these recent FEBID-based superconductors have been published which indicates that their preparation is quite delicate and more effort is needed to establish growth protocols that result in homogenous superconducting materials with reproducible properties.
%
% Superconductivity in doped FEBID structures
%
\subsection{Superconductivity in doped FEBID structures}
So far, the phases are not yet clearly identified which are the substrate of the superconducting condensate in the amorphous materials W-Ga-C-(O), Mo-Ga-C-O and C-Ga-O (FIBID-based), as well as W-C-O and Mo-C-O (FEBID-based). Amorphous Ga, W and Mo, as well as the carbides of W and Mo can show superconductivity in the temperature range reported for the Ga-FIBID and FEBID-based materials reported so far. In this regard an interesting question arises: may doping of charge carriers into an amorphous material, such as W-C-O prepared by FEBID, be sufficient to induce superconductivity by way of driving the system towards a superconductor-insulator transition, see, e.\,g.\ \cite{Sadovskii1997_sc_in_disordered_systems, Feigelman2007_pseudogap_sc_ins_transition, Gantmakher2010_sc_ins_quantum_phase_transition}, on the interplay of Anderson localization and superconductivity in strongly disordered systems and the insulator-superconductor quantum phase transition . Two recent studies indicate that this may be the case.

Magnetization measurements performed on amorphous W-C-O FEBID structures which had been exposed to a sulphur atmosphere at $250^\circ$C for $24\,$h showed indications of superconductivity in both, field- and zero-field cooled curves setting in at temperatures as high as $38\,$K \cite{Felner2012_SC_WCO_S_doped}. The authors estimated a very small shielding fraction of only $0.013\,\%$ but argued that their results prove the effectiveness of sulfur in inducing superconductivity in amorphous carbon. They even suggested that this approach may open new pathways to achieve high-temperature superconductivity in amorphous carbon based materials.

In a very recent study, Porrati and collaborators showed that W-C-O prepared by FEBID employing the precursor W(CO)$_6$ can be driven towards a superconductor-insulator transition by means of Ga doping employing low-dose irradiation with a Ga focused ion beam at $30\,$keV \cite{Porrati2017_SC_W_FEBID_Ga_doping}. By increasing the irradiation dose, a pronounced reduction of the resistivity by more than one order of magnitude was observed in parallel with the occurrence of local superconductivity on the insulating side of the superconductor-insulator transition, indicated by a negative magnetoresistance. On the superconducting side of the superconductor-insulator transition, phase-coherent superconductivity was observed, as indicated by a positive sign of the magnetoresistance.
%
% Challenges and perspectives
%
\subsection{Challenges and perspectives}
Direct-write fabrication of nanostructured superconductors by FEBID can open up a large range of possible application fields, such as for the study of (quantum) phase slips in nanowires \cite{Belkin2015_qps_sc_nanowires}, as inducers for odd-frequency triplet states in proximity junctions with ferromagnets \cite{Bergeret2005_odd_freq_triplet_pairing}, as sensor elements in highly sensitive nanowire bolometers \cite{Natarajan2012_nanowire_sc_bolometers}, or for fundamental studies on the nature of the superconductor-insulator transition in disordered or granular systems \cite{Gantmakher2010_sc_ins_quantum_phase_transition}, to indicate just a few. FEBID-based superconducting materials reported so far represent promising examples. However, with regard to just using them outside the immediate research field of the superconductor-insulator transition, several properties of these materials need to be significantly improved. A reproducible FEBID process to obtain a homogeneous materials that shows a sharp transition into the superconducting state, ideally well above $4.2\,$K, is still sought for.
%
% Alloys and intermetallic compounds
%
\section{Alloys and intermetallic compounds}
The fabrication of multi-component polycrystalline or granular metals by FEBID represents a challenging research approach for the design of novel materials. Multi-component granular metals, i.\,e.\ materials in which metallic nano-crystallites are embedded in an insulating matrix, are particularly interesting, due to their fine tunability in composition and electronic inter-granular tunnel coupling strength. This flexibility is of great advantage for the investigation of binary and ternary alloy systems and to target specific intermetallic compounds.

Currently, there are three different approaches for the fabrication of multi-component FEBID nanostructures, as will be discussed below. These comprise: 1. co-deposition using two different precursors; 2. deposition with one single multi-component (heteronuclear) precursor; 3. intermixing of multilayer nanostructures fabricated with different precursors by low-energy electron irradiation.
%
% Precursor mixing
%
\subsection{Co-deposition using two different precursors}
\begin{figure}
\centering
\includegraphics[width=0.8\textwidth]{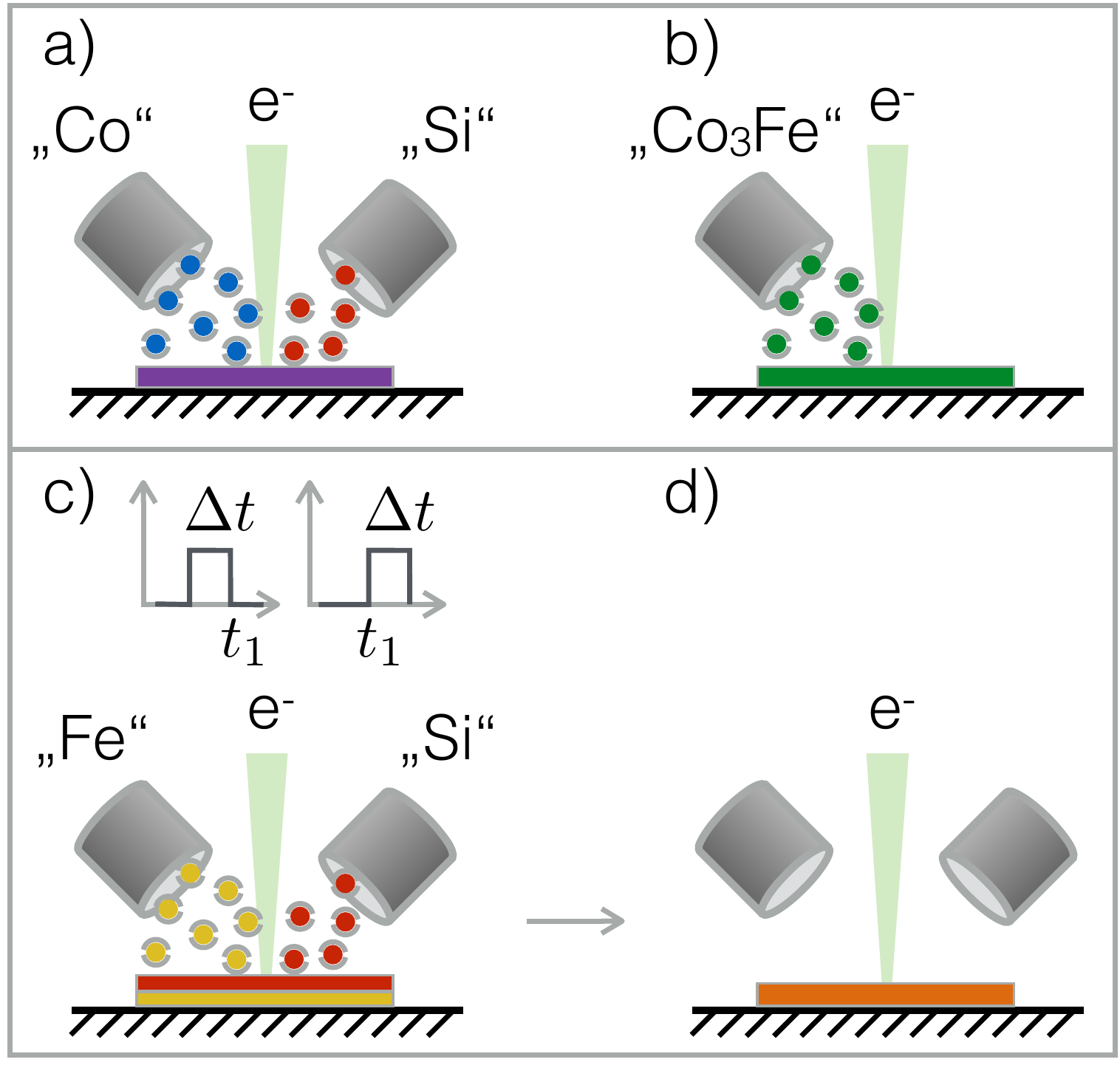}
\caption{(a) Co-deposition with two precursors for the fabrication of a Co-Si alloy. The two precursors (''Co'': Co$_2$(CO)$_8$, ''Si'': neopentasilane) are injected into the SEM by two independent capillaries. (b) Direct deposition of Co$_3$Fe by the use of  the heteronuclear precursor HFeCo$_3$(CO)$_{12}$ (''Co$_3$Fe''). (c) Deposition of a [Fe/Si]$_n$ multilayer by the successive injection of Fe(CO)$_5$ (''Fe'') and neopentasilane as precursors. (d) Low-energy electron irradiation treatment of the multilayer obtained from (c) to obtain a granular FeSi compound by atomic species intermixing.}
\label{fig_febid_alloys_multilayers}
\end{figure}

The co-deposition using two different precursors represents the first and up to now most used approach for the fabrication of multi-component FEBID nanostructures. Technically, two different methods are available. The first consists in employing one single capillary to simultaneously inject two different gases into the SEM \cite{Che2005_FEBID_FePt_holography}. The second consists in using two independent gas injection systems, one for each precursor, see Fig.\,\ref{fig_febid_alloys_multilayers}(a). The latter method is the most present in the literature \cite{Winhold2011_PtSi_alloy, Porrati2012_CoPt_alloy, Porrati2013_CoSi_alloy, Shawrav2014_AuFe_alloy, Porrati2017_FeCoSi_multilayer}. In our group a self-made two-channel gas injection system \cite{Keller2014_master_thesis} allows either the alternating or the simultaneous use of two gases.

Historically, the first material investigated was FePt \cite{Che2005_FEBID_FePt_holography}, which was obtained in an UHV-SEM by mixing Me$_3$CpPt(IV) and Fe(CO)$_5$. An annealing process carried out at $600^\circ$C for two hours was employed to transform the as-grown FePt nano-rods into polycrystalline samples of the L1$_0$ phase. The magnetic characterization of these samples was performed by electron holography. In our group, the PtSi binary system was fabricated by using Me$_3$CpMePt(IV) and neopentasilane, Si$_5$H$_{12}$ \cite{Winhold2011_PtSi_alloy}. In this work, deposits with different composition were characterized by energy dispersive x-ray analysis (EDX), transmission electron microscopy (TEM) and electrical measurements. In particular, for a [Pt]/[Si] ratio of $2/3$, we found evidence for the formation of the Pt$_2$Si$_3$ metastable phase. CoPt was prepared by co-injection of the precursors Co$_2$(CO)$_8$ and Me$_3$CpMePt(IV) \cite{Porrati2012_CoPt_alloy}. Here it was shown that the as-grown amorphous CoPt phase transforms into the L1$_0$ face-centered tetragonal phase by means of a room temperature low-energy electron irradiation treatment. In ref.\ \cite{Porrati2013_CoSi_alloy} we discussed the fabrication of CoSi alloy nanostructures by using Co$_2$(CO)$_8$ and Si$_5$H$_{12}$. In that work, the electrical transport properties of the system were investigated by focusing on the metal-insulator transition, which was reached by tuning the composition of the deposits. In the group of Heinz Wanzenb\"ock, AuFe nano-alloys were prepared by using Me$_2$(tfac)Au(III) and Fe(CO)$_5$ \cite{Shawrav2014_AuFe_alloy}. In this work, the influence of changes of the precursor pressure on the alloy composition was studied. Finally, we note that some work on bimetallic nanostructures, namely AuAg and AuPt, has been carried out by using liquid phase-electron beam induced deposition in the group of Todd Hastings \cite{Bresin2013_bimetallic_liquid_precursor}.

Summarizing the results presented in the aforementioned papers, one can state that the composition of Pt-Si, Co-Pt and Au-Fe binary alloys are characterized by a large amount of carbon, $40\dots 60\,$at.$\%$, and oxygen, $10\dots 20\,$at.$\%$. As a consequence, the electrical properties of these carbon-rich alloys are dominated by electron tunneling between neighboring alloy nano-grains \cite{Winhold2011_PtSi_alloy, Porrati2012_CoPt_alloy}. In Co-Si, the presence of carbon and oxygen is less dominant, since the sum of the concentrations of these two elements is always less than $50\,$at$\%$, so that an assumed metallic Co-Si phase would be expected to lead to a percolating metallic path \cite{Porrati2013_CoSi_alloy}. Therefore, in Co-Si the electrical properties mainly depend on the ratio between [Si] and [Co].

We conclude this paragraph remarking that, given the high carbon and oxygen concentrations present in these alloys, post-growth treatments, like electron irradiation or annealing, are required to obtain pure nano-alloy grains. On the other hand, the presence of carbon in the as-grown deposits may be seen as a benefit, since it allows to tune the microstructure, and thus, the electrical and the magneto-transport properties of the alloys by means of post-growth treatments \cite{Che2005_FEBID_FePt_holography, Porrati2012_CoPt_alloy}. The main advantage of the co-deposition process is its flexibility, which allows to obtain alloys of the form  A$_{1-x}$B$_x$ for any stoichiometric ratio. However, this fabrication approach fails if one of the precursor species dominates adsorption during the deposition process, as in the case of Fe(CO)$_5$ and Si$_5$H$_{12}$. In that case, alternative fabrication strategies have to be employed, as discussed next.
%
% Heteronuclear precursors
%
\subsection{Heteronuclear precursors}
The use of heteronuclear precursors represents the most direct way to fabricate FEBID multi-component nanostructures, see Fig.\,\ref{fig_febid_alloys_multilayers}(b). Usually, FEBID employs homonuclear precursors, as for example Fe(CO)$_5$, Co$_2$(CO)$_8$ or Si$_5$H$_{12}$, whose molecules contain one single metal (or other) atomic species, beyond the presence of hydrogen, carbon and oxygen. Recently, heteronuclear precursors, with two different metal atomic species, started to be investigated \cite{Porrati2015_CoFe_precursor, Kumar2017_FeRu_dissociation_study}. The first precursor used, HFeCo$_3$(CO)$_{12}$, led to Co-Fe magnetic alloys with a metal content of about $80\,$at$\%$ \cite{Porrati2015_CoFe_precursor}. The [Co]/[Fe] ratio was found to be about $3$, which corresponds to the metal stoichiometry of the precursor. In TEM studies the microstructure of the samples was found to be a mixture dominated by bcc Co-Fe with some contributions of an FeCo$_2$O$_4$ spinel oxide phase which was later found to be localized at the surface of Co$_3$Fe nanostructures and is likely a consequence of oxidation under ambient conditions \cite{Keller2017_magnetic_3D}. Micro-Hall magnetometric measurements show that the nanostructures are ferromagnetic up to the highest measured temperature of $250\,$K \cite{Porrati2015_CoFe_precursor}. Recently, another heteronuclear carbonyl precursor, H$_2$FeRu$_3$(CO)$_{13}$, was investigated in a large collaborative effort with regard to its electron dissociation characteristics (gas phase and surface science studies) and its FEBID performance  \cite{Kumar2017_FeRu_dissociation_study}. The metal content found in the deposits obtained under optimized FEBID conditions did not exceed $26\,$at$\%$, a much lower value than the one obtained for the above-mentioned Co$_3$Fe nanostructures. This large difference in the metal content of the deposits for the structurally similar precursors is attributed to their different dissociation characteristics, as found in the gas phase and surface science studies \cite{Kumar2017_FeRu_dissociation_study}.

Clearly, the main advantage of the deposition from heteronuclear precursors is the possibility to target alloys and compounds of interest in a direct way. Furthermore, the composition of the deposits can be tuned, in a limited range, by varying the electron beam deposition parameters. In general, on the one hand the deposition from heteronuclear precursors is very attractive because of its simplicity, on the other hand the precursors currently available are few, greatly limiting the number of target materials.
%
% Intermixing of multilayers
%
\subsection{Intermixing of multilayers}
The third method used for the fabrication of FEBID multi-component deposits consists in the growth of multilayer nanostructures and their subsequent intermixing by a low-energy electron irradiation treatment \cite{Porrati2016_FeSi_alloy, Porrati2017_FeCoSi_multilayer}, see Fig.\,\ref{fig_febid_alloys_multilayers}(c,d). This two-step fabrication method circumvents the limits of the co-deposition process for the case in which the two precursors cannot be used simultaneously. As described in Ref.\ \cite{Porrati2016_FeSi_alloy}, we found that FeSi nano-alloys cannot be fabricated by mixing the precursors Fe(CO)$_5$ and Si$_5$H$_{12}$, probably as a consequence of the specifics of the competition for adsorption sites when these two precursors are simultaneously used during growth. Similarly, in Ref.\ \cite{Porrati2017_FeCoSi_multilayer} the Heusler compound Co$_2$FeSi  could not be prepared by mixing HFeCo$_3$(CO)$_{12}$, Fe(CO)$_5$ and Si$_5$H$_{12}$. To overcome these limits, we deposited [Fe/Si]$_2$, [Fe$_3$/Si]$_2$ and [Co$_2$Fe/Si]$_n$ multilayers and, subsequently, subjected them to a low-energy electron irradiation treatment in order to induce atomic species intermixing. As a result, the compounds FeSi, Fe$_3$Si and Co$_2$FeSi were obtained \cite{Porrati2016_FeSi_alloy, Porrati2017_FeCoSi_multilayer}.

The main advantage of this fabrication method is to provide an alternative approach in those cases for which the co-deposition approach and the fabrication process based on the use of heteronuclear precursors cannot be applied. Moreover, this fabrication method explicitly shows that high-quality nanopatterned multilayers can be fabricated by FEBID \cite{Porrati2017_FeCoSi_multilayer}, which is of potential interest for the fabrication of plasmonic metamaterials and prototype devices in spintronics.
%
% Challenges and perspectives
%
\subsection{Challenges and perspectives}
The possibility to fabricate multi-component materials by FEBID opens the way for the fabrication of a large number of nanostructured alloys and intermetallic compounds. The investigation of entire material classes, as for example Heusler compounds, which for itself encompasses more than $1000$ members, becomes possible at the nanoscale. Two main characteristics distinguish FEBID materials from bulk or thin film samples prepared with other techniques. First, FEBID allows to fabricate one-, two- and three-dimensional nanostructures. Second, the granular microstructure of the material can be tuned with a high degree of precision. This combination makes FEBID multi-component nanostructured materials very attractive for fundamental studies, as for example the investigation of finite-size effects, and, potentially, for applications in spintronics and thermoelectrics.

We conclude remarking that the main part of the precursors potentially interesting for the fabrication of FEBID alloys and intermetallic compounds are not commercially available. Therefore, the synthesis of the precursor, the study of its dissociation behavior, the growth of the deposits and their successive characterization, is a challenging interdisciplinary work involving know-how from the fields of chemistry, physics, surface and materials science. It remains a task for the future to further expand the reservoir of high-potential FEBID precursors.
%
% Metamaterials
%
\section{Metamaterials}
In the narrow sense, research on metamaterials has been largely concerned with negative refraction index materials in optics. However, the idea of rationally designing a unit from selected materials and (periodically) arrange it into an artificial solid is much broader. The properties of that solid are then determined by the structure of the artificial solid and the coupling between its units and can be tailored towards a desired functionality or a particular emerging property. In this regard, several important benefits can be gained from employing FEBID. One can directly take advantage of the particular microstructure of FEBID materials, as exemplified by nano-granular metals which are often obtained when organometallic precursors are used. Else, one can take advantage of the high-resolution writing capability of FEBID to follow the rational design approach towards metamaterials or combine nanostructured FEBID materials with other materials. Finally, these approaches can be combined, as will be exemplified below. It should be noted that the emerging simulation-guided approach for 3D nano-manufacturing, as discussed later, will likely be of large relevance for the future development of optical and magnetic metamaterials.
%
% Ordered and disordered nano-granular metals
%
\subsection{Ordered and disordered nano-granular metals}
\label{sec_nanogranular_metals}
Nano-granular metals, in the ordered form also denoted as nano-dot lattices, are model systems for the study of the interplay of electronic correlation effects, finite size induced quantization of the electronic level structure, and disorder \cite{Beloborodov2007_granular_electronic_systems}. The inter-granular tunnel coupling $g$ (normalized to the conductance quantum $G_0 = 2e^2/h$, including spin degeneracy) is the most important control parameter that governs the electronic properties of this material class. The coupling strength in conjunction with the metallic grain size, that determines the Coulomb charging energy $E_C$ of the grains, and the properties of the insulating matrix, in which the grains are embedded, determine the electronic ground state of a granular metal. In both phases, the weak-coupling or insulating and the strong-coupling or metallic, length scales exist which exceed by far the geometric length scales of grain diameter and grain-to-grain distance. As a result, many-body properties emerge which cause characteristic electronic properties in the respective phases.

\begin{figure}
\centering
\includegraphics[width=0.6\linewidth]{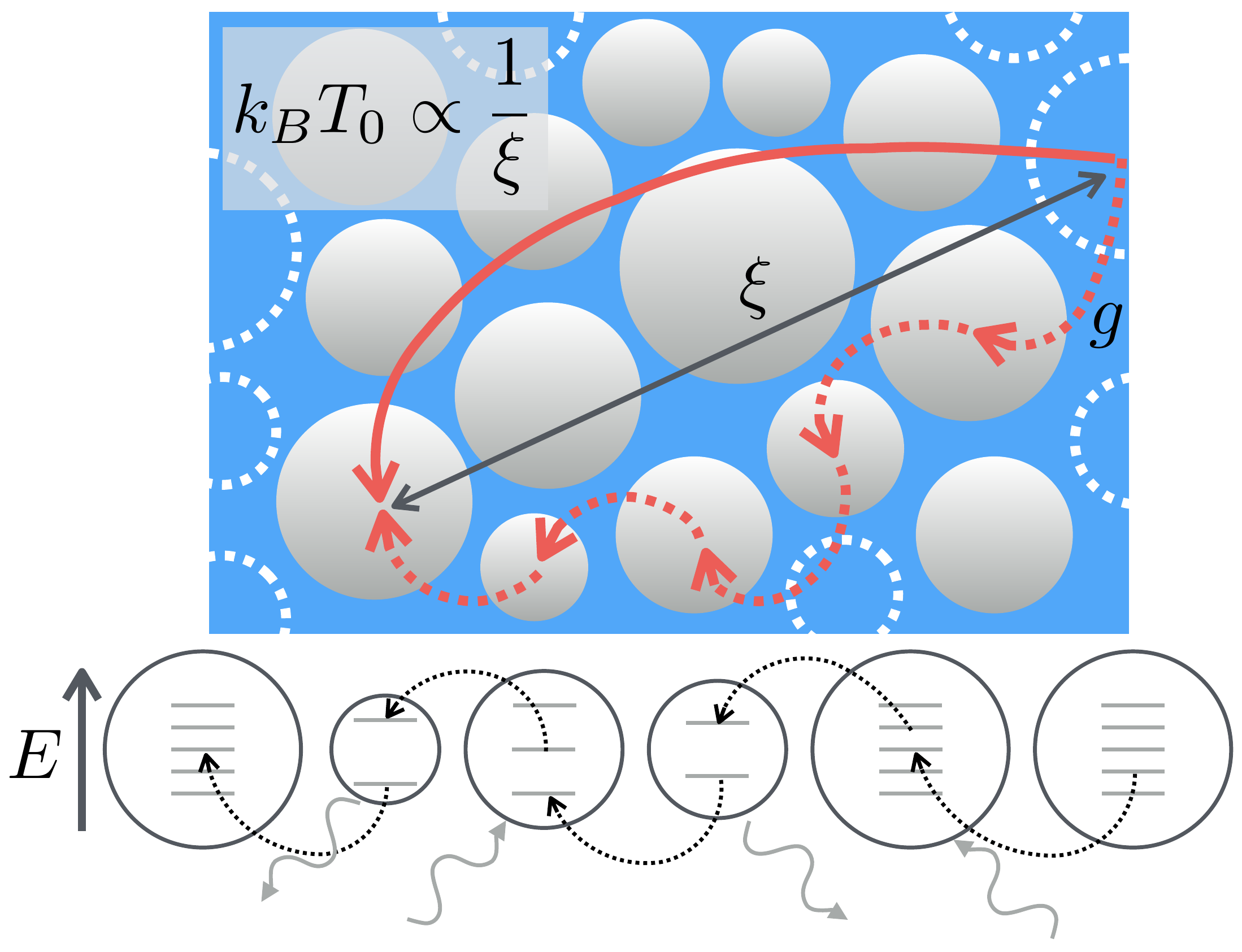}
\caption{Schematic of charge transport in a nano-granular metal in the c-VRH regime. Electron tunneling between metallic grains can occur over larger distances by means of co-tunneling, as indicated in the upper image. A strong reduction of the effective charging energy $E_C$ occurs, as there is no charge accumulation in the grains which take part in the co-tunneling sequence. Over the co-tunneling range $\xi$, that determines the activation temperature $T_0$, the Coulomb interaction between the initial (now positively charged) and final grain (now negatively charged) after the co-tunneling event has occurred is reduced over the distance $\xi$ and by screening effected by the surrounding material. In the lower part of the figure inelastic co-tunneling is indicated, in which the incoming and outgoing charge are on different energy levels. The energy difference is exchanged with the surrounding material via phononic excitations.}
\label{fig_co_tunneling}
\end{figure}

On the weak-coupling side, higher-order tunneling events (sequential inelastic and elastic co-tunneling) start to dominate, as the temperature is reduced leading to a stretched exponential dependence of the conductivity on temperature, the correlated variable range hopping (c-VRH) \cite{Efetov2003_coulomb_effects_granular_metals, Beloborodov2007_granular_electronic_systems}.
\begin{equation}
\sigma(T) = \sigma_0 \exp{\left\{ -\left( \frac{T_0}{T} \right)^{1/2} \right\}} \,.
\label{eq_cVRH_sigma_vs_T}
\end{equation}
This is indicated schematically in Fig.\,\ref{fig_co_tunneling}.

\begin{figure}
\centering
\includegraphics[width=0.6\linewidth]{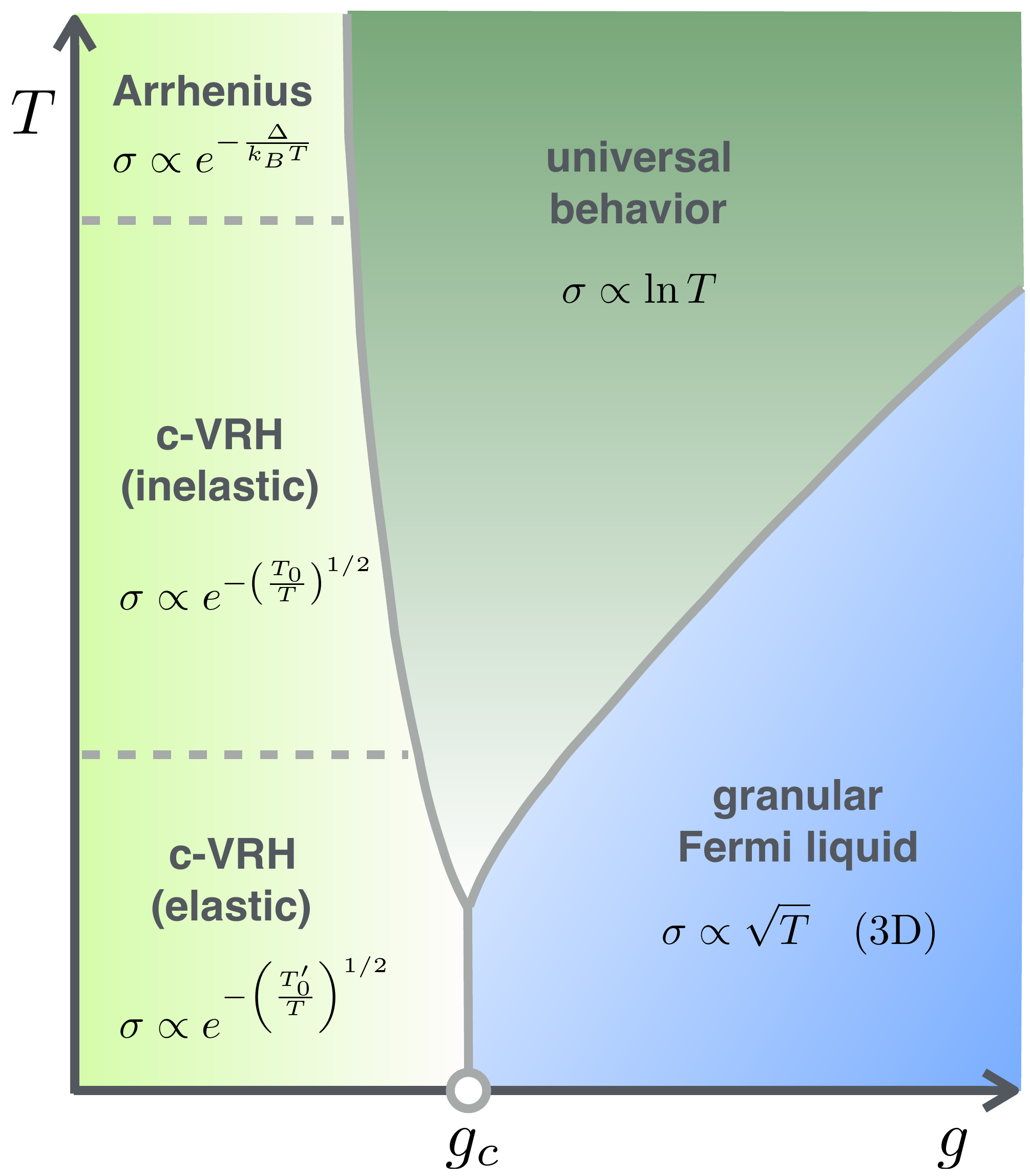}
\caption{Schematic phase diagram of transport regimes of nano-granular metals. Adapted from \cite{Beloborodov2004_granular_fermi_liquid}.}
\label{fig_phase_diagram_granular_metals}
\end{figure}

In the strong-coupling limit a granular Fermi liquid was theoretically predicted \cite{Beloborodov2003_granular_metals_strong_coupling, Beloborodov2004_granular_fermi_liquid} and  experimentally observed in nano-granular Pt prepared by FEBID with the precursor Me$_3$CpMePt(IV) \cite{Sachser2011_PtC_universal_conductance} (see Fig.\,\ref{fig_phase_diagram_granular_metals} for a schematic phase diagram of nano-granular metals). Furthermore, there is growing evidence for the emergence of collective states, indicated by super-ferro- or super-antiferromagnetic behavior in nano-granular Pt in a certain range of the coupling strength $g$ for temperatures below about $20$ to $30\,$K \cite{Porrati2014_PtC_magnetoresistance, Porrati2014_PtC_coupling_Co_nanopillars}. These observations gain additional relevance because of the apparent similarities in the electronic structure of nano-granular metals and electronically correlated organic charge transfer systems of the $\kappa$-[BEDT-TTF]$_2$-X-type (BEDT-TTF: bis-ethylenedithiotetrathiafulvalene, X: halogenide or polymeric anion) close to a superconductor-insulator transition \cite{Diehl2015_pseudogap_kappaET, Guterding2016_sc_order_parameter_kET}. It should also be noted that underdoped high-$T_c$ superconductors show evidence for spontaneous granularity in their electron density at low temperature \cite{Pan2001_STS_underdoped_htsc, Dubi2007_sit_htsc_disordered}. In addition, microstructurally homogeneous, amorphous TiN thin films also show evidence for spontaneous electronic granularity \cite{Baturina2007_localized_sc_TiN}.

Nano-granular Pt, as prepared by FEBID, has been established as a particularly fruitful model system for studies of the coupling strength dependent phase diagram of disordered granular metals (see Fig.\,\ref{fig_phase_diagram_granular_metals}). This is due to the fact that the inter-granular coupling strength can be extremely finely tuned by a simple post-growth electron irradiation treatment combined with \emph{in-situ} conductance monitoring \cite{Porrati2011_PtC_irradiation}. The research that has been done on nano-granular Pt may easily be extended to other nano-granular FEBID structures provided that their insulating matrix shows similar tunability. For as-grown nano-granular Pt the matrix consists mainly of amorphous carbon which tends to graphitize under electron irradiation, thus improving the tunnel-coupling strength between the Pt nano-grains \cite{Porrati2011_PtC_irradiation}. In addition, FEBID has more to offer with regard to ordered nano-granular metals or nano-dot lattices. Employing the precursor W(CO)$_6$ we fabricated ordered 2D square lattices of roughly semi-spherical and metallic W-C-O nano-islands of approximately $20\,$nm diameter and showed that a metal-insulator transition occurs as a function of lattice constant \cite{Sachser2009_2D_nanodot_lattice_transport, Porrati2010_2D_nanodot_lattice_fabrication}. In this case, the metal-poor co-deposit around the metallic nano-islands served as tunneling matrix. It is to be expected that future work will expand the class of tunable nano-granular metals prepared by FEBID.
%
% Proximity-induced superconductivity
%
\subsection{Proximity-induced superconductivity}
The variety of available FEBID materials with tunable structural and magnetic properties in conjunction with its high resolution makes FEBID valuable for the fabrication of hybrid nanostructures and the investigation of emerging effects at interfaces on the meso- and nanoscale. A prominent exemplary phenomenon is the superconducting proximity effect \cite{Buzdin2005_review_fm_proximity} and, more generally, proximity-induced superconductivity \cite{Bergeret2005_odd_freq_triplet_pairing}. In the conventional superconducting proximity effect at a superconductor/ferromagnet (SF) interface, the wave function of the Cooper pairs is singlet as it is formed by two electrons with opposite spins. The exchange field of F tends to align both spins in the same direction which results in a strong pair-breaking effect and causes a rapid exponential oscillatory decay of the superconducting order parameter in F over a distance of the order of $1$\,nm. However, under some circumstances the presence of ferromagnetism may lead to triplet superconducting pairing \cite{Buzdin2005_review_fm_proximity, Bergeret2005_odd_freq_triplet_pairing, Eschrig2008_triplet_supercurrents, Eschrig2015_spin_polarized_supercurrents}. In this case spin-triplet superconducting correlations are extending into F over a proximity length of several microns and this is why the spin-triplet proximity effect at a S/F interface is long-ranged and ''unconventional'' \cite{Wang2010_odd_freq_sc_proximity_nanowire_first}. As theoretically shown by Bergeret \emph{et al.} \cite{Bergeret2001_long_range_proximity_fm}, local inhomogeneity of the magnetization in the vicinity of the S/F interface is necessary for spin-triplet pairing in S/F structures. These inhomogeneities can be either intrinsic to F (domains) and thus modifiable by an external field, or be caused by a material inhomogeneity as a result of experimental manipulations, such as contacting procedures or post-growth processing. The versatility of FEBID for both of these procedures makes it an especially useful experimental technique in this regard.

Sangio \emph{et al.} \cite{Sangiao2011_SC_WGa_andreev_reflection, Sangiao2011_andreev_magnetic_field_sc_fm_febid} demonstrated for the first time functional devices combining magnetic nanostructures grown by FEBID and superconducting nanostructures fabricated by Ga-FIBID. They studied the magnetic-field dependence of the conductance in planar S/F nano-contacts which allowed them to extract the magnetic field dependences of the superconducting gap of the W-based electrode. The value of the superconducting gap extracted from those experiments is in agreement with direct scanning tunneling spectroscopy experiments, emphasizing the possibility of preparing clean nano-contacts by FEBID/FIBID with a single or very few conduction channels \cite{Guillamon2008_tunneling_WGa}. Subsequent experiments on nano-contacts with worse definition or cleanliness gave rise to multi-channel transport \cite{Sharma2014_multi_channel_andreev}.

Kompaniets \emph{et al.} investigated superconducting proximity effects at S/F interfaces in nanowire structures \cite{Kompaniiets2014_proximity_triplet_sc1, Kompaniiets2014_Cu_Co_proximity}. Whereas superconducting correlations induced by a W-FIBID electrode led to up to $30\%$ resistance drops of a $7\,\mu$m-long section of a polycrystalline Co nanowire \cite{Kompaniiets2014_proximity_triplet_sc1}, proximity-induced superconductivity did not become apparent in Co-FEBID nanowire structures \cite{Kompaniiets2014_Cu_Co_proximity}. This may indicate that C ($15\,$at$\%$) and O ($14\,$at$\%$), residual impurities from the FEBID process with Co$_2$(CO)$_8$, are impurities responsible for suppressing the superconducting proximity effect, as these elements may be effective pair-breaking scatterers for triplet Cooper pairs. As possible routes to stimulating long-range spin-triplet superconductivity in Co-based FEBID structures one should mention annealing in a hydrogen atmosphere at elevated substrate temperatures in conjunction with electron irradiation in order to remove undesired C and O impurities \cite{Begun2015_Co_purification_H2}, as well as enhancing the magnetic inhomogeneity on the lateral mesoscale, e.\,g., by formation of a Co/Pt phase with different magnetic properties \cite{Dobrovolskiy2015_CoPt_treatment_H2_O2}.
%
% Fluxonics
%
\subsection{Fluxonics}
Almost all technologically relevant superconductors are superconductors of type II. In a magnetic field $\mathbf{B}$, whose magnitude is between the lower and upper critical field, they are in the mixed state, being penetrated by a flux-line array of Abrikosov vortices, or \emph{fluxons} \cite{Abrikosov1957_flux_lattice, Brandt1995_flux_line_lattice, Moshchalkov2010_sc_nanoscience, Dobrovolskiy2017_fluxonics_review, Woerdenweber2017_sc_at_nanoscale}. Each fluxon carries one magnetic flux quantum, $\Phi_0=2.07\times10^{-15}$~Wb, and the repulsive interaction between vortices makes them to arrange in a triangular lattice with the vortex lattice parameter $a_\bigtriangleup = (2\Phi_0/B\sqrt{3})^{1/2}$, where $B=|\mathbf{B}|$. Each vortex experiences the action of the driving (Lorentz) force induced by the transport current and the pinning force related to the spatial variation of the vortex energy at different locations inside the superconductor. Thus, when the pinning force dominates the Lorentz force, the vortex lattice is locally anchored (pinned) while in the opposite case the vortex lattice moves and the superconducting state is destroyed. For this reason, artificial nanostructures inducing a spatial modulation of the vortex energy are widely used for the manipulation of Abrikosov vortices \cite{Velez2008_vortex_pinning, Dobrovolskiy2017_fluxonics_review}.

\begin{figure}
\centering
\includegraphics[width=0.8\linewidth]{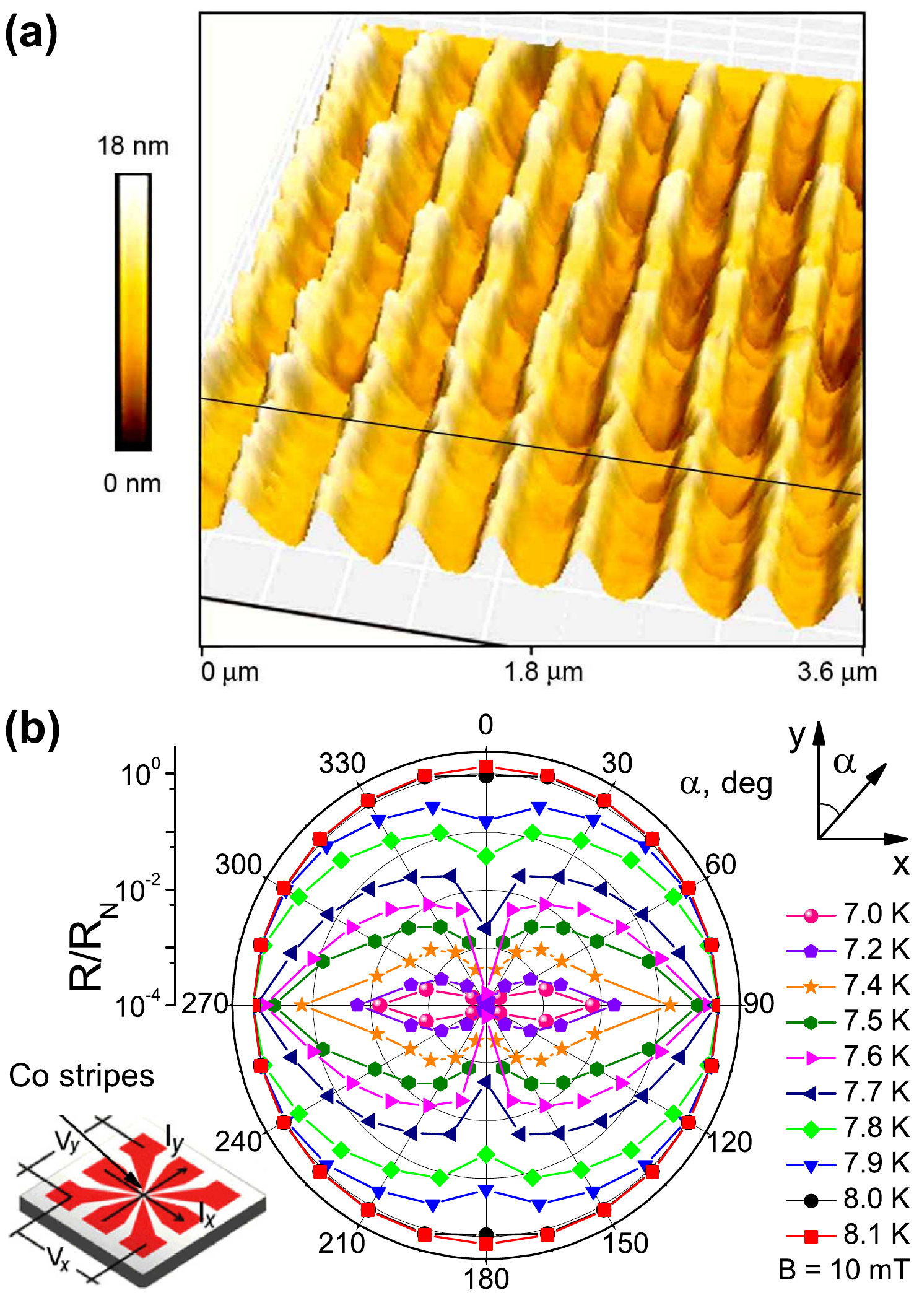}
\caption{(a) Atomic force microscope image of Co stripes deposited by FEBID on the surface of a superconducting Nb film. (b) Polar diagram of the total resistance of the Nb films with the Co stripes measured in the oriented-current geometry. The Co stripes are arranged along the $y$-axis and induce a washboard pinning potential periodic in the $x$-direction. The currents $I_x$ and $I_y$ allow one to pass the transport current $I=\sqrt{I_x^2+I_y^2}$  at any arbitrary angle $\alpha$ with respect to the pinning ''channels''. The measured voltage components $V_x$ and $V_y$ allow one to obtain the polar diagram of the total resistance $R(\alpha)$ of the film for a series of temperatures.}
\label{fig_Co_stripes_vortex_dynamics}
\end{figure}
Our group demonstrated for the first time that nanostructures fabricated by FEBID can be used for the extension of the dissipation-free state of superconductors to higher temperatures and larger currents \cite{Dobrovolskiy2010_guiding_Co_decoration, Dobrovolskiy2011_washboard_pinning, Dobrovolskiy2011_washboard_fabrication}. Fig.\,\ref{fig_Co_stripes_vortex_dynamics}(a) depicts an atomic force microscopy (AFM) image of an array of Co nanostripes (precursor Co$_2$(CO)$_8$) deposited by FEBID on the surface of a superconducting epitaxial Nb film. The film was patterned into an $8$-contact geometry for the oriented-current setup shown in the lower inset of Fig.\,\ref{fig_Co_stripes_vortex_dynamics}(b). The Co-FEBID stripes are aligned along the $y$-axis and induce a pinning potential of the washboard type for Abrikosov vortices, which is periodic in the $x$-direction. In the mixed state, near the superconducting transition temperature $T_c\approx 8.1\,$K, this results in a several orders of magnitude smaller resistance of the superconducting film when the transport current is applied along the stripes (vortices move in the $x$-direction) than perpendicular to the stripes (vortices move in the $y$-direction). Electrical resistance measurements clearly revealed an anisotropic behavior induced by the uniaxial pinning \cite{Dobrovolskiy2010_guiding_Co_decoration} as well as resistance steps ascribed to matching effects between the Co-FEBID stripe periodicity and the vortex lattice periodicity \cite{Dobrovolskiy2011_washboard_pinning}.
%
% Challenges and perspectives
%
\subsection{Challenges and perspectives}
Although the nano-granular nature of many of the materials prepared by FEBID has long been known, the directed use of the tunability of the tunnel coupling strength in conjunction with \emph{in-situ} monitoring is a recent development \cite{Porrati2011_PtC_irradiation}. Several important questions relating to the electronic properties of nano-granular metals might thus become experimentally addressable in future work. Among them are: What is the dependence of the Coulomb charging energy $E_C$ on the coupling strength $g$ as the critical region around the metal-insulator transition is approached and how does it depend on the dimensionality of the nano-granular metal? In which part of the low-temperature transport regimes of nano-granular metals does universal behavior arise and what are the consequences of beginning coherent electron motion over several metal particles in the 1D, 2D and 3D case? Also, tunable nano-granular metals with magnetic nano-grains might help to provide insight into the formation of super-ferromagnetic or super-antiferromagnetic collective states in competition with super-spinglasses, depending on the degree of disorder \cite{Bedanta2007_superferromagnetism_evidence, Morup2010_review_interacting_magnetic_nanoparticles}. Extension of previous work on ordered 2D nano-island lattices to magnetic structures is also highly promising with regard to the controlled formation of such collective phases, as well as in coupling these to other collective phases, like the superconducting. As such, the capability of fabrication of ferromagnetic FEBID structures with high metal content \cite{Begun2015_Co_purification_H2, Dobrovolskiy2015_CoPt_treatment_H2_O2} is of particular advantage for studying the magnetization dynamics of nanomagnets \cite{Lara2014_Co_discs_half_antivortex} and helps to advance investigations of spin-triplet proximity-induced superconductivity in S/F nanostructures. At a later stage, when S/F structures with a large spatial extend of spin-triplet superconducting correlations will be available, this will open a route to studying vortex matter in low-dimensional systems with proximity-induced superconductivity --- an intriguing research line \cite{Kopnin2013_vortex_low_dim_proximity} which lacks experimental investigation so far. While new insights into vortex matter are also expected from studying it in more sophisticated 3D nano-architectures \cite{Thurmer2010_nanomembrane_sc_junctions, Fomin2012_correlated_vortex_tuning}, one should expect that the application of FEBID in the next years will cover an even longer list of sample preparation tasks ranging from the deposition of nanostructures with complex topology to the fabrication of out-of-plane contact electrodes.

With respect to thin-film fluxonic applications, the following advantage of FEBID should be emphasized. A particular feature of structures prepared by conventional lithographic techniques is that they are planar, while FEBID is also suitable for the fabrication of 3D pinning structures. This is especially relevant for the fabrication of asymmetric (ratchet) pinning landscapes for Abrikosov vortices, whose peculiar feature consists in the appearance of a rectified net vortex motion and its reversal \cite{Plourde2009_pinning_anisotropic, Shklovskij2014_vortex_ratchet_asymmetric}. While the functionality of superconductors with linearly-extended asymmetric pinning structures fabricated by FIB milling has been demonstrated for microwave filters \cite{Dobrovolskiy2015_dual_cutoff_filter} and fluxonic metamaterials \cite{Dobrovolskiy2015_ac_microwave_loss_modulation}, we foresee next-generation electronic applications of the fluxonic type realized using re-programmable vortex pinning landscapes fabricated by FEBID. Taking into account the possibility to fabricate 3D magnetic structures by FEBID \cite{Cordoba2016_Fe_Co_3D_nanopillars, Keller2017_magnetic_3D}, one should expect even higher tunability of the strength of vortex pinning induced by them.
%
% Sensor applications
%
\section{Sensor applications}
As has been discussed in section \ref{sec_nanogranular_metals}, charge transport in nano-granular metals depends sensitively on the inter-grain distance, the grain size and the dielectric properties. The corresponding electronic parameters are the inter-granular tunnel coupling $g$, the charging energy $E_C$ and the dielectric function of the matrix material $\epsilon(T, \omega)$. Here $\omega$ denotes the frequency if the granular metal is subject to a harmonic electric field. The nano-granular microstructure of many FEBID materials suggest their use for different sensing tasks, in particular strain and modifications in the dielectric environment, as will be discussed below. In the case of nano-granular magnetic materials, the strongly enhanced anomalous Hall effect in proximity to the metal-insulator transition can be exploited for the realization of magnetic stray field sensors, on which we will also briefly dwell in this section. The relevance of the FEBID approach for theses different sensor application fields mainly derives from the high degree of miniaturization that can be achieved, but also the possibility of fine-tuning the sensor response function by simple modifications of the FEBID process.
%
% Strain sensing
%
\subsection{Strain sensing}
\begin{figure}
\centering
\includegraphics[width=0.8\linewidth]{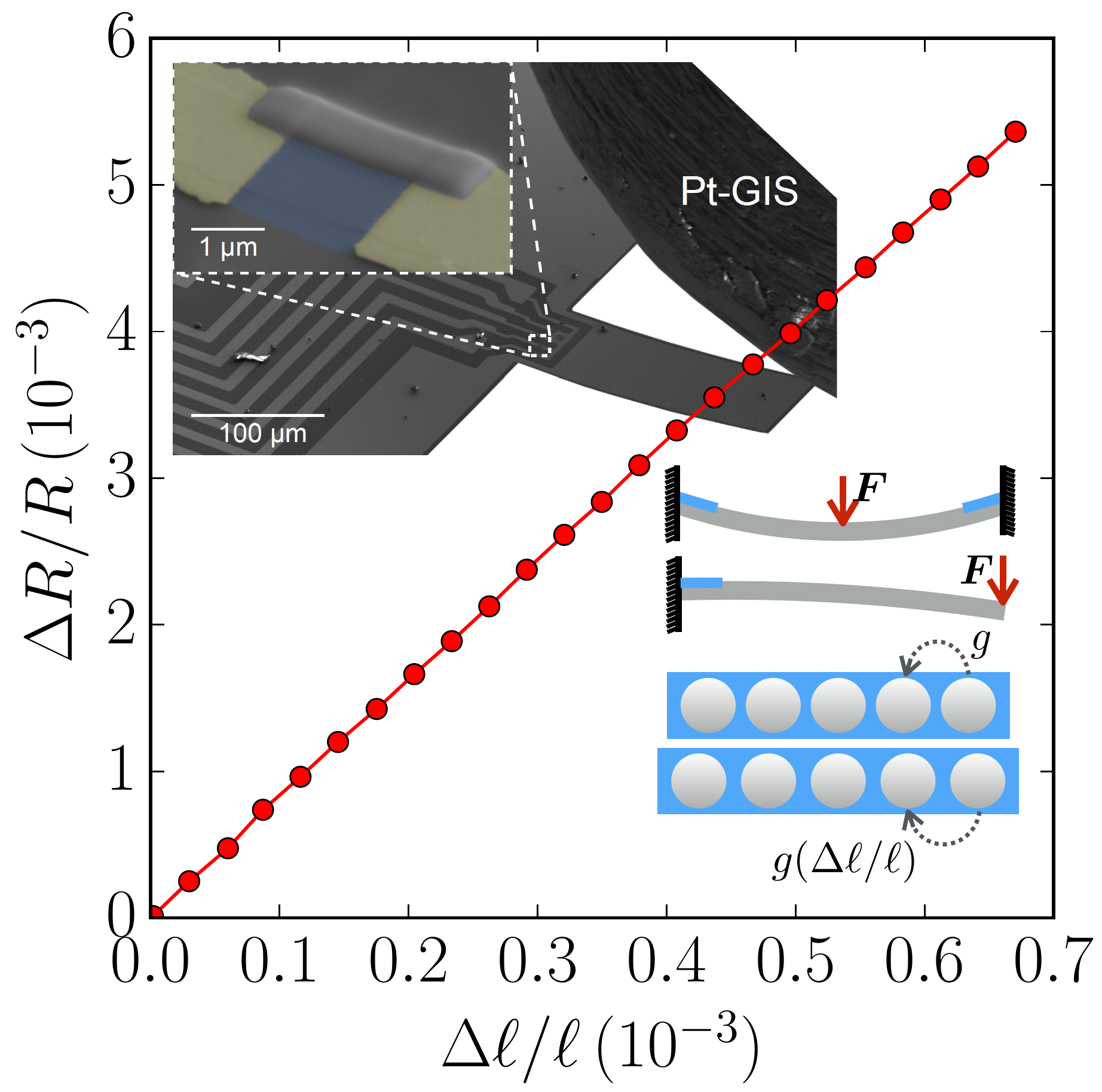}
\caption{Exemplary resistance-strain characteristics of nano-granular Pt sensor element with gauge factor of about $8$ at the bending edge of a soft SiN$_x$ cantilever structure (see SEM image in the upper left). Lower right inset: Schematic of the change in the inter-granular distance and coupling strength $g$ under tensile strain when a force $F$ is applied to a bridge, membrane or cantilever equipped with a nano-granular sensing layer. The sensing layer is deposited at the positions of largest strain.}
\label{fig_strain_sensing}
\end{figure}

Charge transport in nano-granular metals is dominated by tunneling, as has been pointed out in section \ref{sec_nanogranular_metals}. From this it is apparent that nano-granular metals might be suitable materials for strain-sensing, since the tunnel coupling has an exponential dependence on the inter-grain distance which is altered under strain (see also lower right inset of Fig.\,\ref{fig_strain_sensing}). A key parameter that characterizes a strain sensor is its gauge factor $\kappa$ given by
\begin{equation}
\kappa = \frac{\Delta R}{R} \left/ \frac{\Delta\ell}{\ell} \right. \,,
\label{eq_gauge_factor}
\end{equation}
where $\Delta R$ denotes the resistance change under the relative length change $\Delta \ell/\ell$ or strain. In 2010 one of us (MH) could show in a theoretical analysis, based on recent advances in the understanding of the charge transport regimes in nano-granular metals \cite{Beloborodov2007_granular_electronic_systems}, that nano-granular FEBID materials are indeed promising with regard to strain sensing, in particular for use in highly-miniaturized micro-electromechanical (MEMS) sensors \cite{Huth2010_theory_strain_sensing}. In the same year we could experimentally demonstrate that nano-granular Pt FEBID structures are indeed highly linear strain sensing elements that can be quite easily integrated in cantilever sensors \cite{Schwalb2010_strain_sensing}. Fig.\,\ref{fig_strain_sensing} shows the relative resistance change under strain for a soft SiN$_x$ cantilver equipped with a nano-granular Pt sensor.

The possibility to finely tune the inter-granular coupling strength of nano-granular Pt by post-growth electron irradiation (see section \ref{sec_nanogranular_metals}) allows for the optimization of the strain sensor's signal-to-noise ratio which is mandatory for its use in such demanding applications as AFM with self-sensing cantilevers, see \cite{Dukic2016_afm_sensor} for details. In very recent work Mocza\l a and collaborators used the same nano-granular Pt FEBID structures as deflection sensing elements and applied them to read out the resonance frequency of micromechanical SiN$_x$ bridges \cite{Moczala2017_ebid_sensor}. By stress distribution modification employing FIB milling they could increase the deflection detection sensitivity even further.
%
% Gas and dielectric sensing
%
\subsection{Gas and dielectric sensing}
Charge transport in nano-granular metals depends very sensitively on the dielectric environment in which the metallic grains are embedded. In order to see this, we briefly enlarge upon what has already been said concerning nano-granular metals in section \ref{sec_nanogranular_metals}.

The effects of granularity in the charge transport properties are most pronounced at temperatures $T$ above $\Gamma/k_{B}$, where $\Gamma$ denotes the life-time broadening of the energy levels at the Fermi level inside a grain as a consequence of the tunnel-coupling to the neighboring grains. Electron correlations come into play, if one considers that, as an electron is tunneling from one neutral grain to a neighboring grain, a charging energy $E_C$ accrues which amounts approximately to $E_{C}=e^{2}/2C$, where $e$ is the electron charge and $C$ stands for the capacitance
of the grain in its dielectric environment. For a spherical grain, the capacitance is $C=4\pi\epsilon_{0}\epsilon D$ with $\epsilon_{0}$ and $\epsilon$ the dielectric constant of the vacuum and the effective dielectric constant of the granular material, respectively. The energy $E_{C}$ hinders charge transport and eventually leads to a hard or soft energy gap (depending on the degree of disorder) in the density of states at the Fermi level \cite{Beloborodov2004_dos_weak_coupling, Beloborodov2005_dos_close_MIT}. The influence of $E_C$ is strongly visible only in the weak inter-grain coupling $g<0.1$ regime, in which charge transport occurs via c-VRH. In this regime the temperature-dependent conductivity $\sigma(T, \omega=0)$ follows \cite{Beloborodov2007_granular_electronic_systems}
\begin{equation}
\sigma(T) = \sigma_{0}\exp\left\{ -\left(\frac{T_{0}}{T}\right)^{1/2}\right\} \quad\textnormal{with}\quad k_{B}T{}_{0}\approx e^{2}/4\pi\epsilon_{0}\epsilon\xi(T) \,,
\label{eq_cVRH}
\end{equation}
in which the activation temperature $T_0$ ($k_B$: Boltzmann temperature) depends on the relevant type of sequential co-tunneling governing this transport regime via the wave function attenuation length or co-tunneling range $\xi$. $\xi$ takes on different functional forms in the elastic and inelastic co-tunneling channel, see \cite{Beloborodov2007_granular_electronic_systems} for details. The dielectric sensing effect is based on the fact that changes of the effective dielectric properties of the environment of the metallic grains are directly reflected in changes of the charging energy $E_{C}$ which depends on $\epsilon(T,\omega)$.

\begin{figure}
\centering
\includegraphics[width=0.8\linewidth]{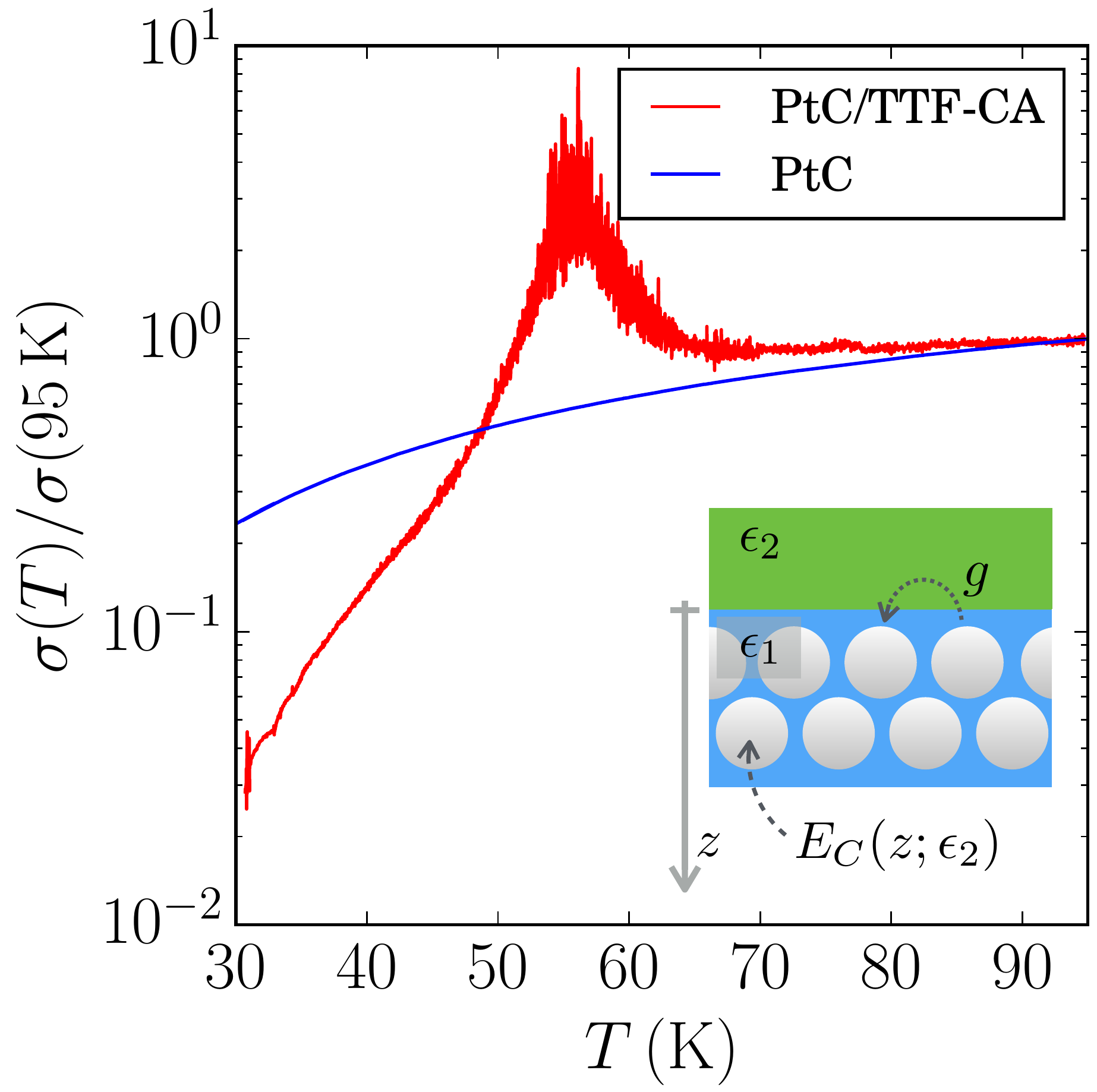}
\caption{Conductance vs.\ temperature behavior of nano-granular Pt FEBID sensing layer as part of a bilayer structure (red curve) consisting of nano-granular Pt (effective dielectric constant $\epsilon_1$, see inset) and an insulating dielectric top layer (dielectric constant $\epsilon_2$, see inset; here TTT-p-chloranil). The conductance increase is caused by $E_C$ renormalization effects induced by the ferroelectric transition of TTF-p-chloranil at about $56\,$K \cite{Huth2014_sensor_ttfca}. The blue curve shows the temperature-dependent conductance of a reference Pt nano-granular metal layer prepared under identical conditions but without TTF-p-chloranil top layer.}
\label{fig_dielectric_sensing}
\end{figure}

We consider a bilayer-system consisting of a nano-granular metal layer of typical thickness $5$ to $20\,$nm on which a non-conducting top layer to be dielectrically characterized is deposited (see schematic in Fig.\,\ref{fig_dielectric_sensing}). The activation temperature $T_0$, characterizing the nano-granular metal in the weak-coupling regime, depends on the wave function attenuation length $\xi(T)$ which itself is governed by the charging energy $E_{C}$ of the metallic grains in the following form (inelastic co-tunneling regime) \cite{Beloborodov2007_granular_electronic_systems}
\begin{equation}
\xi(T)=2D/\ln\left[E_{C}^{2}/16\pi g\left(k_{B}T\right)^{2}\right]\,.
\label{eq_attenuation_length_ineleastic_co_tunneling}
\end{equation}
Apparently, the activation temperature $T_{0}$ depends on two accounts on the dielectric environment of the metallic grains. First, it is inversely proportional to the effective dielectric constant $\epsilon$ of the surrounding medium. Second, changes of the charging energy $E_{C}$ caused by the surrounding medium will directly modify the attenuation length and, thus, the activation temperature. In recent work one of us (MH) developed a model to account for the conductance changes induced in the nano-granular metal by the dielectric properties of the top layer. On the mean field level, the model describes the change of
the effective dielectric constant experienced by the metallic grains at various distances to the interface of a nano-granular metal / dielectric layer heterostructure \cite{Huth2014_diel_sensing_theory}.

The first experimental evidence for the dielectric sensing effects was demonstrated by Kolb and collaborators \citep{Kolb2013_H2O_sensing} from the group of Harald Plank. In this work it was shown that strong conductance changes in $5$ to $20\,$ nm thick nano-granular Pt FEBID layers occur under adsorption and desorption of water with sub-monolayer sensitivity \citep{Kolb2013_H2O_sensing}. These observations eventually lead to the development of the mean-field theory of the sensing mechanism mentioned above \cite{Huth2014_diel_sensing_theory}. This could then also be applied to understand the observed  conductance modulation in nano-granular Pt FEBID sensing layers on top of which a thin film of the insulating, organic ferroelectric TTF-p-chloranil (TTF: tetrathiafulvalene) was deposited \cite{Huth2014_sensor_ttfca}. TTF-p-chloranil layers (under tensile strain) show a paraelectric-to-ferroelectric phase transition at about $56\,$K, at which the real part of the dielectric constant exhibits a pronounced maximum. As shown in Fig.\,\ref{fig_dielectric_sensing}, this causes a strong increase in the conductance of the nano-granular Pt sensing layer, which is due to the reduction of the charging energy of the Pt nano-grains caused by the enhanced dielectric screening effect of the ferroelectric at the phase transition, see \cite{Huth2014_sensor_ttfca} for details.
%
% Magnetic sensing
%
\subsection{Magnetic sensing}
Several complementary techniques are available for the quantitative detection of local magnetic stray fields, each of which is characterized by specific strengths and weaknesses. The most relevant quantities for magnetic stray field sensors used in scanning probe microscopy applications are the smallest detectable field change and the lateral resolution. Established sensor materials in this regard are two-dimensional electron gases (2DEG), e.\,g.\ based on GaAs/AlGaAs heterostructures, and semimetals, typically Bi. The stray field is measured by detecting the associated Hall voltage $V_H$ induced in a miniaturized Hall cross which is scanned at a small distance (typically below $1\,\mu$m) over the magnetic structure to be mapped \cite{Bending1999_principles_local_hall_sensors}. The thermal voltage noise $V_{th} = \sqrt{4k_BTR\Delta f}$ of such a Hall cross sets a lower limit to the detectable field change $\delta B_{min}$. Here $\Delta f$ denotes the frequency bandwidth. Given the signal-to-noise ratio $\textnormal{SNR}$ \cite{Bending1999_principles_local_hall_sensors}
\begin{equation}
\textnormal{SNR} = \frac{V_H}{V_{th}} = \frac{R_HIB}{\sqrt{4k_BTR\Delta f}} \,,
\label{eq_SNR_hall_cross}
\end{equation}
$\delta B_{min}$ is obtained by setting $\textnormal{SNR}=1$ and solving for $B$
\begin{equation}
\delta B_{min} = \frac{\sqrt{4k_BTR\Delta f}}{R_HI} \,.
\label{eq_bmin_Bi}
\end{equation}
This limit can only be reached in ac measurements, as the voltage noise spectrum typically shows a $1/f$-characteristic, so that the minimal voltage noise is only reached above a temperature- and current-dependent frequency value.

Besides 2DEGs and semimetals, granular ferromagnetic materials are a promising material class for sensitive magnetic stray field measurements. Soon after the discovery of the Hall effect it was found that the Hall coefficient is strongly enhanced in iron sheet metal as compared to non-ferromagnetic metals. The Hall resistivity $\rho_H$ was found to have two contributions. The first contribution relates to the conventional Hall effect $\rho_{OH}$ and is governed by the charge carrier density. The second contribution $\rho_{EHE}$ is denoted as extraordinary (EHE) or anomalous and relates to the ferromagnet's magnetization $M(T, H)$
\begin{equation}
\rho_H = \rho_{OH} + \rho_{EHE} = \mu_0 \left[ R_0H + R_SM(T, H) \right] \,.
\end{equation}
Here $H$ denotes the field component perpendicular to the sample surface and the applied current. $M$ accordingly represents the magnetization component in field direction, so that the corresponding magnetic flux density component $B$ inside the sample is $B = \mu_0 (H + M)$. $R_0$ and $R_S$ are material constants. The enhanced Hall effect in ferromagnets is a consequence of material-intrinsic spin-orbit effects which are reflected in the electronic band structure, as well as chiral scattering contributions from non-magnetic impurities (extrinsic effects). The intrinsic effects are related to the geometric phase of the charge carrying electronic states, see a recent review by Nagaosa for more details \cite{Nagaosa2010_review_anomalous_hall_effect}.

\begin{figure}
\centering
\includegraphics[width=0.6\linewidth]{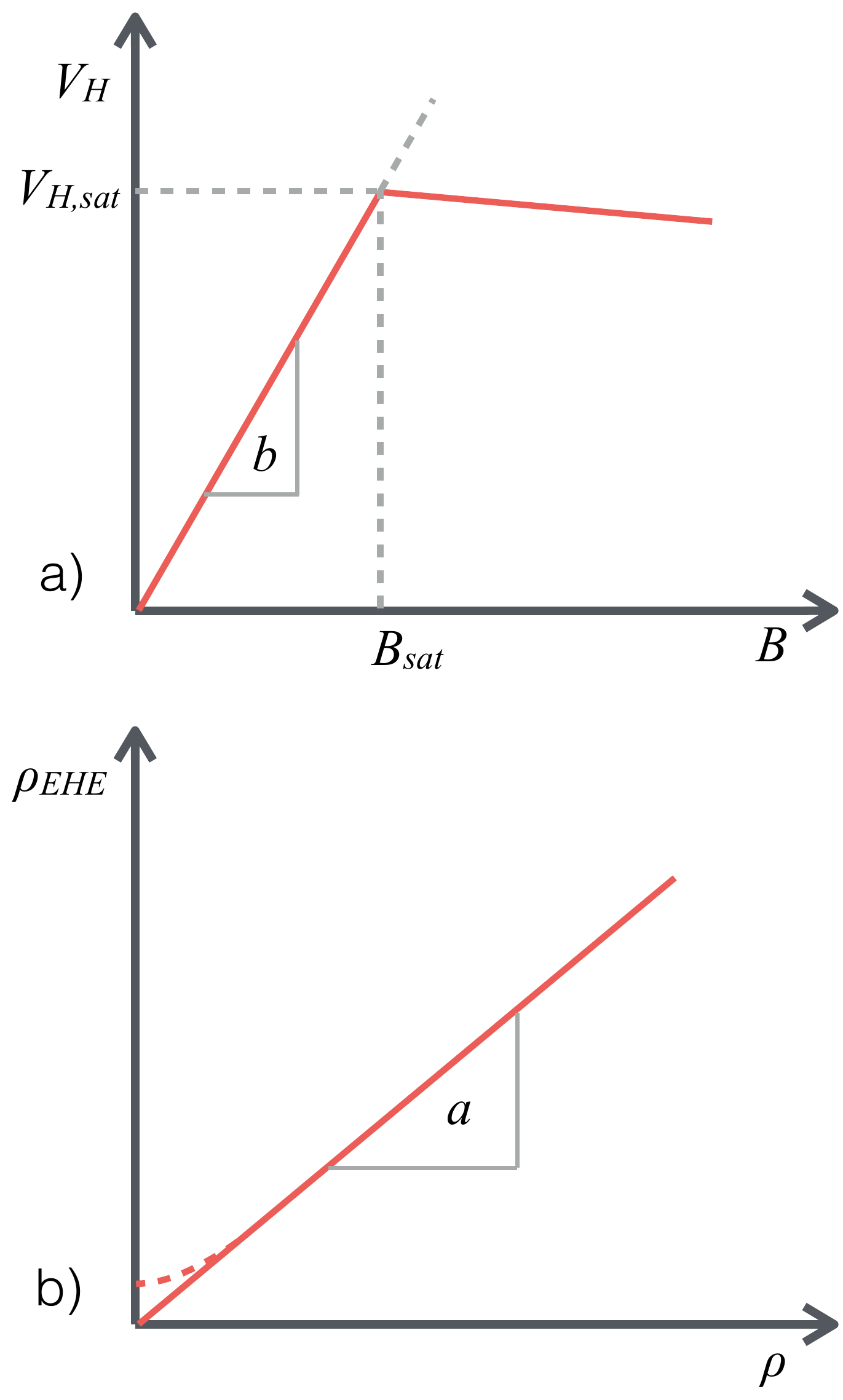}
\caption{(a) Hall voltage vs.\ applied magnetic field for a granular ferromagnet at $T > T_B$. (b) Relation between the extraordinary Hall resistivity and the longitudinal resistivity of a granular ferromagnet.}
\label{fig_schematic_anomalous_hall_effect}
\end{figure}

In order to use the enhanced Hall effect in ferromagnets, two further observations are crucial. First, for granular ferromagnets, i.\,e.\ magnetic grains embedded in an insulating matrix, $\rho_{EHE}$ is found to be even larger than in bulk ferromagnets. For a material- and microstructure-specific volume fraction of the metal grains the anomalous Hall effect can be enhanced by up to three orders of magnitude, which is also denoted as giant anomalous Hall effect \cite{Denardin2003_review_giant_hall_effect_granular_ferromagnets}. Second, due to the granular microstructure the metallic grains tend to remain in a super-paramagnetic state for temperatures above a characteristic blocking temperature $T_B$. In this state magnetic hysteresis effects do not occur. Fig.\,\ref{fig_schematic_anomalous_hall_effect} gives a schematic overview of the response of a Hall sensor based on a granular ferromagnet for $T > T_B$. Also shown is the typically linear dependence of the extraordinary Hall resistivity on the longitudinal resistivity, which is useful for estimating the achievable SNR for a given geometry of a Hall cross (see Eq.\,\ref{eq_SNR_hall_cross}). For magnetic flux density values $B > B_{sat}$, i.\,e.\ above magnetic saturation, the slope of the Hall voltage is only determined by the ordinary Hall effect. However, typical values for $B_{sat}$ are in the $1\,$T-range for granular ferromagnetic Hall structures prepared by FEBID, so that the useful range of magnetic stray field detection is large.

Research into granular ferromagnetic Hall sensor for magnetic stray field detection based on nano-scale FEBID structures was pioneered by the group of Ivo Utke \cite{Boero2005_CoC_Hall_sensor, Gabureac2010_granular_hall_sensors} and later applied to the detection of suspended superparamagnetic beads \cite{Gabureac2013_gfm_magnetic_bead}.  In this work the precursor Co$_2$(CO)$_8$ was used. For Hall-cross sensor areas down to $100\times 100\,$nm$^2$ the smallest detectable field change was found to be $B_{min} \approx 5\,\mu$T/$\sqrt{\textnormal{Hz}}$ at room temperature \cite{Gabureac2010_granular_hall_sensors}. This value is about a factor of $10$ lower than what has been achieved with semimetallic Bi Hall sensors \cite{Sandhu2004_Bi_hall_sensors}. Frequency-dependent voltage noise measurement showed a $1/f$-characteristic running into the Johnson noise limit at about $1\,$kHz at the largest currents the sensors could sustain without being damaged \cite{Boero2005_CoC_Hall_sensor}. Within the magnetic bead detection and tracing experiments a lateral resolution of $230\,$nm$/\sqrt{\mathrm{Hz}}$ was achieved at a field resolution of $300\,\mu$T$/\sqrt{\mathrm{Hz}}$ \cite{Gabureac2013_gfm_magnetic_bead}.

A very systematic analysis of the specific Hall resistivity for Fe FEBID structures was performed by Rosa C\'ordoba and collaborators employing the precursor Fe$_2$(CO)$_9$ \cite{Cordoba2012_GAHE_Fe}. By admixture of variable amounts of water during the depositon process the Fe-to-oxocarbide ratio of the deposits was carefully controlled and the development of the anomalous Hall effect was investigated in dependence of the longitudinal resistivity. An almost linear dependence was found up to anomalous Hall resistivity values of more than $100\,\mu\Omega$cm.
%
% Challenges and perspectives
%
\subsection{Challenges and perspectives}
Strain sensing based on nano-granular FEBID is most promising if applied in very small resonating structures for which the high-resolution direct-write capabilities of FEBID are essentially without alternative \cite{Dukic2016_afm_sensor}. On the other hand, this requires additional development in high-speed detection electronics, in particular for AFM applications, as the resonance frequencies of sub-micron sized structures can extend into the several $10\,$MHz regime.

With regard to the novel dielectric sensing approach, an extension into ac driven systems could be promising. One may expect a universal frequency dependence of the real part of the conductivity, which is long known for a broad range of different  disordered systems \cite{Dyre2000_ac_conductance_universality}. This universality has recently been shown for a nano-granular metal \cite{Bakkali2016_universality_ac_response_granular_metals} but not yet for nano-granular FEBID materials.

Nano-granular ferromagnetic Hall sensors feature a very high sensitivity for magnetic stray fields and show good down-scaling behavior. Nevertheless, they still have to be shown to work reliably in actual sensor devices.

In all cases discussed in this section, the sensing mechanism relied on a particular property of the nano-granular microstructure of the deposits. The longterm stability of the electric and magnetic properties of nano-granular FEBID materials is an issue on which not much has been worked on so far, but see \cite{Winhold2015_phd_thesis} with regard to the transient behavior in the electrical conductance of nano-granular Pt up to the time scale of months. Further work towards a better understanding of aging effects and the development of reliable stabilization procedures will certainly be needed, if nano-granular FEBID sensors are to be applied in real devices in the future.
%
% 3D FEBID structures
%
\section{3D FEBID structures}
Its 3D writing capability is probably one of the prime disciplines of FEBID. Nevertheless, the fabrication of even moderately complex 3D shapes is not straightforward. Over the last two decades, several works dedicated to 3D growth have been published. In early work by Hans Koops and collaborators, photonic crystals \cite{Koops2001_FEBID_photonic_crystals} and field emitters and electron optics structures were addressed \cite{Kretz1994_field_emitter_structures, Koops1995_Pt_tips_FEBID, Floreani2001_field_emitters_FEBID}. Following this, suspended FEBID structures were demonstrated, see e.\,g.\ \cite{Gazzadi2007_suspended_FEBID_structures}, and plasmonic structures came more into focus \cite{Hoeflich2011_plasmonics_Au_3D, Esposito2015_helix_FEBID_plasmonics}, as well as simple 3D structures, such as nano-pillars, with application in nanomagnetism \cite{Pacheco2013_magnetics_Co_3D_spirals, Cordoba2016_Fe_Co_3D_nanopillars, Navarro2017_Co_3D_pillars}. Recently a nanospray liquid precursor FEBID process with strongly enhanced growth rates has been demonstrated by Andrei Fedorov's group \cite{Fisher2015_nanospray_liquid_precursor_3d_FEBID}. By this process nominally pure Ag pillar structures have been fabricated. Future work will have to show whether sub-micron sized deposits are feasible. 

The advent of simulation-assisted 3D growth goes back to very recent joint work of the groups of Harald Plank and Philipp Rack with leading part in the simulations by Jason Fowlkes \cite{Fowlkes2016_febid_3D_simulation}. This work was not yet aimed at functional 3D FEBID structures, but very soon several follow up works focused on 3D pure metal deposits by either a laser-assisted approach \cite{Lewis2017_purification_3D} or post-growth purification \cite{Winkler2017_3D_plasmonic}. Complex nanomagnetic 3D structures have also very recently been fabricated \cite{Keller2017_magnetic_3D}.
%
% Simulating 3D growth
%
\subsection{Simulating 3D growth}
How to simulate the FEBID process at several levels of complexity? It must be stressed at the beginning that it is still a far way to go with regard to a full simulation of all sub-process which are relevant in FEBID, as this represents a multi-scale problem on both, time (femtoseconds to seconds) and length scales (fraction of nanometers to many micrometers).

On the \emph{ab-initio} level several works have recently focused on the adsorption and stability of selected precursor materials on amorphous SiO$_2$ using density functional theory (DFT) \cite{Muthukumar2014_adsorption_several_dft}. Si/SiO$_2$ is by far the most often used substrate material. One important example is therefore the intrinsic instability of the precursor Co$_2$(CO)$_8$ under adsorption onto SiO$_2$ with non-hydroxylated dangling bonds \cite{Muthukumar2012_dissociation_co2co8}. This has been experimentally observed and also holds relevance for Fe(CO)$_5$, a precursor which is frequently used for the deposition of magnetic nanostructures with high metal content \cite{Muthukumar2012_dissociation_co2co8, Vollnhals2013_ebisa_Fe}. A DFT approach was also used in a theoretical electronic structure investigation in the W-C-O phase field \cite{Muthukumar2012_tungsten_mit_uspex} with respect to the metal-insulator transition found in FEBID material obtained from the precursor W(CO)$_6$ \cite{Huth2009_WCO_deposits_MIT}. Very recent work has used a novel molecular dynamics approach to describe electron irradiation-driven transformations of molecular structures (IDMD -- irradiation driven molecular dynamics) with relevance for FEBID, in particular focusing on W(CO)$_6$ \cite{Sushko2016_MD_simulation_WCO6}.  One has to keep in mind that molecular dynamics of FEBID processes is probably feasible into the few microseconds time range, which is by far not sufficient to describe 3D growth. This is why other approaches to FEBID simulation do not address the microscopic details of the electron-induced chemical transformation processes, but rather tackle the problem of how to predict FEBID growth rates and shape evolution of FEBID structures using an effective theory. Such an effective approach requires to combine the spatial and energy distribution of the primary, backscattered and, especially, secondary electrons with the precursor coverage on the growth front. The interaction of an electron beam with a solid can be efficiently modeled by the Monte Carlo (MC) method, as has been recognized early on \cite{Joy1991_monte_carlo_sem}. From this the required spatial and energy distribution of the secondary electrons can be derived as input for a reaction-diffusion equation describing the spatial and temporal evolution of the precursor density on the growth front of the developing deposit. This latter part, commonly known as the FEBIP continuum model, has recently been reviewed by Milos Toth and colleagues for the 2D case \cite{Toth2015_review_cont_models} and is able to simulate both, deposition and electron-induced etching, can handle complex adsorption processes \cite{Bishop2012_activated_chemisorption}, and can deal with etch processes that proceed through multiple reaction pathways with several reaction products present at the substrate surface. The combination of MC simulations of the electron distribution for a Gaussian beam shape with the reaction-diffusion equation has been pioneered by the Rack group \cite{Smith2007_MC_modeling_febid} and was recently refined by Jason Fowlkes and collaborators to the first simulation-guided approach for 3D nano-manufacturing with very impressive examples of experimental realization by the Plank group \cite{Fowlkes2016_febid_3D_simulation}. This may well prove to be the start signal for future activity towards 3D fabrication of sophisticated, functional nanostructures by FEBID. It is therefore appropriate to briefly summarize the main ideas of this development. For details we refer to the work by Fowlkes and collaborators \cite{Fowlkes2016_febid_3D_simulation}.

\begin{figure}
\centering
\includegraphics[width=0.8\textwidth]{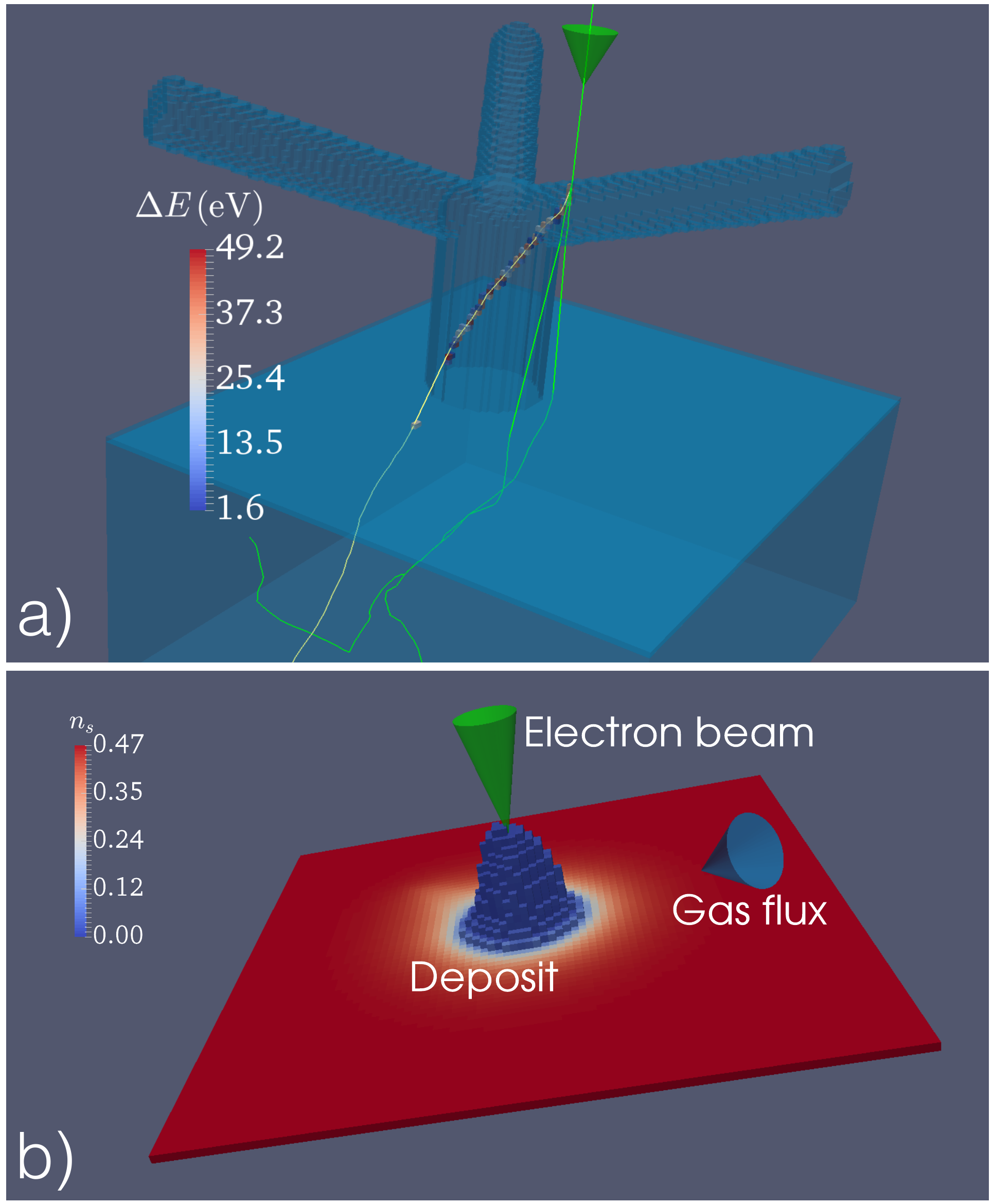}
\caption{Schematic of FEBID growth process based on MC simulation of the electron trajectory and energy loss and numerical solution of the reaction-diffussion for the precursor density $n_s$. (a) Primary electrons with $20\,$keV (green cone and line) enter the deposit and experience a series of elastic and inelastic scattering events. Shown are three exemplary trajectories, for one of which the energy loss associated with the inelastic processes is shown as color-coded volume voxels. Secondary electrons are generated along the trajectories. If close enough to the deposit or substrate surface, they contribute to the electron flux $\phi$ which controls precursor dissociation. (b) Surface position dependent precursor density $n_s$ (see color code) for a pillar-like deposit created by a Gaussian beam, as indicated. The gas flux direction is shown by the blue cone. The shadowing effect created by the deposit is visible as reduced precursor coverage.}
\label{fig_3d_modeling}
\end{figure}
Let us assume that FEBID growth has progressed to a deposit geometry as indicated in Fig.\,\ref{fig_3d_modeling}(a). The electrons of the primary beam enter the deposit, experience many elastic and inelastic collision events, thereby generating secondary electrons, eventually leave the deposit and re-enter at another position on the deposit or proceed into the substrate. All generated secondary electrons which are within the energy-dependent escape depth to the surface of the deposit or the substrate can induce at a given surface position $(x_s, y_s, z_s)$ precursor dissociation with a rate proportional to the local precursor density (coverage) $n_s(x_s , y_s , z_s, t)$. Here we assume that the maximum coverage is one monolayer and consider $n_s$ to be normalized to the one-monolayer density $n_{ML}$, i.\,e.\ $n_s$ takes on values between $0$ and $1$. The dissociation rate is also proportional to the dissociation cross section $\sigma(E)$ which is, in general, energy-dependent and becomes large only for low energies \cite{Thorman2015_dissociation_case_studies}. For lack of detailed data of the energy dependence for almost all of the precursors used in FEBID, it is customary to replace $\sigma(E)$ by an energy-averaged value $\sigma$. The precursor density is furthermore governed by the interplay of the three elemental processes of precursor adsorption, diffusion and desorption. The secondary electron distribution $\phi(x_s, y_s, z_s, t)$ at all surface positions $(x_s, y_s, z_s)$ obtained from the MC simulation now enters the reaction-diffusion equation for the precursor density which reads
\begin{equation}
\frac{\partial n_s}{\partial t} = D\left( \frac{\partial^2 n_s}{\partial x^2} + \frac{\partial^2 n_s}{\partial y^2} \right) - \sigma\phi(x_s, y_s, z_s, t)n_s + \frac{\Phi}{n_{ML}} \left( 1 - n_s \right) - \frac{n_s}{\tau}\,.
\label{eq_reaction_diffusion_febid}
\end{equation}
$\tau$ is the temperature-dependent average residence time of a precursor molecule. $\Phi(x_s, y_s, z_s, t)$ represents the precursor flux provided by the gas injection system and can -- for flat deposit shapes -- be estimated from kinetic gas theory for a given precursor pressure at the substrate surface. However, for sufficiently accurate simulations of 3D FEBID a more sophisticated approach is necessary. Experimental studies have demonstrated that the precursor coverage depends sensitively on the direction of the gas flux vector, which is determined by the geometrical arrangement of the gas injector capillary \cite{Winkler2014_gas_flux_influence}. This has been simulated in good quantitative agreement with experimental results by Friedli and collaborators \cite{Friedli2009_gas_flux_simulation}. The gas flux distribution $\Phi(x_s, y_s, z_s, t)$ in Eq.\,\ref{eq_reaction_diffusion_febid} therefore has two components. The first component, $\Phi_1(t)$, which may depend on time but not on the surface position, is the result of precursor adsorption governed by the precursor pressure above the surface. The second component, $\Phi_2(x_s, y_s, z_s, t)$ depends on the surface position for two reasons \cite{Fowlkes2016_febid_3D_simulation}. For one, depending on the orientation of the surface normal at this position with regard to the gas flux vector, the flux will be higher or lower. Secondly, the flux may be reduced by shadowing effects caused by the current 3D deposit shape. This effect is exemplified in Fig.\,\ref{fig_3d_modeling}(b) that shows the result of the numerical solution of Eq.\,\ref{eq_reaction_diffusion_febid} for a stationary, Gauss-shaped beam and a directed precursor flux component, as indicated. The actual time-evolution of the growth front can directly by obtained from the known precursor density at a given time and position using the following relationship between the local height increase $\Delta h_\perp$ (perpendicular to the local surface) and $n_s(x_s, y_s, z_s, t)$
\begin{equation}
\Delta h_\perp(x_s, y_s, z_s, t) = V_D\sigma\phi(x_s, y_s, z_s, t)n_sn_{ML}\Delta t
\label{eq_relationship_growthrate_precursor_density}
\end{equation}
within a time step $\Delta t$. Here $V_D$ denotes the volume deposited for each dissociated precursor molecule. In the next sub-section we will provide examples for 3D FEBID for which the simulation-guided approach was applied.
%
% Examples for simulation-guided 3D FEBID growth
%
\subsection{Examples for simulation-guided 3D FEBID growth}
By way of introducing a hybrid Monte Carlo / continuum simulation and comparison with experimental FEBID results using the precursor Me$_3$CpMePt(IV), Jason Fowlkes and collaborators established the foundation for the simulation-guided 3D nanofabrication with FEBID \cite{Fowlkes2016_febid_3D_simulation}. In follow-up work a software tool for 3D computer-aided design was presented, which generates the beam parameters necessary for FEBID by both, simulation and experiment \cite{Fowlkes2017_upcoming_cad_tool}. First application in functional 3D nano-architectures was demonstrated by Robert Winkler and collaborators in the field of resonant optics \cite{Winkler2017_3D_plasmonic}. Following direct-write fabrication of 3D structures using the precursor Me$_2$(acac)Au(III), a water-assisted purification process was applied, which left the 3D structures mostly intact, see also section \ref{sec_post_growth_purification}. By electron energy loss spectroscopy (EELS) on tetrapod Au structures the plasmonic activity of these structures was clearly demonstrated.

A highly interesting example of correlation microscopy employing 3D FEBID in conjunction with SEM and \emph{in-situ} AFM characterization was recently published by the groups of Harald Plank (FEBID) and Georg Fantner (high-speed AFM) \cite{Yang2017_3d_FEBID_and_AFM}. In this work the evolution of the mechanical strength during the growth of Pt-based 3D FEBID structures was investigated by employing a slice-by-slice approach in which the sequence of FEBID growth, followed by AFM characterization was repeated many times over.

\begin{figure}
\centering
\includegraphics[width=0.95\textwidth]{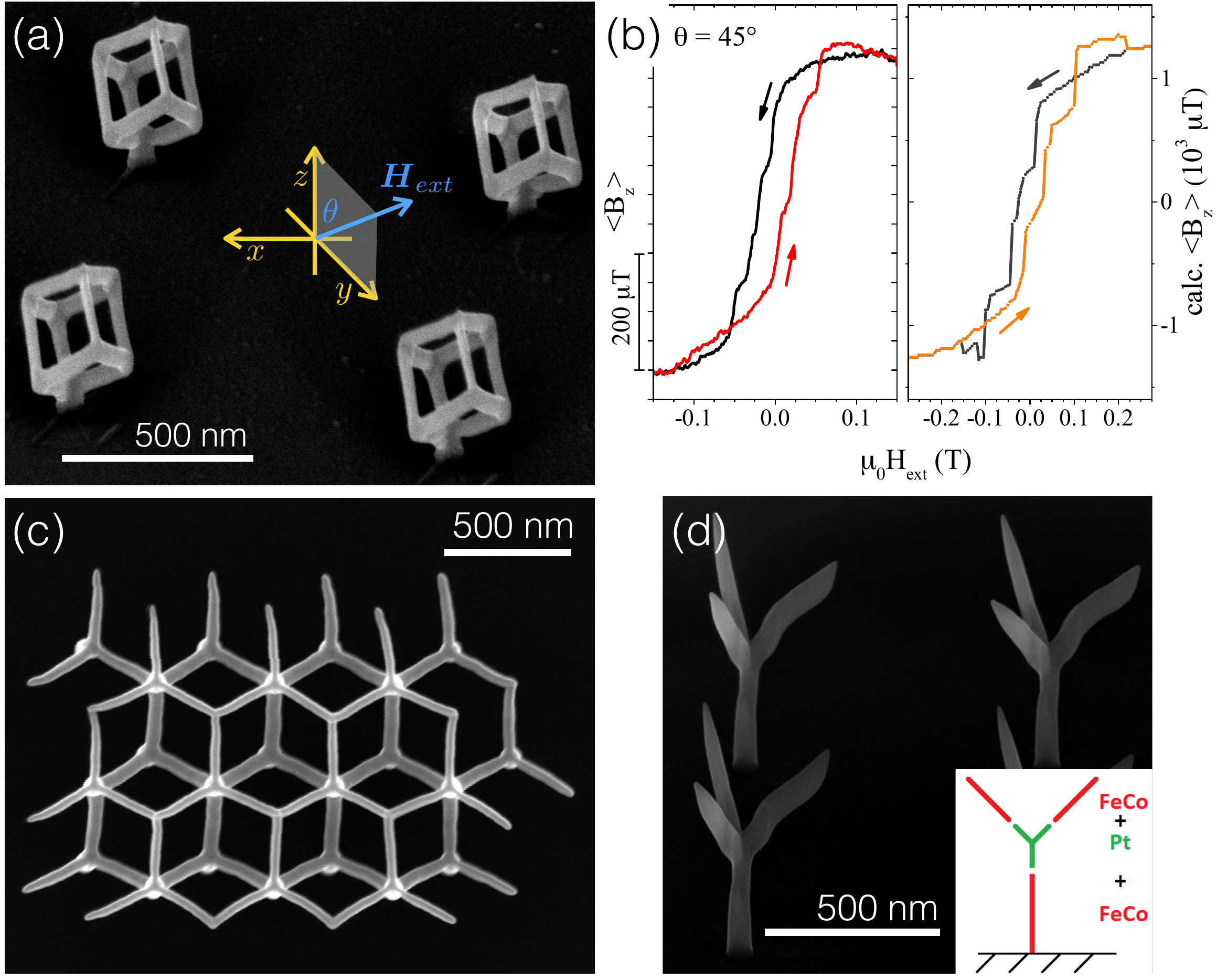}
\caption{(a) $2\times 2$-array of Co$_3$Fe nano-cubes on micro-Hall sensor. (b) Left: Magnetic stray field as measured at $30\,$K by micro-Hall magnetometry of nano-cubes in an applied magnetic field under $\theta=45^\circ$ with regard to the surface normal; see inset in (a). Right: Result of $T=0$ micromagnetic simulation. (c) Top view of 3D array of Co$_3$Fe nano-trees (diamond structure). (d) $2\times 2$-array of Co$_3$Fe nano-trees in which the vertex segment consists of nano-granular Pt, as indicated in the inset. See \cite{Keller2017_magnetic_3D} for details.}
\label{fig_3d_nanogmagnetism}
\end{figure}

As the last example for application of the simulation-guided approach we briefly dwell on 3D nanomagnetic structures, a research field of rapidly growing interest \cite{Pacheco2017_3d_nanomagnetism}. In very recent work Lukas Keller and collaborators demonstrated the direct-write fabrication of freestanding ferromagnetic 3D Co$_3$Fe nano-architectures (precursor HCo$_3$Fe(CO)$_{12}$) with focus on the consequences of frustrated magnetic interactions \cite{Keller2017_magnetic_3D}. In particular, nano-cube and nano-tree structures were chosen, as these imply magnetic vertex segments in which three or four magnetic edges of roughly cylindrical shape join (see Fig.\,\ref{fig_3d_nanogmagnetism}(a) and (d)). Individual nano-cube and nano-tree Co$_3$Fe structures were carefully magnetically characterized by employing micro-Hall sensing in conjunction with micromagnetic and macro-spin simulations (see Fig.\,\ref{fig_3d_nanogmagnetism}(c)). By carefully adapting the writing strategy and with support by a software tool that generates suitable pattern files for FEBID \cite{Keller2017_pattern_generator}, it was possible to write 3D arrays of Co$_3$Fe nano-trees forming a diamond-like lattice which is very interesting for future work on artificial 3D spin-ice systems. As an additional novel aspect in this work, Keller \emph{et al.} demonstrated the combination of ferromagnetic 3D elements with other 3D elements of different chemical composition and intrinsic material properties, namely nano-granular Pt on the vertex segments positions in nano-trees whose edges were made from Co$_3$Fe alloy (see Fig.\,\ref{fig_3d_nanogmagnetism}(c)).
%
% Challenges and perspectives
%
\subsection{Challenges and perspectives}
Simulation-assisted 3D FEBID growth in conjunction with the developed software tools to generate suitable pattern files will likely boost future activity in the fabrication of 3D nano-architectures for several application fields, such as resonant optics and nanomagnetism, to name just two. This may turn out to be one of the most promising recent developments in FEBID with a view to expanding both, the community of FEBID users as well as the application fields in which FEBID has a chance to bequeath a sustainable impression in the coming years. Nevertheless, there is plenty of room for future work on optimizing the 3D writing strategies, e.\,g.\ with regard to shape control of the different structural units that make a 3D element, the reliable fabrication of larger array structures, and the precision with which different FEBID materials can be combined in the 3D heterostructures.  
%
% Concluding remarks
%
\section{Concluding remarks}
Chances have never been so good for FEBID to become the method of choice for nanofabrication in an ever broader range of research and application fields. The development of various processes for clean metal nanofabrication, the combination of FEBID with area-selective atomic layer deposition, as well as the introduction of well-defined protocols for 2D and true 3D nanofabrication of structures for resonant optics and nanomagnetism will strongly increase the awareness about the unique usefulness of this direct-write approach. Nevertheless, significant challenges have to be met in several ongoing FEBID research fields. E.\,g., a FEBID process still has to be developed that yields a homogeneous superconducting material positioned far on the superconducting side of the superconductor-insulator transition. Also, with regard to sensor applications based on nano-granular materials, long-term stability is an important issue which needs to be addressed.

Based on the present status of the field, possible directions for future development could be 3D FEBID nano-architectures for resonant optics and nanomagnetism, as well as a broadening of the application of FEBID for the fabrication of artificial solids and metamaterials. Due to the extremely high flexibility and resolution capabilities of FEBID, in conjunction with the ever growing reservoir of suitable precursors for a wide range of material properties of the deposits, one can expect many more research and application fields for FEBID to open up in the coming years, many of which are presently not anticipated.
%
% Acknowledgements
%
\section*{Acknowledgements}
M.~H.\ acknowledges financial support by the Deutsche Forschungsgemeinschaft (DFG) under grant No.\ HU 752/11-1, through the priority program 1928 (Coordination Networks: Building Blocks for Functional Systems) under grant HU 752/12-1, and through the Collaborative Research Centre SFB/TR\,49 (Condensed Matter Systems with Variable Many-Body Interactions). O.~D.\ acknowledges financial support by the DFG under grant DO 1511/3-1. This work was conducted within the framework of the COST Action CM1301 (CELINA). Part of the work on hybride superconductor/ferromagnet nanostructures was conducted within the framework of the COST Action CA16218 (NANOCOHYBRI).
%
% references
%
\section*{References}
\bibliographystyle{unsrt}
\bibliography{mee_2016_rev1}

\begin{thebibliography}{100}

\bibitem{Randolph2006_febid_review}
S.~J. Randolph, J.~D. Fowlkes, and P.~D. Rack.
\newblock Focused, nanoscale electron-beam-induced deposition and etching.
\newblock {\em Critical Reviews in Solid State and Materials Sciences},
  31(3):55--89, 2006.

\bibitem{Utke2008_febid_review}
Ivo Utke, Patrik Hoffmann, and John Melngailis.
\newblock Gas-assisted focused electron beam and ion beam processing and
  fabrication.
\newblock {\em Journal of Vacuum Science \& Technology B: Microelectronics and
  Nanometer Structures Processing, Measurement, and Phenomena},
  26(4):1197--1276, 2008.

\bibitem{Huth2012_febid_review}
Michael Huth, Fabrizio Porrati, Christian Schwalb, Marcel Winhold, Roland
  Sachser, Maja Dukic, Jonathan Adams, and Georg Fantner.
\newblock Focused electron beam induced deposition: A perspective.
\newblock {\em Beilstein Journal of Nanotechnology}, 3:597--619, 2012.

\bibitem{Edinger2014_mask_repair}
K.~Edinger, K.~Wolff, H.~Steigerwald, N.~Auth, P.~Spies, J.~Oster,
  H.~Schneider, M.~Budach, T.~Hofmann, and M.~Waiblinger.
\newblock Bringing mask repair to the next level.
\newblock {\em Proc. SPIE}, 9235:92350R--92350R--9, 2014.

\bibitem{Makise2014_ebid_superconductivity}
Kazumasa Makise, Kazutaka Mitsuishi, Masayuki Shimojo, and Bunju Shinozaki.
\newblock Microstructural analysis and transport properties of {MoO} and {MoC}
  nanostructures prepared by focused electron beam-induced deposition.
\newblock {\em Scientific Reports}, 4:5740, 2014.

\bibitem{Sengupta2015_W_ebid_superconductivity}
Shamashis Sengupta, Chuan Li, Cedric Baumier, Alik Kasumov, S.~Gu{\'e}ron,
  H.~Bouchiat, and F.~Fortuna.
\newblock Superconducting nanowires by electron-beam-induced deposition.
\newblock {\em Applied Physics Letters}, 106(4):042601, 2015.

\bibitem{Winhold2014_Pb_superconductor}
M.~Winhold, P.~M. Weirich, C.~H. Schwalb, and M.~Huth.
\newblock Superconductivity and metallic behavior in {Pb$_x$C$_y$O$_\delta$}
  structures prepared by focused electron beam induced deposition.
\newblock {\em Applied Physics Letters}, 105(16):162603, 2014.

\bibitem{DeTeresa2016_ebid_magnetic_review}
J~M~De Teresa, A~Fern\'andez-Pacheco, R~C\'ordoba, L~Serrano-Ram\'on,
  S~Sangiao, and M~R Ibarra.
\newblock Review of magnetic nanostructures grown by focused electron beam
  induced deposition ({FEBID}).
\newblock {\em Journal of Physics D: Applied Physics}, 49(24):243003, 2016.

\bibitem{Che2005_FEBID_FePt_holography}
R.~C. Che, M.~Takeguchi, M.~Shimojo, W.~Zhang, and K.~Furuya.
\newblock Fabrication and electron holography characterization of {FePt} alloy
  nanorods.
\newblock {\em Applied Physics Letters}, 87(22):223109, 2005.

\bibitem{Winhold2011_PtSi_alloy}
Marcel Winhold, Christian~H. Schwalb, Fabrizio Porrati, Roland Sachser,
  Achilleas~S. Frangakis, Britta K{\"a}mpken, Andreas Terfort, Norbert Auner,
  and Michael Huth.
\newblock Binary {Pt-Si} nanostructures prepared by focused
  electron-beam-induced deposition.
\newblock {\em ACS Nano}, 5(12):9675--9681, 2011.
\newblock PMID: 22050515.

\bibitem{Porrati2012_CoPt_alloy}
F~Porrati, E~Begun, M~Winhold, Ch~H Schwalb, R~Sachser, A~S Frangakis, and
  M~Huth.
\newblock Room temperature {L1$_0$} phase transformation in binary {CoPt}
  nanostructures prepared by focused-electron-beam-induced deposition.
\newblock {\em Nanotechnology}, 23(18):185702, 2012.

\bibitem{Porrati2013_CoSi_alloy}
F.~Porrati, B.~K{\"a}mpken, A.~Terfort, and M.~Huth.
\newblock Fabrication and electrical transport properties of binary {Co-Si}
  nanostructures prepared by focused electron beam-induced deposition.
\newblock {\em Journal of Applied Physics}, 113(5):053707, 2013.

\bibitem{Shawrav2014_AuFe_alloy}
Mostafa~Moonir Shawrav, Domagoj Beli{\'c}, Marco Gavagnin, Stefan Wachter,
  Markus Schinnerl, Heinz~D. Wanzenboeck, and Emmerich Bertagnolli.
\newblock Electron beam-induced {CVD} of nanoalloys for nanoelectronics.
\newblock {\em Chemical Vapor Deposition}, 20(7-8-9):251--257, 2014.

\bibitem{Porrati2015_CoFe_precursor}
F~Porrati, M~Pohlit, J~M{\"u}ller, S~Barth, F~Biegger, C~Gspan, H~Plank, and
  M~Huth.
\newblock Direct writing of {CoFe} alloy nanostructures by focused electron
  beam induced deposition from a heteronuclear precursor.
\newblock {\em Nanotechnology}, 26(47):475701, 2015.

\bibitem{Porrati2016_FeSi_alloy}
F.~Porrati, R.~Sachser, G.~C. Gazzadi, S.~Frabboni, and M.~Huth.
\newblock Fabrication of {FeSi} and {Fe$_3$Si} compounds by electron beam
  induced mixing of {[Fe/Si]$_2$} and {[Fe$_3$/Si]$_2$} multilayers grown by
  focused electron beam induced deposition.
\newblock {\em Journal of Applied Physics}, 119(23):234306, 2016.

\bibitem{Porrati2017_FeCoSi_multilayer}
Fabrizio Porrati, Roland Sachser, Gian-Carlo Gazzadi, Stefano Frabboni, and
  Michael Huth.
\newblock Alloy multilayers and {Heusler} nanostructures by direct-write
  approach.
\newblock {\em Nanotechnology (accepted)}, 2017.

\bibitem{Dobrovolskiy2015_CoPt_treatment_H2_O2}
Oleksandr~V. Dobrovolskiy, Maksym Kompaniiets, Roland Sachser, Fabrizio
  Porrati, Christian Gspan, Harald Plank, and Michael Huth.
\newblock Tunable magnetism on the lateral mesoscale by post-processing of
  {Co/Pt} heterostructures.
\newblock {\em Beilstein Journal of Nanotechnology}, 6:1082--1090, 2015.

\bibitem{Schwalb2010_strain_sensing}
Christian~H. Schwalb, Christina Grimm, Markus Baranowski, Roland Sachser,
  Fabrizio Porrati, Heiko Reith, Pintu Das, Jens M{\"u}ller, Friedemann
  V{\"o}lklein, Alexander Kaya, and Michael Huth.
\newblock A tunable strain sensor using nanogranular metals.
\newblock {\em Sensors}, 10(11):9847--9856, 2010.

\bibitem{Huth2014_diel_sensing_theory}
Michael Huth, Florian Kolb, and Harald Plank.
\newblock Dielectric sensing by charging energy modulation in a nano-granular
  metal.
\newblock {\em Applied Physics A}, 117(4):1689--1696, 2014.

\bibitem{Huth2014_sensor_ttfca}
M~Huth, A~Rippert, R~Sachser, and L~Keller.
\newblock Probing near-interface ferroelectricity by conductance modulation of
  a nano-granular metal.
\newblock {\em Materials Research Express}, 1(4):046303, 2014.

\bibitem{Dukic2016_afm_sensor}
Maja Dukic, Marcel Winhold, Christian~H. Schwalb, Jonathan~D. Adams, Vladimir
  Stavrov, Michael Huth, and Georg~E. Fantner.
\newblock Direct-write nanoscale printing of nanogranular tunnelling strain
  sensors for sub-micrometre cantilevers.
\newblock {\em Nature Communications}, 7:12487 EP --, 09 2016.

\bibitem{Moczala2017_ebid_sensor}
M.~Mocza{\l}a, K.~Kwoka, T.~Piasecki, P.~Kunicki, A.~Sierakowski, and
  T.~Gotszalk.
\newblock Fabrication and characterization of micromechanical bridges with
  strain sensors deposited using focused electron beam induced technology.
\newblock {\em Microelectronic Engineering}, 176:111 -- 115, 2017.
\newblock Micro- and Nano-Fabrication.

\bibitem{Utke2012_book}
Ivo Utke, Stanislav Moshkalev, and Phillip Russell, editors.
\newblock {\em Nanofabrication Using Focused Ion and Electron Beams: Principles
  and Applications}.
\newblock Oxford Series in Nanomanufacturing, 2012.

\bibitem{Fowlkes2016_febid_3D_simulation}
Jason~D. Fowlkes, Robert Winkler, Brett~B. Lewis, Michael~G. Stanford, Harald
  Plank, and Philip~D. Rack.
\newblock Simulation-guided {3D} nanomanufacturing via focused electron beam
  induced deposition.
\newblock {\em ACS Nano}, 10(6):6163--6172, 2016.
\newblock PMID: 27284689.

\bibitem{Winkler2017_3D_plasmonic}
Robert Winkler, Franz-Philipp Schmidt, Ulrich Haselmann, Jason~D. Fowlkes,
  Brett~B. Lewis, Gerald Kothleitner, Philip~D. Rack, and Harald Plank.
\newblock Direct-write {3D} nanoprinting of plasmonic structures.
\newblock {\em ACS Applied Materials \& Interfaces}, 9(9):8233--8240, 2017.
\newblock PMID: 28269990.

\bibitem{Donev2009_Pt_liquid_precursor}
Eugenii~U. Donev and J.~Todd Hastings.
\newblock Electron-beam-induced deposition of platinum from a liquid precursor.
\newblock {\em Nano Letters}, 9(7):2715--2718, 2009.
\newblock PMID: 19583284.

\bibitem{Bresin2013_bimetallic_liquid_precursor}
Matthew Bresin, Adam Chamberlain, Eugenii~U. Donev, Chandan~B. Samantaray,
  Gregory~S. Schardien, and J.~Todd Hastings.
\newblock Electron-beam-induced deposition of bimetallic nanostructures from
  bulk liquids.
\newblock {\em Angewandte Chemie}, 125(31):8162--8165, 2013.

\bibitem{Thorman2015_dissociation_case_studies}
Rachel~M. Thorman, Ragesh Kumar~T. P., D.~Howard Fairbrother, and Oddur
  Ing{\'o}lfsson.
\newblock The role of low-energy electrons in focused electron beam induced
  deposition: four case studies of representative precursors.
\newblock {\em Beilstein Journal of Nanotechnology}, 6:1904--1926, 2015.

\bibitem{Muthukumar2011_wco6_adsorption_dft}
Kaliappan Muthukumar, Ingo Opahle, Juan Shen, Harald~O. Jeschke, and Roser
  Valent\'i.
\newblock Interaction of {W(CO)$_6$} with {SiO$_2$} surfaces: A density
  functional study.
\newblock {\em Phys. Rev. B}, 84:205442, Nov 2011.

\bibitem{Muthukumar2012_dissociation_co2co8}
Kaliappan Muthukumar, Harald~O. Jeschke, Roser Valent{\'\i}, Evgeniya Begun,
  Johannes Schwenk, Fabrizio Porrati, and Michael Huth.
\newblock Spontaneous dissociation of {Co$_2$(CO)$_8$} and autocatalytic growth
  of {Co} on {SiO$_2$}: A combined experimental and theoretical investigation.
\newblock {\em Beilstein Journal of Nanotechnology}, 3:546--555, 2012.

\bibitem{Walz2010_ebisa}
Marie-Madeleine Walz, Michael Schirmer, Florian Vollnhals, Thomas Lukasczyk,
  Hans-Peter Steinr{\"u}ck, and Hubertus Marbach.
\newblock Electrons as invisible ink: Fabrication of nanostructures by local
  electron beam induced activation of {SiO$_x$}.
\newblock {\em Angewandte Chemie International Edition}, 49(27):4669--4673,
  2010.

\bibitem{Botman2009_purification_first_review}
A~Botman, J~J~L Mulders, and C~W Hagen.
\newblock Creating pure nanostructures from electron-beam-induced deposition
  using purification techniques: a technology perspective.
\newblock {\em Nanotechnology}, 20(37):372001, 2009.

\bibitem{Roberts2012_purification_pulsed_laser}
Nicholas~A. Roberts, Jason~D. Fowlkes, Gregory~A. Magela, and Philip~D. Rack.
\newblock Enhanced material purity and resolution via synchronized laser
  assisted electron beam induced deposition of platinum.
\newblock {\em Nanoscale}, 5:408, 2012.

\bibitem{Henry2016_gasjet_pure_tungsten}
Matthew~R. Henry, Songkil Kim, and Andrei~G. Fedorov.
\newblock High purity tungsten nanostructures via focused electron beam induced
  deposition with carrier gas assisted supersonic jet delivery of
  organometallic precursors.
\newblock {\em The Journal of Physical Chemistry C}, 120(19):10584--10590,
  2016.

\bibitem{Weirich2013_ga_first_pub}
Paul~M. Weirich, Marcel Winhold, Christian~H. Schwalb, and Michael Huth.
\newblock In situ growth optimization in focused electron-beam induced
  deposition.
\newblock {\em Beilstein Journal of Nanotechnology}, 4:919--926, 2013.

\bibitem{Winhold2014_ga_modeling}
M.~Winhold, P.M. Weirich, C.H. Schwalb, and M.~Huth.
\newblock Modeling the in-situ conductance optimization process in focused
  electron-beam-induced deposition.
\newblock {\em Microelectronic Engineering}, 121:42 -- 46, 2014.

\bibitem{Hussey2004_mott_ioffe_regel}
N.~E. Hussey, K.~Takenaka, and H.~Takagi.
\newblock Universality of the {Mott-Ioffe-Regel} limit in metals.
\newblock {\em Philosophical Magazine A}, 84(27):2847 -- 2864, 2004.

\bibitem{Villamor2015_Pt_purification_direct_O2}
E~Villamor, F~Casanova, P~H~F Trompenaars, and J~J~L Mulders.
\newblock Embedded purification for electron beam induced {Pt} deposition using
  {MeCpPtMe$_3$}.
\newblock {\em Nanotechnology}, 26(9):095303, 2015.

\bibitem{Shawrav2016_Au_with_H2O}
Mostafa~M. Shawrav, Philipp Taus, Heinz~D. Wanzenboeck, M.~Schinnerl,
  M.~St{\"o}ger-Pollach, S.~Schwarz, A.~Steiger-Thirsfeld, and Emmerich
  Bertagnolli.
\newblock Highly conductive and pure gold nanostructures grown by electron beam
  induced deposition.
\newblock {\em Scientific Reports}, 6:34003, 09 2016.

\bibitem{Lavrijsen2011_Fe_febid_pure}
R~Lavrijsen, R~C{\'o}rdoba, F~J Schoenaker, T~H Ellis, B~Barcones, J~T
  Kohlhepp, H~J~M Swagten, B~Koopmans, J~M~De Teresa, C~Mag{\'e}n, M~R Ibarra,
  P~Trompenaars, and J~J~L Mulders.
\newblock {Fe:O:C} grown by focused-electron-beam-induced deposition: magnetic
  and electric properties.
\newblock {\em Nanotechnology}, 22(2):025302, 2011.

\bibitem{Cordoba2016_Fe_nonacarbonyl}
R.~C\'ordoba, D.-S. Han, and B.~Koopmans.
\newblock Manipulating the switching in modulated iron nanowires grown by
  focused electron beam induced deposition.
\newblock {\em Microelectronic Engineering}, 153:60 -- 65, 2016.
\newblock Micro- and Nanofabrication 2015.

\bibitem{Ramon2011_Co_febid_clean_highres}
Luis Serrano-Ram\'on, Rosa C\'ordoba, Luis~Alfredo Rodr\'iguez, C\'esar
  Mag\'en, Etienne Snoeck, Christophe Gatel, In\'es Serrano, Manuel~Ricardo
  Ibarra, and Jos\'e~Mar\'ia De~Teresa.
\newblock Ultrasmall functional ferromagnetic nanostructures grown by focused
  electron-beam-induced deposition.
\newblock {\em ACS Nano}, 5(10):7781--7787, 2011.
\newblock PMID: 21939205.

\bibitem{Hoeflich2017_Ag_metal}
Katja H\"oflich, Jakub Jurczyk, Yucheng Zhang, Marcos~V. Puydinger~dos Santos,
  Maximilian G\"otz, Carlos Guerra-Nu\~nez, James~P. Best, Czeslaw Kapusta, and
  Ivo Utke.
\newblock Direct electron beam writing of silver-based nanostructures.
\newblock {\em ACS Applied Materials \& Interfaces}, 9(28):24071--24077, 2017.
\newblock PMID: 28631921.

\bibitem{Mulders2014_review_purification}
J.~J.~L. Mulders.
\newblock Purity and resistivity improvements for electron-beam-induced
  deposition of {Pt}.
\newblock {\em Applied Physics A}, 117(4):1697 -- 1704, 2014.

\bibitem{Mehendale2015_Au_purification_O2}
S.~Mehendale, J.~H.~L. Mulders, and P.~H.~F. Trompenaars.
\newblock Purification of {Au} {EBID} structures by electron beam
  post-irradiation under oxygen flux at room temperature.
\newblock {\em Microelectronic Engineering}, 141:207 -- 210, 2015.

\bibitem{Roberts2013_laebid_tungsten}
Nicholas~A Roberts, Carlos~M Gonzalez, Jason~D Fowlkes, and Philip~D Rack.
\newblock Enhanced by-product desorption via laser assisted electron beam
  induced deposition of {W(CO)$_6$} with improved conductivity and resolution.
\newblock {\em Nanotechnology}, 24(41):415301, 2013.

\bibitem{Porrati2011_Fe_nanowires_ebisa}
F~Porrati, R~Sachser, M-M Walz, F~Vollnhals, H-P Steinr{\"u}ck, H~Marbach, and
  M~Huth.
\newblock Magnetotransport properties of iron microwires fabricated by focused
  electron beam induced autocatalytic growth.
\newblock {\em Journal of Physics D: Applied Physics}, 44(42):425001, 2011.

\bibitem{Vollnhals2013_ebisa_Fe}
Florian Vollnhals, Tom Woolcot, Marie-Madeleine Walz, Steffen Seiler,
  Hans-Peter Steinr{\"u}ck, Geoff Thornton, and Hubertus Marbach.
\newblock Electron beam-induced writing of nanoscale iron wires on a functional
  metal oxide.
\newblock {\em The Journal of Physical Chemistry C}, 117(34):17674--17679,
  2013.
\newblock PMID: 24159366.

\bibitem{Porrati2011_PtC_irradiation}
F.~Porrati, R.~Sachser, C.~H. Schwalb, A.~S. Frangakis, and M.~Huth.
\newblock Tuning the electrical conductivity of {Pt}-containing granular metals
  by postgrowth electron irradiation.
\newblock {\em Journal of Applied Physics}, 109(6):063715, 2011.

\bibitem{Plank2011_PtC_irradiation}
Harald Plank, Gerald Kothleitner, Ferdinand Hofer, Stephan~G. Michelitsch,
  Christian Gspan, Andreas Hohenau, and Joachim Krenn.
\newblock Optimization of postgrowth electron-beam curing for focused
  electron-beam-induced {Pt} deposits.
\newblock {\em Journal of Vacuum Science \& Technology B, Nanotechnology and
  Microelectronics: Materials, Processing, Measurement, and Phenomena},
  29(5):051801, 2011.

\bibitem{Sachser2011_PtC_universal_conductance}
Roland Sachser, Fabrizio Porrati, Christian~H. Schwalb, and Michael Huth.
\newblock Universal conductance correction in a tunable strongly coupled
  nanogranular metal.
\newblock {\em Phys. Rev. Lett.}, 107:206803, Nov 2011.

\bibitem{Mehendale2013_PtC_purification_O2}
S~Mehendale, J~J~L Mulders, and P~H~F Trompenaars.
\newblock A new sequential {EBID} process for the creation of pure {Pt}
  structures from {MeCpPtMe$_3$}.
\newblock {\em Nanotechnology}, 24(14):145303, 2013.

\bibitem{Plank2014_PtC_purification_O2}
Harald Plank, Joo~Hyon Noh, Jason~D. Fowlkes, Kevin Lester, Brett~B. Lewis, and
  Philip~D. Rack.
\newblock Electron-beam-assisted oxygen purification at low temperatures for
  electron-beam-induced {Pt} deposits: Towards pure and high-fidelity
  nanostructures.
\newblock {\em ACS Applied Materials \& Interfaces}, 6(2):1018--1024, 2014.
\newblock PMID: 24377304.

\bibitem{Lewis2015_purification_pt_oxygen_modeling}
Brett~B. Lewis, Michael~G. Stanford, Jason~D. Fowlkes, Kevin Lester, Harald
  Plank, and Philip~D. Rack.
\newblock Electron-stimulated purification of platinum nanostructures grown via
  focused electron beam induced deposition.
\newblock {\em Beilstein Journal of Nanotechnology}, 6:907--918, 2015.

\bibitem{Geier2014_Pt_purification_H2O}
Barbara Geier, Christian Gspan, Robert Winkler, Roland Schmied, Jason~D.
  Fowlkes, Harald Fitzek, Sebastian Rauch, Johannes Rattenberger, Philip~D.
  Rack, and Harald Plank.
\newblock Rapid and highly compact purification for focused electron beam
  induced deposits: A low temperature approach using electron stimulated
  {H$_2$O} reactions.
\newblock {\em The Journal of Physical Chemistry C}, 118(25):14009--14016,
  2014.

\bibitem{Begun2015_Co_purification_H2}
E~Begun, O~V Dobrovolskiy, M~Kompaniiets, R~Sachser, Ch~Gspan, H~Plank, and
  M~Huth.
\newblock Post-growth purification of {Co} nanostructures prepared by focused
  electron beam induced deposition.
\newblock {\em Nanotechnology}, 26(7):075301, 2015.

\bibitem{Sachser2014_PtC_purification_pulsed_O2}
Roland Sachser, Heiko Reith, Daniel Huzel, Marcel Winhold, and Michael Huth.
\newblock Catalytic purification of directly written nanostructured {Pt}
  microelectrodes.
\newblock {\em ACS Applied Materials \& Interfaces}, 6(18):15868--15874, 2014.
\newblock PMID: 25111450.

\bibitem{Mackus2010_Pt_asald_first}
A.~J.~M. Mackus, J.~J.~L. Mulders, M.~C.~M. van~de Sanden, and W.~M.~M.
  Kessels.
\newblock Local deposition of high-purity {Pt} nanostructures by combining
  electron beam induced deposition and atomic layer deposition.
\newblock {\em Journal of Applied Physics}, 107(11):116102, 2010.

\bibitem{Mackus2012_Pt_asad_resolution}
A.~J.~M. Mackus, S.~A.~F. Dielissen, J.~J.~L. Mulders, and W.~M.~M. Kessels.
\newblock Nanopatterning by direct-write atomic layer deposition.
\newblock {\em Nanoscale}, 4:4477--4480, 2012.

\bibitem{Mackus2013_Pt_asald_highres}
A.~J.~M. Mackus, N.~F.~W. Thissen, J.~J.~L. Mulders, P.~H.~F. Trompenaars,
  M.~A. Verheijen, A.~A. Bol, and W.~M.~M. Kessels.
\newblock Direct-write atomic layer deposition of high-quality {Pt}
  nanostructures: Selective growth conditions and seed layer requirements.
\newblock {\em The Journal of Physical Chemistry C}, 117(20):10788--10798,
  2013.

\bibitem{Altonen2003_Pt_ald}
Titta Aaltonen, Mikko Ritala, Timo Sajavaara, Juhani Keinonen, and Markku
  Leskel{\"a}.
\newblock Atomic layer deposition of platinum thin films.
\newblock {\em Chemistry of Materials}, 15(9):1924--1928, 2003.

\bibitem{Diprima2017_asald_monitoring}
Giorgia Di~Prima, Roland Sachser, Peter Gruszka, Marc Hanefeld, Thomas
  Halbritter, Alexander Heckel, and Michael Huth.
\newblock In-situ conductance monitoring of pt thin film growth by
  area-selective atomic layer deposition.
\newblock {\em Nanotechnology (submitted)}, 2017.

\bibitem{Sadki2004_SC_WGa_FIB_first}
E.~S. Sadki, S.~Ooi, and K.~Hirata.
\newblock Focused-ion-beam-induced deposition of superconducting nanowires.
\newblock {\em Applied Physics Letters}, 85(25):6206--6208, 2004.

\bibitem{Li2008_SC_WGa_tuning}
Wuxia Li, J.~C. Fenton, Yiqian Wang, D.~W. McComb, and P.~A. Warburton.
\newblock Tunability of the superconductivity of tungsten films grown by
  focused-ion-beam direct writing.
\newblock {\em Journal of Applied Physics}, 104(9):093913, 2008.

\bibitem{Guillamon2008_tunneling_WGa}
I.~Guillam\'on, H.~Suderow, S.~Vieira, A.~Fern\'andez-Pacheco, J.~Ses\'e,
  R.~C\'ordoba, J.~M.~De Teresa, and M.~R. Ibarra.
\newblock Nanoscale superconducting properties of amorphous {W}-based deposits
  grown with a focused-ion-beam.
\newblock {\em New Journal of Physics}, 10(9):093005, 2008.

\bibitem{Guillamon2014_vortex_matter_WGa}
I.~Guillam\'on, H.~Suderow, P.~Kulkarni, S.~Vieira, R.~C\'ordoba, J.~Ses\'e,
  J.M.~De Teresa, M.R. Ibarra, Gorky Shaw, and S.S. Bannerjee.
\newblock Nanostructuring superconducting vortex matter with focused ion beams.
\newblock {\em Physica C: Superconductivity and its Applications}, 503:70 --
  74, 2014.

\bibitem{Sangiao2011_SC_WGa_andreev_reflection}
S.~Sangiao, L.~Morell\'on, M.R. Ibarra, and J.M.~De Teresa.
\newblock Ferromagnet--superconductor nanocontacts grown by focused
  electron/ion beam techniques for current-in-plane {Andreev} reflection
  measurements.
\newblock {\em Solid State Communications}, 151(1):37 -- 41, 2011.

\bibitem{Wang2010_odd_freq_sc_proximity_nanowire_first}
Jian Wang, Meenakshi Singh, Mingliang Tian, Nitesh Kumar, Bangzhi Liu, Chuntai
  Shi, J.~K. Jain, Nitinand Samarth, T.~E. Mallouk, and M.~H.~W. Chan.
\newblock Interplay between superconductivity and ferromagnetism in crystalline
  nanowires.
\newblock {\em Nature Physics}, 6:389, 2010.

\bibitem{Kompaniiets2014_proximity_triplet_sc1}
M.~Kompaniiets, O.~V. Dobrovolskiy, C.~Neetzel, F.~Porrati, J.~Br{\"o}tz,
  W.~Ensinger, and M.~Huth.
\newblock Long-range superconducting proximity effect in polycrystalline {Co}
  nanowires.
\newblock {\em Applied Physics Letters}, 104(5):052603, 2014.

\bibitem{Weirich2014_SC_MoGaCO}
P.~M. Weirich, C.~H. Schwalb, M.~Winhold, and M.~Huth.
\newblock Superconductivity in the system {Mo$_x$C$_y$Ga$_z$O$_\delta$}
  prepared by focused ion beam induced deposition.
\newblock {\em Journal of Applied Physics}, 115(17):174315, 2014.

\bibitem{Dhakal2010_SC_CGaO}
Pashupati Dhakal, G.~McMahon, S.~Shepard, T.~Kirkpatrick, J.~I. Oh, and M.~J.
  Naughton.
\newblock Direct-write, focused ion beam-deposited, 7 {K} superconducting
  {C-Ga-O} nanowire.
\newblock {\em Applied Physics Letters}, 96(26):262511, 2010.

\bibitem{Sadovskii1997_sc_in_disordered_systems}
Michael~V. Sadovskii.
\newblock Superconductivity in disordered systems.
\newblock {\em Physics Reports}, 282:225, 1997.

\bibitem{Feigelman2007_pseudogap_sc_ins_transition}
M.~V. Feigel'man, L.~B. Ioffe, V.~E. Kravtsov, and E.~A. Yuzbashyan.
\newblock Eigenfunction fractality and pseudogap state near the
  superconductor-insulator transition.
\newblock {\em Phys. Rev. Lett.}, 98:027001, Jan 2007.

\bibitem{Gantmakher2010_sc_ins_quantum_phase_transition}
Vsevolod~F Gantmakher and Valery~T Dolgopolov.
\newblock Superconductor--insulator quantum phase transition.
\newblock {\em Physics-Uspekhi}, 53(1):1, 2010.

\bibitem{Felner2012_SC_WCO_S_doped}
I.~Felner, O.~Wolf, and O.~Millo.
\newblock High-temperature superconductivity in sulfur-doped amorphous carbon
  systems.
\newblock {\em Journal of Superconductivity and Novel Magnetism}, 25(1):7--10,
  Jan 2012.

\bibitem{Porrati2017_SC_W_FEBID_Ga_doping}
F~Porrati, L~Keller, C~Gspan, H~Plank, and M~Huth.
\newblock Electrical transport properties of {Ga} irradiated {W}-based granular
  nanostructures.
\newblock {\em Journal of Physics D: Applied Physics}, 50(21):215301, 2017.

\bibitem{Belkin2015_qps_sc_nanowires}
A.~Belkin, M.~Belkin, V.~Vakaryuk, S.~Khlebnikov, and A.~Bezryadin.
\newblock Formation of quantum phase slip pairs in superconducting nanowires.
\newblock {\em Phys. Rev. X}, 5:021023, Jun 2015.

\bibitem{Bergeret2005_odd_freq_triplet_pairing}
F.~S. Bergeret, A.~F. Volkov, and K.~B. Efetov.
\newblock Odd triplet superconductivity and related phenomena in
  superconductor-ferromagnet structures.
\newblock {\em Rev. Mod. Phys.}, 77:1321--1373, Nov 2005.

\bibitem{Natarajan2012_nanowire_sc_bolometers}
Chandra~M Natarajan, Michael~G Tanner, and Robert~H Hadfield.
\newblock Superconducting nanowire single-photon detectors: physics and
  applications.
\newblock {\em Superconductor Science and Technology}, 25(6):063001, 2012.

\bibitem{Keller2014_master_thesis}
Lukas Keller.
\newblock {\em Characterization of cooperative transport phenomena in the
  inhomogeneous systems FeSi and WGa prepared by focused electron beam induced
  deposition}.
\newblock Master thesis, Goethe-University, Frankfurt, 2014.

\bibitem{Kumar2017_FeRu_dissociation_study}
Ragesh T.~P. Kumar, Paul Weirich, Lukas Hrachowina, Marc Hanefeld, Ragnar
  Bjornsson, Helgi~Rafn Hrodmarsson, Sven Barth, D.~Howard Fairbrother, Michael
  Huth, and Oddur Ing\'olfsson.
\newblock Electron interactions with the heteronuclear carbonyl precursor
  {H$_2$FeRu$_3$(CO)$_{13}$}: from fundamental gas phase and surface science
  studies to focused electron beam induced deposition.
\newblock {\em Beilstein Journal of Nanotechnology (submitted)}, 2017.

\bibitem{Keller2017_magnetic_3D}
Lukas Keller, Mohanad K.~I. Al~Mamoori, Jonathan Pieper, Christian Gspan, Irina
  Stockem, Christian Schr\"oder, Sven Barth, Robert Winkler, Harald Plank,
  Merlin Pohlit, Jens M\"uller, and Michael Huth.
\newblock Direct-write of free-form {3D} nanostructures with controlled
  magnetic frustration.
\newblock {\em Nature Communications}, 2017 (submitted).

\bibitem{Beloborodov2007_granular_electronic_systems}
I.~S. Beloborodov, A.~V. Lopatin, V.~M. Vinokur, and K.~B. Efetov.
\newblock Granular electronic systems.
\newblock {\em Rev. Mod. Phys.}, 79:469--518, Apr 2007.

\bibitem{Efetov2003_coulomb_effects_granular_metals}
K.~B. Efetov and A.~Tschersich.
\newblock Coulomb effects in granular materials at not very low temperatures.
\newblock {\em Phys. Rev. B}, 67:174205, May 2003.

\bibitem{Beloborodov2003_granular_metals_strong_coupling}
I.~S. Beloborodov, K.~B. Efetov, A.~V. Lopatin, and V.~M. Vinokur.
\newblock Transport properties of granular metals at low temperatures.
\newblock {\em Phys. Rev. Lett.}, 91:246801, Dec 2003.

\bibitem{Beloborodov2004_granular_fermi_liquid}
I.~S. Beloborodov, A.~V. Lopatin, and V.~M. Vinokur.
\newblock Universal description of granular metals at low temperatures:
  Granular {Fermi} liquid.
\newblock {\em Phys. Rev. B}, 70:205120, Nov 2004.

\bibitem{Porrati2014_PtC_magnetoresistance}
F~Porrati, R~Sachser, and M~Huth.
\newblock Magnetoresistance of granular {Pt-C} nanostructures close to the
  metal-insulator transition.
\newblock {\em Journal of Physics: Condensed Matter}, 26(8):085302, 2014.

\bibitem{Porrati2014_PtC_coupling_Co_nanopillars}
F~Porrati, E~Begun, R~Sachser, and M~Huth.
\newblock Spin-dependent transport between magnetic nanopillars through a
  nano-granular metal matrix.
\newblock {\em Journal of Physics D: Applied Physics}, 47(49):495001, 2014.

\bibitem{Diehl2015_pseudogap_kappaET}
Sandra Diehl, Torsten Methfessel, Ulrich Tutsch, Jens M{\"u}ller, Michael Lang,
  Michael Huth, Martin Jourdan, and Hans-Joachim Elmers.
\newblock Disorder-induced gap in the normal density of states of the organic
  superconductor {$\kappa$-(BEDT-TTF)$_2$Cu[N(CN)$_2$]Br}.
\newblock {\em Journal of Physics: Condensed Matter}, 27(26):265601, 2015.

\bibitem{Guterding2016_sc_order_parameter_kET}
Daniel Guterding, Sandra Diehl, Michaela Altmeyer, Torsten Methfessel, Ulrich
  Tutsch, Harald Schubert, Michael Lang, Jens M\"uller, Michael Huth, Harald~O.
  Jeschke, Roser Valent\'{\i}, Martin Jourdan, and Hans-Joachim Elmers.
\newblock Evidence for eight-node mixed-symmetry superconductivity in a
  correlated organic metal.
\newblock {\em Phys. Rev. Lett.}, 116:237001, Jun 2016.

\bibitem{Pan2001_STS_underdoped_htsc}
S.~H. Pan, J.~P. O'Neal, R.~L. Badzey, C.~Chamon, H.~Ding, J.~R. Engelbrecht,
  Z.~Wang, H.~Eisaki, S.~Uchida, A.~K. Gupta, K.~W. Ng, E.~W. Hudson, K.~M.
  Lang, and J.~C. Davis.
\newblock Microscopic electronic inhomogeneity in the high-tc superconductor
  {Bi$_2$Sr$_2$CaCu$_2$O$_{8+x}$}.
\newblock {\em Nature}, 413(6853):282--285, 09 2001.

\bibitem{Dubi2007_sit_htsc_disordered}
Yonatan Dubi, Yigal Meir, and Yshai Avishai.
\newblock Nature of the superconductor-insulator transition in disordered
  superconductors.
\newblock {\em Nature}, 449(7164):876--880, 10 2007.

\bibitem{Baturina2007_localized_sc_TiN}
T.~I. Baturina, A.~Yu. Mironov, V.~M. Vinokur, M.~R. Baklanov, and C.~Strunk.
\newblock Localized superconductivity in the quantum-critical region of the
  disorder-driven superconductor-insulator transition in tin thin films.
\newblock {\em Phys. Rev. Lett.}, 99:257003, Dec 2007.

\bibitem{Sachser2009_2D_nanodot_lattice_transport}
Roland Sachser, Fabrizio Porrati, and Michael Huth.
\newblock Hard energy gap and current-path switching in ordered two-dimensional
  nanodot arrays prepared by focused electron-beam-induced deposition.
\newblock {\em Phys. Rev. B}, 80:195416, Nov 2009.

\bibitem{Porrati2010_2D_nanodot_lattice_fabrication}
F~Porrati, R~Sachser, M~Strauss, I~Andrusenko, T~Gorelik, U~Kolb,
  L~Bayarjargal, B~Winkler, and M~Huth.
\newblock Artificial granularity in two-dimensional arrays of nanodots
  fabricated by focused-electron-beam-induced deposition.
\newblock {\em Nanotechnology}, 21(37):375302, 2010.

\bibitem{Buzdin2005_review_fm_proximity}
A.~I. Buzdin.
\newblock Proximity effects in superconductor-ferromagnet heterostructures.
\newblock {\em Rev. Mod. Phys.}, 77:935--976, Sep 2005.

\bibitem{Eschrig2008_triplet_supercurrents}
Matthias Eschrig and Tomas L\"ofwander.
\newblock Triplet supercurrents in clean and disordered half-metallic
  ferromagnets.
\newblock {\em Nature Phys.}, 4(2):138--143, Jan 2008.

\bibitem{Eschrig2015_spin_polarized_supercurrents}
Matthias Eschrig.
\newblock Spin-polarized supercurrents for spintronics: a review of current
  progress.
\newblock {\em Rep. Prog. Phys.}, 78(10):104501, 2015.

\bibitem{Bergeret2001_long_range_proximity_fm}
F.~Bergeret, A.~Volkov, and K.~Efetov.
\newblock Long-range proximity effects in superconductor-ferromagnet
  structures.
\newblock {\em Phys. Rev. Lett.}, 86(18):4096--4099, Apr 2001.

\bibitem{Sangiao2011_andreev_magnetic_field_sc_fm_febid}
S.~Sangiao, J.~M. De~Teresa, M.~R. Ibarra, I.~Guillam\'on, H.~Suderow,
  S.~Vieira, and L.~Morell\'on.
\newblock Andreev reflection under high magnetic fields in
  ferromagnet-superconductor nanocontacts.
\newblock {\em Phys. Rev. B}, 84:233402, Dec 2011.

\bibitem{Sharma2014_multi_channel_andreev}
N.~Sharma, P.~Vugts, C.~Daniels, W.~Keuning, J.~T. Kohlhepp, O.~Kurnosikov, and
  B.~Koopmans.
\newblock Multi-channel {Andreev} reflection in {Co-W} nanocontacts fabricated
  using focused electron/ion beam induced deposition.
\newblock {\em Nanotechnol.}, 25(49):495201, 2014.

\bibitem{Kompaniiets2014_Cu_Co_proximity}
M.~Kompaniiets, O.~V. Dobrovolskiy, C.~Neetzel, E.~Begun, F.~Porrati,
  W.~Ensinger, and M.~Huth.
\newblock Proximity-induced superconductivity in crystalline {Cu} and {Co}
  nanowires and nanogranular {Co} structures.
\newblock {\em J. Appl. Phys.}, 116(7):073906--1--10, 2014.

\bibitem{Abrikosov1957_flux_lattice}
A.~A. Abrikosov.
\newblock On the magnetic properties of superconductors of the second group.
\newblock {\em Sov. Phys. JETP.}, 5:1174--1182, 1957.

\bibitem{Brandt1995_flux_line_lattice}
E.~H. Brandt.
\newblock The flux-line lattice in superconductors.
\newblock {\em Rep. Progr. Phys.}, 58(11):1465--1594, 1995.

\bibitem{Moshchalkov2010_sc_nanoscience}
V.~V. Moshchalkov, R.~W\"ordenweber, and M.~Lang, editors.
\newblock {\em Nanoscience and Engineering in Superconductivity}.
\newblock Springer-Verlag, Berlin Heidelberg, 2010.

\bibitem{Dobrovolskiy2017_fluxonics_review}
O.~V. Dobrovolskiy.
\newblock Abrikosov fluxonics in washboard nanolandscapes.
\newblock {\em Physica C}, 533:80--90, 2017.

\bibitem{Woerdenweber2017_sc_at_nanoscale}
R.~W\"ordenweber, V.~Moshchalkov, S.~Bending, and F.~Tafuri, editors.
\newblock {\em Superconductors at the Nanoscale: From Basic Research to
  Applications}.
\newblock Walter De Gruyter Inc., Berlin, 2017.

\bibitem{Velez2008_vortex_pinning}
M.~Velez, J.~I. Martin, J.~E. Villegas, A.~Hoffmann, E.~M. Gonzalez, J.~L.
  Vicent, and I.~K. Schuller.
\newblock Superconducting vortex pinning with artificial magnetic
  nanostructures.
\newblock {\em J. Magn. Magnet. Mat.}, 320(21):2547--2562, 2008.

\bibitem{Dobrovolskiy2010_guiding_Co_decoration}
O.~V. Dobrovolskiy, M.~Huth, and V.~A. Shklovskij.
\newblock Anisotropic magnetoresistive response in thin {Nb} films decorated by
  an array of {Co} stripes.
\newblock {\em Supercond. Sci. Technol.}, 23(12):125014--1--5, 2010.

\bibitem{Dobrovolskiy2011_washboard_pinning}
Oleksandr~V. Dobrovolskiy, Evgeniya Begun, Michael Huth, Valerij~A. Shklovskij,
  and Menachem~I. Tsindlekht.
\newblock Vortex lattice matching effects in a washboard pinning potential
  induced by {Co} nanostripe arrays.
\newblock {\em Physica C}, 471(15-16):449--452, 2011.

\bibitem{Dobrovolskiy2011_washboard_fabrication}
Oleksandr~V. Dobrovolskiy, Michael Huth, and Valerij~A. Shklovskij.
\newblock Fabrication of artificial washboard pinning structures in thin
  niobium films.
\newblock {\em J. Supercond. Nov. Magnet.}, 24:375--380, 2011.

\bibitem{Bedanta2007_superferromagnetism_evidence}
S.~Bedanta, T.~Eim\"uller, W.~Kleemann, J.~Rhensius, F.~Stromberg,
  E.~Amaladass, S.~Cardoso, and P.~P. Freitas.
\newblock Overcoming the dipolar disorder in dense {CoFe} nanoparticle
  ensembles: Superferromagnetism.
\newblock {\em Phys. Rev. Lett.}, 98:176601, Apr 2007.

\bibitem{Morup2010_review_interacting_magnetic_nanoparticles}
Steen M{\o}rup, Mikkel~Fougt Hansen, and Cathrine Frandsen.
\newblock Magnetic interactions between nanoparticles.
\newblock {\em Beilstein Journal of Nanotechnology}, 1:182--190, 2010.

\bibitem{Lara2014_Co_discs_half_antivortex}
A.~Lara, O.~V. Dobrovolskiy, J.~L. Prieto, M.~Huth, and F.~G. Aliev.
\newblock Magnetization reversal assisted by half antivortex states in
  nanostructured circular cobalt disks.
\newblock {\em Appl. Phys. Lett.}, 105(18):182402, 2014.

\bibitem{Kopnin2013_vortex_low_dim_proximity}
N.~B. Kopnin, I.~M. Khaymovich, and A.~S. Melnikov.
\newblock Vortex matter in low-dimensional systems with proximity-induced
  superconductivity.
\newblock {\em J. Exp. Theor. Phys.}, 117:418--438, 2013.

\bibitem{Thurmer2010_nanomembrane_sc_junctions}
D.~J. Thurmer, C.~C.~B. Bufon, Ch. Deneke, and O.~G. Schmidt.
\newblock Nanomembrane-based mesoscopic superconducting hybrid junctions.
\newblock {\em Nano Lett.}, 10(9):3704--3709, 2010.
\newblock PMID: 20687521.

\bibitem{Fomin2012_correlated_vortex_tuning}
V.~M Fomin, R.~O. Rezaev, and O.~G. Schmidt.
\newblock Tunable generation of correlated vortices in open superconductor
  tubes.
\newblock {\em Nano Lett.}, 12(3):1282--1287, 2012.

\bibitem{Plourde2009_pinning_anisotropic}
B.~L.~T. Plourde.
\newblock Nanostructured superconductors with asymmetric pinning potentials:
  Vortex ratchets.
\newblock {\em IEEE Trans. Appl. Supercond.}, 19:3698--3714, Oct 2009.

\bibitem{Shklovskij2014_vortex_ratchet_asymmetric}
V.~A. Shklovskij, V.~V. Sosedkin, and O.~V. Dobrovolskiy.
\newblock Vortex ratchet reversal in an asymmetric washboard pinning potential
  subject to combined dc and ac stimuli.
\newblock {\em J. Phys.: Cond. Matt.}, 26(2):025703, 2014.

\bibitem{Dobrovolskiy2015_dual_cutoff_filter}
Oleksandr~V. Dobrovolskiy and Michael Huth.
\newblock Dual cut-off direct current-tunable microwave low-pass filter on
  superconducting {Nb} microstrips with asymmetric nanogrooves.
\newblock {\em Appl. Phys. Lett.}, 106(14):142601--1--5, 2015.

\bibitem{Dobrovolskiy2015_ac_microwave_loss_modulation}
Oleksandr~V. Dobrovolskiy, Michael Huth, and Valerij~A. Shklovskij.
\newblock Alternating current-driven microwave loss modulation in a fluxonic
  metamaterial.
\newblock {\em Appl. Phys. Lett.}, 107(16):162603--1--5, 2015.

\bibitem{Cordoba2016_Fe_Co_3D_nanopillars}
Rosa Cord\'oba, Nidhi Sharma, Sebastian K\"olling, Paul~M Koenraad, and Bert
  Koopmans.
\newblock High-purity {3D} nano-objects grown by focused-electron-beam induced
  deposition.
\newblock {\em Nanotechnology}, 27(35):355301, 2016.

\bibitem{Huth2010_theory_strain_sensing}
Michael Huth.
\newblock Granular metals: From electronic correlations to strain-sensing
  applications.
\newblock {\em Journal of Applied Physics}, 107(11):113709, 2010.

\bibitem{Beloborodov2004_dos_weak_coupling}
I.~S. Beloborodov, A.~V. Lopatin, G.~Schwiete, and V.~M. Vinokur.
\newblock Tunneling density of states of granular metals.
\newblock {\em Phys. Rev. B}, 70:073404, Aug 2004.

\bibitem{Beloborodov2005_dos_close_MIT}
I.~S. Beloborodov, A.~V. Lopatin, and V.~M. Vinokur.
\newblock Coulomb effects and hopping transport in granular metals.
\newblock {\em Phys. Rev. B}, 72:125121, Sep 2005.

\bibitem{Kolb2013_H2O_sensing}
Florian Kolb, Kerstin Schmoltner, Michael Huth, Andreas Hohenau, Joachim Krenn,
  Andreas Klug, Emil J~W List, and Harald Plank.
\newblock Variable tunneling barriers in {FEBID} based {PtC} metal-matrix
  nanocomposites as a transducing element for humidity sensing.
\newblock {\em Nanotechnology}, 24(30):305501, 2013.

\bibitem{Bending1999_principles_local_hall_sensors}
S.~J. Bending.
\newblock Local magnetic probes of superconductors.
\newblock {\em Advances in Physics}, 48:449--535, 1999.

\bibitem{Nagaosa2010_review_anomalous_hall_effect}
N.~Nagaosa, J.~Sinova, S.~Onoda, A.~H. MacDonald, and N.~P. Ong.
\newblock Anomalous {Hall} effect.
\newblock {\em Rev. Mod. Phys.}, 82:1539, 2010.

\bibitem{Denardin2003_review_giant_hall_effect_granular_ferromagnets}
J.~C. Denardin, M.~Knobel, X.~X. Zhang, and A.~B. Pakhomov.
\newblock Giant {Hall} effect in superparamagnetic granular films.
\newblock {\em Journal of Magnetism and Magnetic Materials}, 262, 2003.

\bibitem{Boero2005_CoC_Hall_sensor}
G.~Boero, I.~Utke, T.~Bret, N.~Quack, M.~Todorova, S.~Mouaziz, P.~Kejik,
  J.~Brugger, R.~S. Popovic, and P.~Hoffmann.
\newblock Submicrometer {Hall} devices fabricated by focused
  electron-beam-induced deposition.
\newblock {\em Applied Physics Letters}, 86(4):042503, 2005.

\bibitem{Gabureac2010_granular_hall_sensors}
M.~Gabureac, L.~Bernau, I.~Utke, and G.~Boero.
\newblock Granular {Co-C} nano-{Hall} sensors by focused-beam-induced
  deposition.
\newblock {\em Nanotechnology}, 21:115503, 2010.

\bibitem{Gabureac2013_gfm_magnetic_bead}
M.~S. Gabureac, L.~Bernau, G.~Boero, and I.~Utke.
\newblock Single superparamagnetic bead detection and direct tracing of bead
  position using novel nanocomposite nano-{Hall} sensors.
\newblock {\em IEEE Transactions on Nanotechnology}, 12(5):668--673, Sept 2013.

\bibitem{Sandhu2004_Bi_hall_sensors}
A.~Sandhu, K.~Kurosawa, M.~Dede, and A.~Oral.
\newblock 50 nm hall sensors for room temperature scanning {Hall} probe
  microscopy.
\newblock {\em Japanse Journal of Applied Physics}, 43:777, 2004.

\bibitem{Cordoba2012_GAHE_Fe}
R~C\'ordoba, R~Lavrijsen, A~Fern\'andez-Pacheco, M~R Ibarra, F~Schoenaker,
  T~Ellis, B~Barcones-Campo, J~T Kohlhepp, H~J~M Swagten, B~Koopmans, J~J~L
  Mulders, and J~M~De Teresa.
\newblock Giant anomalous {Hall} effect in {Fe}-based microwires grown by
  focused-electron-beam-induced deposition.
\newblock {\em Journal of Physics D: Applied Physics}, 45(3):035001, 2012.

\bibitem{Dyre2000_ac_conductance_universality}
Jeppe~C. Dyre and Thomas~B. Schr\/oder.
\newblock Universality of ac conduction in disordered solids.
\newblock {\em Rev. Mod. Phys.}, 72:873, 2000.

\bibitem{Bakkali2016_universality_ac_response_granular_metals}
Hicham Bakkali, Manuel Dominguez, Xavier Batlle, and Am{\'\i}lcar Labarta.
\newblock Universality of the electrical transport in granular metals.
\newblock {\em Scientific Reports}, 6:29676 EP --, 07 2016.

\bibitem{Winhold2015_phd_thesis}
Marcel Winhold.
\newblock {\em Focused electron-beam-induced deposition -- From process
  optimization to cantilever-based strain-sensing applications using
  nanogranular tunneling resistors}.
\newblock PhD thesis, Goethe University, Frankfurt am Main, Germany, 2015.

\bibitem{Koops2001_FEBID_photonic_crystals}
H.W.P Koops, O.E Hoinkis, M.E.W Honsberg, R~Schmidt, R~Blum, G~B{\"o}ttger,
  A~Kuligk, C~Liguda, and M~Eich.
\newblock Two-dimensional photonic crystals produced by additive
  nanolithography with electron beam-induced deposition act as filters in the
  infrared.
\newblock {\em Microelectronic Engineering}, 57:995 -- 1001, 2001.
\newblock Micro- and Nano-Engineering 2000.

\bibitem{Kretz1994_field_emitter_structures}
J.~Kretz, M.~Rudolph, M.~Weber, and H.W.P. Koops.
\newblock Three-dimensional structurization by additive lithography, analysis
  of deposits using {TEM} and {EDX}, and application to field-emitter tips.
\newblock {\em Microelectronic Engineering}, 23(1):477 -- 481, 1994.

\bibitem{Koops1995_Pt_tips_FEBID}
H.~W.~P. Koops, A.~Kaya, and M.~Weber.
\newblock Fabrication and characterization of platinum nanocrystalline material
  grown by electron‐beam induced deposition.
\newblock {\em Journal of Vacuum Science \& Technology B: Microelectronics and
  Nanometer Structures Processing, Measurement, and Phenomena},
  13(6):2400--2403, 1995.

\bibitem{Floreani2001_field_emitters_FEBID}
F.~Floreani, H.W. Koops, and W.~Els{\"a}{\ss}er.
\newblock Operation of high power field emitters fabricated with electron beam
  deposition and concept of a miniaturised free electron laser.
\newblock {\em Microelectronic Engineering}, 57:1009 -- 1016, 2001.
\newblock Micro- and Nano-Engineering 2000.

\bibitem{Gazzadi2007_suspended_FEBID_structures}
G~C Gazzadi, S~Frabboni, and C~Menozzi.
\newblock Suspended nanostructures grown by electron beam-induced deposition of
  {Pt} and {TEOS} precursors.
\newblock {\em Nanotechnology}, 18(44):445709, 2007.

\bibitem{Hoeflich2011_plasmonics_Au_3D}
Katja H\"oflich, Ren~Bin Yang, Andreas Berger, Gerd Leuchs, and Silke
  Christiansen.
\newblock The direct writing of plasmonic gold nanostructures by
  electron-beam-induced deposition.
\newblock {\em Advanced Materials}, 23(22-23):2657--2661, 2011.

\bibitem{Esposito2015_helix_FEBID_plasmonics}
Marco Esposito, Vittorianna Tasco, Massimo Cuscun{\`a}, Francesco Todisco,
  Alessio Benedetti, Iolena Tarantini, Milena~De Giorgi, Daniele Sanvitto, and
  Adriana Passaseo.
\newblock Nanoscale {3D} chiral plasmonic helices with circular dichroism at
  visible frequencies.
\newblock {\em ACS Photonics}, 2(1):105--114, 2015.

\bibitem{Pacheco2013_magnetics_Co_3D_spirals}
Amalio Fern{\'a}ndez-Pacheco, Luis Serrano-Ram{\'o}n, Jan~M. Michalik,
  M.~Ricardo Ibarra, Jos{\'e}M. De~Teresa, Liam O'Brien, Doroth{\'e}e Petit,
  Jihyun Lee, and Russell~P. Cowburn.
\newblock Three dimensional magnetic nanowires grown by focused electron-beam
  induced deposition.
\newblock {\em Scientific Reports}, 3:1492 EP --, 03 2013.

\bibitem{Navarro2017_Co_3D_pillars}
Javier Pablo-Navarro, D\'edalo Sanz-Hern\'andez, C\'esar Mag\'en, Amalio
  Fern\'andez-Pacheco, and Jos\'e~Mar\`ia de~Teresa.
\newblock Tuning shape, composition and magnetization of {3D} cobalt nanowires
  grown by focused electron beam induced deposition ({FEBID}).
\newblock {\em Journal of Physics D: Applied Physics}, 50(18):18LT01, 2017.

\bibitem{Fisher2015_nanospray_liquid_precursor_3d_FEBID}
Jeffrey~S. Fisher, Peter~A. Kottke, Songkil Kim, and Andrei~G. Fedorov.
\newblock Rapid electron beam writing of topologically complex {3D}
  nanostructures using liquid phase precursor.
\newblock {\em Nano Letters}, 15(12):8385--8391, 2015.
\newblock PMID: 26561872.

\bibitem{Lewis2017_purification_3D}
Brett~B. Lewis, Robert Winkler, Xiahan Sang, Pushpa~R. Pudasaini, Michael~G.
  Stanford, Harald Plank, Raymond~R. Unocic, Jason~D. Fowlkes, and Philip~D.
  Rack.
\newblock {3D} nanoprinting via laser-assisted electron beam induced
  deposition: growth kinetics, enhanced purity, and electrical resistivity.
\newblock {\em Beilstein Journal of Nanotechnology}, 8:801--812, 2017.

\bibitem{Muthukumar2014_adsorption_several_dft}
Kaliappan Muthukumar, Roser Valent{\'\i}, and Harald~O. Jeschke.
\newblock Dynamics of tungsten hexacarbonyl, dicobalt octacarbonyl, and their
  fragments adsorbed on silica surfaces.
\newblock {\em The Journal of Chemical Physics}, 140(18):184706, 2014.

\bibitem{Muthukumar2012_tungsten_mit_uspex}
Kaliappan Muthukumar, Roser Valent{\'\i}, and Harald~O Jeschke.
\newblock Simulation of structural and electronic properties of amorphous
  tungsten oxycarbides.
\newblock {\em New Journal of Physics}, 14(11):113028, 2012.

\bibitem{Huth2009_WCO_deposits_MIT}
M~Huth, D~Klingenberger, Ch~Grimm, F~Porrati, and R~Sachser.
\newblock Conductance regimes of {W}-based granular metals prepared by electron
  beam induced deposition.
\newblock {\em New Journal of Physics}, 11(3):033032, 2009.

\bibitem{Sushko2016_MD_simulation_WCO6}
Gennady~B. Sushko, Ilia~A. Solov'yov, and Andrey~V. Solov'yov.
\newblock Molecular dynamics for irradiation driven chemistry:application to
  the {FEBID} process.
\newblock {\em The European Physical Journal D}, 70(10):217, 2016.

\bibitem{Joy1991_monte_carlo_sem}
David~C. Joy.
\newblock An introduction to {Monte} {Carlo} simulations.
\newblock {\em Scanning Microscopy}, 5:329 -- 337, 1991.

\bibitem{Toth2015_review_cont_models}
Milos Toth, Charlene Lobo, Vinzenz Friedli, Aleksandra Szkudlarek, and Ivo
  Utke.
\newblock Continuum models of focused electron beam induced processing.
\newblock {\em Beilstein Journal of Nanotechnology}, 6:1518--1540, 2015.

\bibitem{Bishop2012_activated_chemisorption}
James Bishop, Charlene~J. Lobo, Aiden Martin, Mike Ford, Matthew Phillips, and
  Milos Toth.
\newblock Role of activated chemisorption in gas-mediated electron beam induced
  deposition.
\newblock {\em Phys. Rev. Lett.}, 109:146103, Oct 2012.

\bibitem{Smith2007_MC_modeling_febid}
D~A Smith, J~D Fowlkes, and P~D Rack.
\newblock A nanoscale three-dimensional {Monte} {Carlo} simulation of
  electron-beam-induced deposition with gas dynamics.
\newblock {\em Nanotechnology}, 18(26):265308, 2007.

\bibitem{Winkler2014_gas_flux_influence}
Robert Winkler, Jason Fowlkes, Aleksandra Szkudlarek, Ivo Utke, Philip~D. Rack,
  and Harald Plank.
\newblock The nanoscale implications of a molecular gas beam during electron
  beam induced deposition.
\newblock {\em ACS Applied Materials \& Interfaces}, 6(4):2987--2995, 2014.
\newblock PMID: 24502299.

\bibitem{Friedli2009_gas_flux_simulation}
V~Friedli and I~Utke.
\newblock Optimized molecule supply from nozzle-based gas injection systems for
  focused electron- and ion-beam induced deposition and etching: simulation and
  experiment.
\newblock {\em Journal of Physics D: Applied Physics}, 42(12):125305, 2009.

\bibitem{Fowlkes2017_upcoming_cad_tool}
Jason~D. Fowlkes.
\newblock Computer-aided design for {3D} nanoprinting using electrons.
\newblock {\em ACS Applied Materials and Interfaces (to be submitted)}, 2017.

\bibitem{Yang2017_3d_FEBID_and_AFM}
Chen Yang, Robert Winkler, Maja Dukic, Jie Zhao, Harald Plank, and Georg.~E.
  Fantner.
\newblock Probing the morphology and evolving dynamics of {3D} printed
  nanostructures using high-speed atomic force microscopy.
\newblock {\em ACS Applied Materials \& Interfaces}, 9(29):24456--24461, 2017.
\newblock PMID: 28699346.

\bibitem{Pacheco2017_3d_nanomagnetism}
Amalio Fern\'andez-Pacheco, Robert Streubel, Olivier Fruchart, Riccardo Hertel,
  Peter Fischer, and Russell~P. Cowburn.
\newblock Three-dimensional nanomagnetism.
\newblock {\em Nature Communications}, 8:15756 EP --, 06 2017.

\bibitem{Keller2017_pattern_generator}
Lukas Keller and Michael Huth.
\newblock Pattern generation for direct writing threedimensional nanoscale
  structures via focused electron beam induced deposition.
\newblock {\em Microelectronic Engineering (to be submitted)}, 2017.

\end{thebibliography}
\newpage
%
% figures
%
%\begin{figure}
%\centering
%\includegraphics[width=\textwidth]{fig1.pdf}
%\caption{blah.}
%\label{fig1}
%\end{figure}
%
\end{document}